\documentclass[ ]{copernicus2}

\usepackage{url,lineno}
\usepackage{color}
\usepackage[T1]{fontenc}

\begin{document}

\title{Collisionless Magnetic Reconnection in Space Plasmas%
}

\author[1,2]{R. A. Treumann\thanks{Visiting the International Space Science Institute, Bern, Switzerland}}
\author[3]{W. Baumjohann}

\affil[1]{Department of Geophysics and Environmental Sciences, Munich University, Munich, Germany}
\affil[2]{Department of Physics and Astronomy, Dartmouth College, Hanover NH 03755, USA}
\affil[3]{Space Research Institute, Austrian Academy of Sciences, Graz, Austria}

\runningtitle{Collisionless Reconnection }

\runningauthor{R. A. Treumann and W. Baumjohann}

\correspondence{R. A.Treumann\\ (rudolf.treumann@geophysik.uni-muenchen.de)}

\received{ }
\revised{ }
\accepted{ }
\published{ }


\firstpage{1}

\maketitle

\section*{Abstract}
Magnetic reconnection, the merging of oppositely directed magnetic fields that leads to field reconfiguration, plasma heating, jetting and acceleration, is one of the most celebrated processes in collisionless plasmas. It requires the violation of the frozen-in condition which ties gyrating charged particles to the magnetic field inhibiting diffusion. Ongoing reconnection has been identified in near-Earth space as  being responsible for the excitation of substorms, magnetic storms, generation of field aligned currents and their consequences, the wealth of auroral phenomena. Its theoretical understanding is now on the verge of being completed. Reconnection takes place in thin current sheets. Analytical concepts proceeded gradually down to the microscopic scale, the scale of the electron skin depth or inertial length, recognizing that current layers that thin do preferentially undergo spontaneous reconnection. Thick current layers start reconnecting when being forced by plasma inflow to thin. For almost half a century the physical mechanism of reconnection has remained a mystery. Spacecraft in situ observations in combination with sophisticated numerical simulations in two and three dimensions recently clarified the mist, finding that reconnection produces a specific structure of the current layer inside the electron inertial (also called electron diffusion) region around the reconnection site, the X line. Onset of reconnection is attributed to pseudo-viscous contributions of the electron pressure tensor aided by electron inertia and drag, creating a complicated structured electron current sheet, electric fields, and an electron exhaust extended along the current layer. We review the general background theory and  recent developments in numerical simulation on collisionless reconnection. It is impossible to cover the entire field of reconnection in a short space-limited review.  The presentation necessarily remains cursory, determined by our taste, preferences, and knowledge. Only a small amount of observations is included in order to support the few selected numerical simulations. 

\introduction
Magnetic reconnection in plasmas -- sometimes also spelled reconnexion -- was proposed as early as in 1946 as a possibly viable physical process \citep{giovanelli1946} which could trigger the enormous energy release of $\Delta W\sim10^{25}-10^{26}$ J in a single solar flare.  Evidence for the truth of this suggestion has accumulated over the decades \citep[for a general older reference cf., e.g.,][for recent arguments based on observation and theory]{priest2000,cassak2008}. It is also believed that reconnection is responsible for the spectacular disturbances in Earth's magnetosphere, substorms \citep[as for recent evidence cf., e.g.,][]{angelopoulos2008} and magnetic storms, causing aurorae and being a last chain element in hazardous space weather events. More recent observations in the solar wind, at the magnetopause and in the magnetosphere, have been reviewed by \citet{paschmann2008} and \citet{paschmann2013}. Reconnection is considered as being one of the main mechanisms of transforming magnetic energy stored in electric current sheets into kinetic plasma energy, the counterpart of the magnetic dynamo, which may happen allover in astrophysics and also in laboratory plasmas. 

Giovanelli's proposal was revived one decade later \citep[in a 1957 conference talk by][]{sweet1958} and was given its fluid theoretical physical justification by \citet{parker1957} who defined a so-called (stationary) \emph{reconnection rate} as the Alfv\'enic Mach number 
\begin{equation}\label{eq-sweetparker}
\mathcal{M}_{SP}=V_{in}/V_A=d/L_\mathit{SP}=1/\sqrt{S},
\end{equation}
of the (Giovanelli-)Sweet-Parker model as the ratio of velocity $V_{in}$ of plasma inflow into the reconnection region to the Alfv\'enic outflow velocity  $V_A$ from the reconnection region. From material flux conservation this rate equals the ratio of thickness $d$ of the reconnection layer to its length $L_\mathit{SP}$. More generally, a reconnection rate {would be the amount of magnetic flux $\Delta\Phi/\Delta t=(\mathrm{d/d}t)\int |\Delta\mathbf{B}|\cdot\mathrm{d}\mathbf{F}=-\oint\mathbf{E}_\mathit{rec}\cdot\mathrm{d}\mathbf{s}$ that is exchanged in the merging of the oppositely directed magnetic fields $|\Delta\mathbf{B}|=|\mathbf{B}_2- \mathbf{B}_1|$ across the surface of contact $\mathrm{d}\mathbf{F}$ of the fields, with circumference $\mathbf{s}$, within a reconnection time $\Delta t=\tau_\mathit{rec}=V_\mathit{in}/d$. Normalizing to stationary flux, $(\Delta B)L_\mathit{SP}d$, and Alfv\'enic outflow time, $\tau_A=L_\mathit{SP}/V_A$, reduces it to the Mach number $\mathcal{M}_\mathit{SP}$ respectively the ratio $d/L_\mathit{SP}$ which}, in the Sweet-Parker model Eq. (\ref{eq-sweetparker}), {with current density $\mathbf{J}$, is}  expressed as the inverse square root of the Lundqvist number, 
\begin{equation}\label{eq-lund}
S=\mu_0\sigma LV_A\equiv D_A/D_\sigma, \qquad L=L_\mathit{SP}
\end{equation}
which is the dimensionless ratio of the respective Alfv\'en and collisional diffusivities, 
\begin{equation}
D_A= V_A^2\tau_A, \qquad D_\sigma=(\mu_0\sigma)^{-1}.
\end{equation}
In this model the reconnection rate is a constant. $\mathbf{E}_\mathit{rec}$ is the induction electric field equivalently generated in the reconnection process (for further discussion see below). The Alfv\'en speed $V_A=B/\sqrt{\mu_0m_iN}$ is given in terms of magnetic field strength $B$, plasma density $N$, and ion mass $m_i$. $L$ is the length along the magnetic field $\mathbf{B}$ over that reconnection occurs, in this case $L=L_{SP}$, and $\sigma=\epsilon_0\omega_e^2/\nu$ is the \emph{homogeneous} electrical conductivity of the plasma expressed in terms of electron plasma $\omega_e=\sqrt{e^2N/\epsilon_0m_e}$ and binary collision frequencies $\nu$, respectively.  $e$ is elementary charge, and $m_e$ electron mass. $\tau_A=L/V_A$ is the time an Alfv\'en wave needs to travel the distance $L$ along $\mathbf{B}$. In fully ionized plasma the collision frequency $\nu\simeq\omega_e/N\lambda_D^3$ refers to Coulomb collisions between ions and electrons. It is given in terms of the thermal fluctuation level $W_\mathit{th}=\epsilon_0\langle|\mathbf{E}_\mathit{th}(\mathbf{k})|^2\rangle/e\approx NT_e/\lambda_D^3$ of plasma oscillations. $\lambda_D$ is the Debye length. In collisionless plasmas, $\nu\to0$; otherwise, when wave-particle interactions are taken into account, $\nu\to\nu_a$ is replaced by some \emph{anomalous} collision frequency $\nu_a\simeq  \omega_e\left(W_w/NT_e\right)$, with  plasma wave saturation level $W_w=\epsilon_0\langle|\mathbf{E}_w(\mathbf{k})|^2\rangle/2$.  {Here, $\mathbf{E}_\mathit{th}(\mathbf{k}), \mathbf{E}_\mathit{w}(\mathbf{k})$ are the respective spontaneously emitted amplitudes of electric field fluctuations at temperature $T$, and the unstably excited wave spectral amplitudes, both being functions of wavenumber $\mathbf{k}$. The angular brackets indicate averaging with respect to space-time and spectral distribution.} In weakly ionized plasmas, collisions with neutrals, as well as other processes like charge exchange, excitations, recombination etc., must in addition be taken into account.

Since inflow speeds $V_\mathit{in}$ are naturally  small (not at least for the reason of avoiding the formation of shock waves, though there is not any reason to assume that reconnection could \emph{not also} take place inside shocks!) and conductivities $\sigma$ are very high in dilute high temperature plasmas like those expected in solar flares -- theoretically, here, for vanishing collision frequencies, $\sigma\to\infty$ -- making $S$ a very large number, Sweet-Parker reconnection rates turn out very small; moreover, for finite resistivity they lead to elongated reconnection regions. Physically this is so, because all plasma and magnetic flux tubes in Sweet-Parker reconnection have to cross the current layer before flowing out from it along $L=L_{SP}$. Therefore, $L_{SP}$ is necessarily very long, about as long as the entire current sheet. 
 \begin{figure*}[t!]
\centerline{\includegraphics[width=1.0\textwidth,clip=]{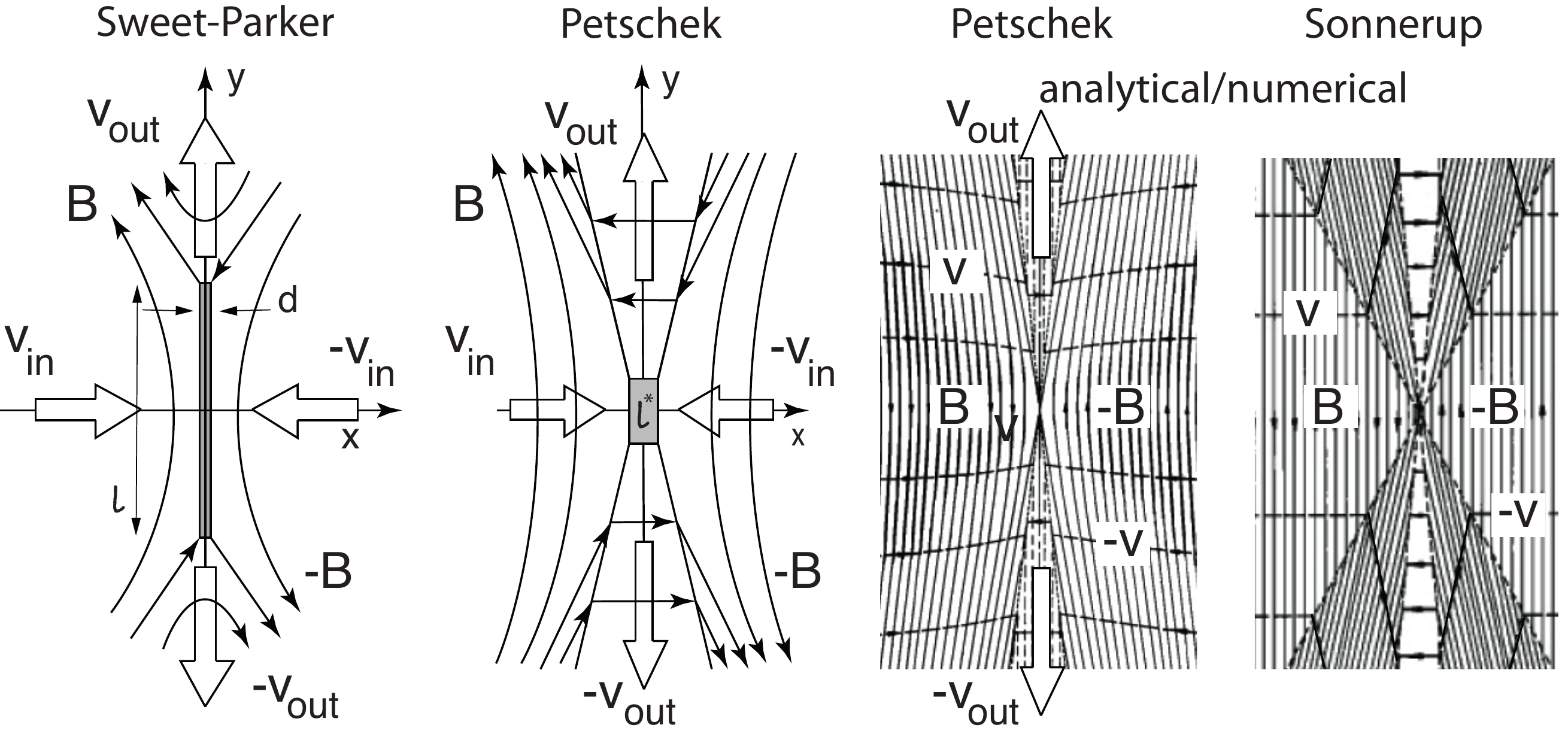} 
}
\caption[]
{The two basic fluid models of reconnection. \emph{Left}: Sweet-Parker's long thin current sheet model with all plasma flow across the sheet. \emph{Centre}: Petschek's small (unspecified) diffusion region/slow shock model. The proposed slow shocks are the four separatirces emanating from the corners of the diffusion region of lengths $\ell^*$ in this figure. \emph{Right}: The semianalytical solutions of Petschek's and Sonnerup's extension of Petschek's model which includes another discontinuity.}\label{fig-SPPS}
\end{figure*}

In order to develop a feeling for the scales, consider two examples: the solar corona and Earth's magnetospheric tail. In the solar corona we have $N\sim 10^{15} - 10^{16}$ m$^{-3}$, $B\sim 10^{-3}-10^{-2}$ T, $T_e\sim 200$ eV, yielding $V_A\approx 10^6$ ms$^{-1}$. Assuming Coulomb collisions and $L\approx 10^7-10^8$ m for the current sheet length, the Lundqvist number becomes $S\sim 10^{12}$, and the Alfv\'en time $\tau_A\approx 10-100$ s. This yields $\mathcal{M}_\mathit{SP}\sim 10^{-6}$. This is very small. Expressed as typical reconnection time one has $\tau_\mathit{rec}\sim\tau_A\sqrt{S}\approx 10^7-10^8$ s, much longer than a typical explosion time of a flare, which is estimated to $\sim 10^3$ s. 

In the magnetotail we have $N\sim 10^{-6}$ m$^{-3}$, $B\sim 10^{-8}$ T, {$0.3<T_e< 1.0$ keV \citep{artemyev2011},} $V_A\sim 2\times10^4$ ms$^{-1}$.  With length $L\sim 6\times10^6$ m and assuming Coulomb collisions, the Lundqvist number becomes $S\sim 10^8-10^9$. Thus $\mathcal{M}_{SP}\approx 10^{-5}-10^{-4}$, as well very small. With these numbers, the typical reconnection time is $\tau_\mathit{rec}\approx 10^5$ s, compared to a substorm growth time of $\sim 600$ s, again far away from any reality.

For these reasons \citet{petschek1964} modified the Sweet-Parker model of reconnection to account for a finite electrical conductivity only in a narrow region of space. This small high-resistive reconnection site locally affects just a narrow flow channel of length $L=L_P$. In order to deviate the flow and fields outside the reconnection site, it becomes the source of four slow shocks \citep[in more sophisticated models like that of][a whole compound of additional discontinuities]{sonnerup1970}  {forming separatrices through which all remaining flow  can pass \emph{outside} the resistive reconnection site \citep[in particular, in the collisionless regime of interest here,][reported the formation of such compounds in collisionless particle-in-cell simulations; cf. the simulation section below]{liu2011a,liu2011b}. During passage of these shocks (or compounds of discontinuities)}, this external flow enters the domain of the fast flow that is ejected from the reconnection site. In this way, the plasma inflow becomes accelerated into a plasma jet without ever having crossed the resistive spot. Accounting for this modification, the square root $\sqrt{S}\to\ln S$ in the Sweet-Parker reconnection rate is replaced with the logarithm of $S$,  yielding substantially higher reconnection rates
\begin{equation}\label{eq-petschek}
\mathcal{M}_P\ \simeq\ \pi/8\ln S.
\end{equation}
Applied to solar flares this yields a reconnection efficiency $\mathcal{M}_{SP}\sim 10^{-2}$ of 1\% and much better reconnection times which are only one order of magnitude too long. In the magnetosphere one would have $\mathcal{M}_{SP}\sim 0.1$ corresponding to a reconnection efficiency of 10\% and grossly not unreasonable reconnection times.

Since Petschek reconnection rates $\mathcal{M}_P$ depend logarithmically on $\sigma$, their range of variation is very narrow. They are about constant, close to being independent of the properties of the interacting flows. \citet{vasyliunas1975} has shown that, in the collisionless case, the maximum Petschek reconnection rate can be expressed more precisely as
\begin{equation}
\mathcal{M}_P\ \lesssim\ \pi/8\ln (\mathcal{M}_PL_P/\lambda_e)\  \lesssim 1.
\end{equation}
In this expression the inertial length $\lambda_e=c/\omega_e$ plays the role of an `inertial resistivity'. Here the rate appears on both sides but can be considered on the right of being practically constant. The Petschek reconnection rate is then determined by the ratio $L/\lambda_e$. Of course, the limitation on $\mathcal{M}_P<1$ implies that $L/\mathrm{e}>\lambda_e/\mathcal{M}_P$ is still substantially larger than the inertial length, also in the Petschek model. This leads to the following scaling of the lengths of the reconnection region for the Sweet-Parker and Petschek models:
\begin{equation}
L_{SP}\ \gg\!\! >\ L_P\ \gg\ \lambda_e.
\end{equation}
Refinements of Petschek's flow model have been provided by \citet{sonnerup1970} and several others who added more discontinuities to the slow shock in order to smooth the flow transition from the inflow to the outflow region (as shown in {Fig. \ref{fig-SPPS}} for Sonnerup's extension). {\citet{erkaev2000}, matching an incompressible steady-state external symmetric Petschek solution to the diffusion region, obtained an expression for the reconnection rate which incorporates both the Sweet-Parker and Petschek regimes. Petschek's model in this case reduces to the limit of a highly localized resistivity.}

The formation of slow shocks during reconnection {was question already early on by} \citet{syrovatskii1966,syrovatskii1967,syrovatskii1971} who argued that, in fluid theory at low resistivity, plasma motion near the X point should prevent the localization of the diffusion region required in Petschek's model. Rather long structured current sheets (of Sweet-Parker type) should form whose length is determined entirely by the external boundary conditions. This conjecture has been confirmed by {\citet{biskamp1984}} in numerical fluid simulations showing the formation of a reverse current near the reconnection site which prevents slow shock formation at small resistivity. Only for highly (presumably anomalously) resistive systems, where large current densities at the X point are avoided, Petschek-like slow shock formation becomes probable \citep{biskamp1994}. \citet{krauss1999} showed that the separatrices do not obey Rankine-Hugoniot conditions of slow shocks.  \citet{yin2007a,yin2007b} have revived the idea of slow shocks as being a consequence of kinetic Alfv\'en waves generated in a plasma by ion-ion beam interaction. Under such conditions these authors found in particle-in-cell simulations that the electron temperature anisotropy caused by acceleration of electrons in the parallel electric field of the kinetic Alfv\'en wave leads to formation of highly oblique slow shocks at low electron $\beta$ conditions which apply outside the reconnection site, a case barely realized in the accessible space plasma. {Recent particle-in-cell simulations have not confirmed the formation of slow shocks either; they instead show the formation of separatrices emanating from the localized reconnection site due to particle effects absent in Petschek's and Sweet-Parker's models producing clusters of discontinuities  \citep{liu2011a,liu2011b}. These latter findings are strongly supported by spacecraft observations in solar wind reconnection exhausts analyzed by \citet{teh2009} to reconstruct the discontinuities, finding no evidence for slow shocks rather than compounds of discontinuities. The transitions from the exhaust to the region outside interpreted as discontinuities turned out to be composites of various kinds of discontinuities all located in the same spatial domain and exhibiting properties of slow and intermediate modes.}

Formally, the reconnection rate may also be represented as ratio $E_\mathit{rec}/E_\perp$ of a reconnection electric field $E_\mathit{rec}\approx j/\sigma$ inside the reconnection ``diffusion'' region to the external convection field $E_\perp \sim V_\mathit{in}B$.
In the Sweet-Parker model with diffusion region width $d$, current density $j\approx B/\mu_0 d$, and formally $E_\mathit{rec}= -V_\mathit{rec}B$, one has  $V_\mathit{rec}\approx (\mu_0d\sigma)^{-1}\approx V_Ad/L_{SP}$. Eq. (\ref{eq-lund}) gives $L_{SP}/d\approx \sqrt{S}$. Thus $V_\mathit{rec}\approx V_A/\sqrt{S}$, yielding for $E_\mathit{rec}\approx E_\|$:
\begin{equation}
E_\|/E_\perp\approx V_\mathit{rec}/V_\mathit{in}\approx V_A/V_\mathit{in}\sqrt{S}=\Big(\mathcal{M}_A\sqrt{S}\Big)^{-1}.
\end{equation}
This field is directed along the current sheet. Since $d/L_{SP}\ll1$ and $d/L_{SP}\approx \tan\theta_{SP}\sim \theta_{SP}$ is the opening angle of the accelerated outflow cone, this field is \emph{almost} parallel to the magnetic field, as indicated by the suffix $\|$, albeit only adjacent to the small diffusion region around the X point where it represents  an \emph{equivalent} parallel electric field $E_\|$ that is generated in the reconnection process by annihilation of some magnetic flux. Its effect is the acceleration of the reconnected plasma along the current sheet away from the reconnection site. In the Petschek model a similar expression holds when replacing $\sqrt{S}\to (8/\pi)\ln(\mathcal{M}_AL_P/\lambda_e)$. There, however, the parallel electric field $E_\|$ is mainly directed along the pair of separatrices emanating from the X point. It eliminates the plasma from this region and causes a narrow density depletion which is typical for a separatrix.  Outside the difusion region only the reconnection field $E_\mathit{rec}$ remains and is by no means anymore parallel to the field; rather it points in the direction of the current and is sustained by the generalized Ohm's law. \citet{boozer2013} recently proved mathematically that magnetic field lines preserve their identity only when the electric potential difference $\Delta U=\int_0^L\mathrm{d}sE_\|(s)\equiv 0$ vanishes identically along a field line, which is another expression for the fact that reconnection, i.e. violation of this condition that causes $\Delta U\neq 0$ and leads to field line reorganization and change of field line identity, can happen only if and if a net parallel electric field $E_\|$ is generated by the physical process of reconnection along the length $L$ of the reconnection site. According to \citet{boozer2013} this violation is most probable in three dimensions which are most sensitive to dependence on initial conditions; it causes field line entanglement. 

Here, an important comment is in place. Both the Sweet-Parker and Petschek models assume implicitly that the reconnection process is self-regulatory in the sense that reconnection \emph{determines} the inflow velocity. This frequently leads to confusion, for it applies only to plasmas at rest. If all quantities, including the length $L$ of the reconnection region, are prescribed, with the exception of $V_{in}$, then the reconnection rate does indeed fix $V_\emph{in}$. In the Sweet-Parker model this is the diffusion speed \emph{into} the reconnection region based on the conductivity of the plasma. Similar for Petschek's model. Hence, these models just provide \emph{scaling relations} and no real physical processes.

On the other hand, if two plasmas encounter each other, which is the usual case in nature, in solar flares, at the magnetopause, in the magnetotail and many other places, then the inflow velocity $V_\mathit{in}$ is prescribed. Both models leave no or little freedom in this case. Given the inflow velocity, magnetic field, and conductivity (or inertial length) fixes the reconnection rate and just determines the lengths of the reconnection sites. 
\begin{figure*}[t!]
\centerline{\includegraphics[width=0.975\textwidth,clip=]{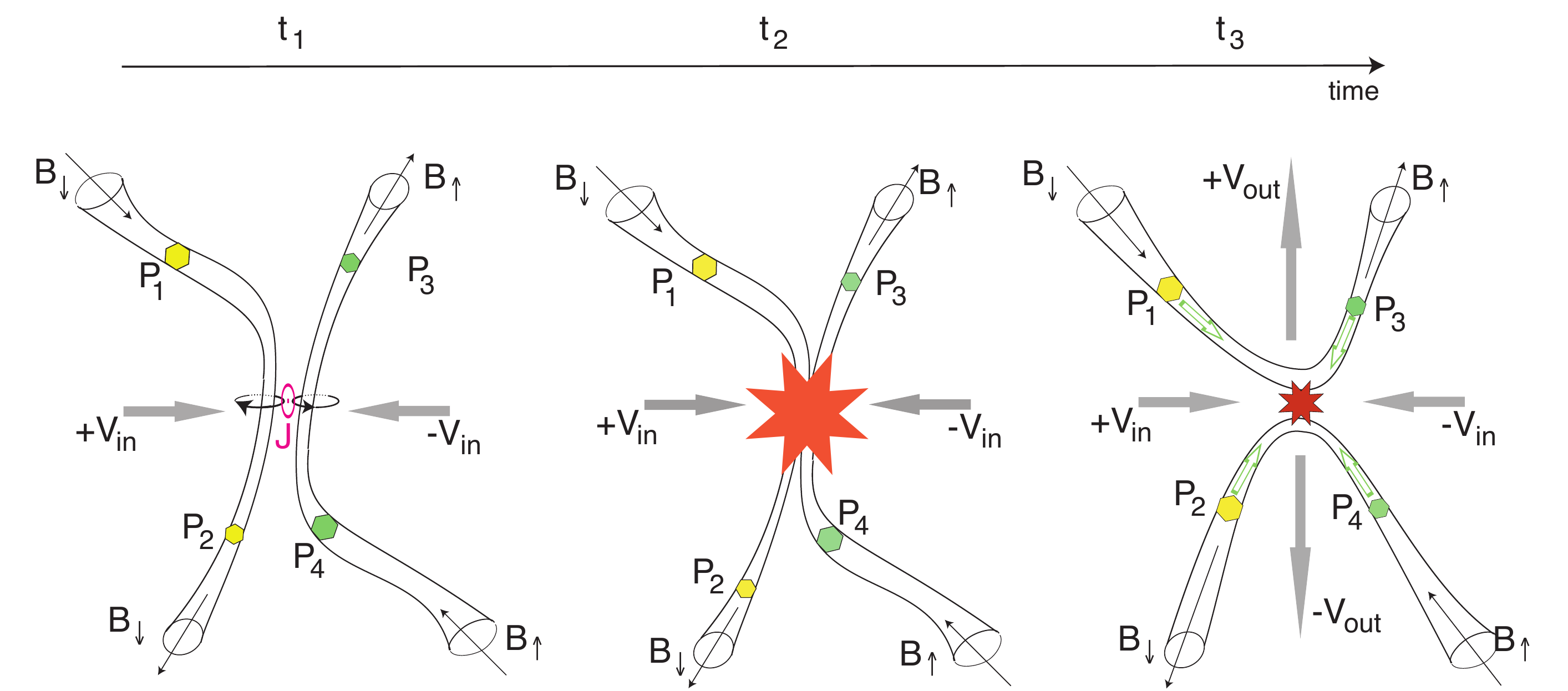}
}
\caption[ ]
{Schematic of the interaction of two flux tubes of different radii leading to reconnection when approaching each other with inflow speed $V_\mathit{in}$. The tubes carry magnetic fields $B_\downarrow, B_\uparrow$ that, over a certain length along the flux tubes, are oppositely directed. When the tubes approach each other at time $t_1$, a narrow current layer of length with current density $J$ occurs between them. Once contacting at time $t_2$, the antiparallel fluxes annihilate (as indicated symbolically by the red star). The excess magnetic energy is consumed in the reorganization of the magnetic fields into two new, heavily kinked flux tubes at time $t_3$, as well as in heating of the plasma inside the volume of contact (small red star). The magnetic stresses stored in the kinked tubes relax by accelerating the plasma that is attached to the tubes attributing it an outflow velocity $V_\mathit{out}> V_\mathit{in}$ in the orthogonal direction. As a by-product, the reorganization of the tubes enables the initially separated different plasmas $P_1$ and $P_3$ and $P_2$ and $P_4$ to mix by moving along the reconnected tubes (open green arrows).}\label{fig-rec1}
\end{figure*}

Unlike its physical mechanism which, even today, more than half a century after its theoretical investigation by Giovanelli, Sweet, Parker and Petschek, still remains to be poorly, at least incompletely and unsatisfactorily understood, the very picture of reconnection is simple and quite suggestive as a profound change of magnetic topology. In this picture, reconnection consists in the \emph{merging} of two \emph{oppositely} directed magnetic field lines -- or, better, flux tubes containing oppositely directed magnetic fluxes. The contacting oppositely directed magnetic fluxes annihilate along their common meeting length $L$. Their excess magnetic energies are transformed into kinetic energy of the plasma, which becomes both heated and accelerated. The heating and acceleration can be understood as the transition from a state of higher, i.e. simpler order to a state of higher disorder, i.e. a more complicated (or turbulent) order accompanied by thermalization. The higher initial order consists in the presence of a well defined current layer of width $d$ which initially separates the oppositely directed undisturbed flux tubes. When the current layer is disturbed in some location with the two flux tubes getting into mutual contact, spontaneous annihilation sets on, and the current layer becomes locally disrupted by the breaking-through and merging/reconnecting magnetic field lines/flux tubes as schematically shown in Fig. \ref{fig-rec1}. In this process the magnetic fields reorganize in a new way that is subject to its own dynamics depending on the global magnetic field configuration and local plasma properties. It it important to understand that this dynamics may be very different for unlike initial settings. 

In each case, when reconnection takes place in one or the other way, a certain amount of magnetic excess energy is released. Initially, the excess energy was stored in the magnetic configuration in either of the equivalent forms of anti-parallel magnetic topology, current density of the current layer that separates the antiparallel fields, or the energy in the plasma flow which transports the antiparallel magnetic field towards each other until reconnecting. These forms correspond to different views of the reconnection process as either magnetic energy release, transverse current instability and pinching/filamentation (most easily understood physically when realizing that parallel currents attract each other and, therefore, an extended current sheet might readily decay to form a chain of current filaments), or head-on penetration of counter streaming magnetized plasmas causing magnetic boundary layers, plasma mixing and turbulence. Of these views, however, the change in magnetic topology as result of reconnection is the most intuitive, and for this reason one may speak of \emph{magnetic} energy release even though the source of energy, as for instance in solar wind-magnetosphere interaction, may be found in the mechanical energy of counter-streaming plasma flows.  

The amount of released magnetic energy can be estimated easily from the antiparallel magnetic field components  $B_\uparrow,B_\downarrow$. Since electrodynamics (in the absence of magnetic monopoles and magnetization currents) from $\nabla\cdot\mathbf{B}=0$ categorically requires continuity of the magnetic field, only equal amounts of antiparallel magnetic fields, i.e. magnetic flux density will be annihilated, and one may write $B_\uparrow=-B_\downarrow=\Delta B$. The available magnetic energy density is thus $w_\mathit{rec}=2(\Delta B)^2/\mu_0$. Multiplying with the volume $\mathcal{V}_B=\pi L(R^2_1+R_2^2)$ of the two reconnecting flux tubes of radius $R_1,R_2$ and common length of contact $L$ (with $L$ being equal to the above mentioned length of the reconnection region along the antiparallel fields), the total released energy becomes $W_\mathit{rec}=2\pi(R^2_1+R^2_2)L(\Delta B)^2/\mu_0$. With this energy released during reconnection time $\tau_\mathit{rec}$, the reconnection power becomes $\mathcal{P}_\mathit{rec}=W_\mathit{rec}/\tau_\mathit{rec}$. Conversely, the action  exerted on the plasma is of order $\mathcal{J}_\mathit{rec}=W_\mathit{rec}\tau_\mathit{rec}$. Part of it is used to fractionally heat the plasma, increasing plasma temperature and causing particle acceleration, {part is stored in the magnetic stresses due to the newly generated curvature of the reconnected magnetic fields}. Relaxation and stretching of the bent magnetic flux tubes gives rise to bulk plasma jetting by adding momentum to the magnetically frozen plasma component. The latter processes are complex and involved and require a kinetic treatment. Any fluid or, in particular, magnetohydrodynamic theory, which ignores the different dynamics of the electron and ion components, does not cover them properly. 

This simple and intuitive topological picture has been applied by \citet{dungey1961} to geometrically describe the generation of the observed plasma convection patterns in Earth's magnetosphere by \emph{merging} the antiparallel components of the geomagnetic and interplanetary (solar wind) magnetic fields at the dayside boundary of the magnetosphere, the magnetopause. After this merging the solar wind convectively transports the merged magnetic flux tubes downtail along the magnetopause. This field line motion stirs the plasma inside the magnetosphere. Subsequent dis-connection of the geomagnetic field from the interplanetary field and \emph{re-connexion} of the two separate sections of the geomagnetic field in the center of the tail continuously restores the original geomagnetic field configuration. 

Dungey's model superseded any other model of diffusive coupling becoming the canonical paradigm of magnetic energy release by reconnection. This paradigm found its application wherever oppositely directed magnetic fields in plasmas come into contact in space, astrophysics, and also in the laboratory. The widespread application of this topological reconnection model is reflected in the large number of reviews, starting with \citet{vasyliunas1975}, which are available in the published literature \citep[for two recent reviews the reader is directed to][]{zweibel2009, yamada2010}, books \citep[for a purely MHD treatment see, e.g.,][for a mostly mathematical exercise]{priest2000,biskamp2000} and book chapters on reconnection \citep[for instance in:][]{kivelson1995,treumann1997,baumjohann2012}.

The present review focusses on collisionless reconnection in space. It follows a more physical approach trying to keep balance between theory and observation. Observations are taken mainly from space plasmas for the reason that the scales of space plasmas are large enough to be considered approximately smooth. Laboratory plasmas \citep{yamada2010} have the advantage that experiments can be set up at well controlled initial conditions; their disadvantages consist in the small scales and high dimensionality of any laboratory installation which unavoidably demands dealing \emph{ab initio} with magnetically complex configurations, presence of boundaries, incomplete ionization, neutral gas effects, temperature gradients, inhomogeneities, and non-negligible initial resistivity. They are thus difficult to analyze. Space plasmas, in the regions of interest (solar corona, solar wind, magnetopause, magnetotail), are collisionless, large scale, with boundaries being located fairly remote; in many cases they are very close to two-dimensionality, are sufficiently uniform magnetically as well as thermodynamically. Their problems consist in their high time-variability, the impossibility of preparing a controlled initial state, and last not least, the relatively sporadic availability of data. Their main deficiency is that they are subject to passive point-like observations only taken in small regions of space and at stochastically distributed times when the spacecraft incidentally passes close to a reconnection site. The point-like character of the observations is imposed by the miniscule scale of the spacecraft  in comparison  to all spatial scales in dilute space plasmas. Though at first glance this seems of advantage, the high time variability of space plasmas together with the notorious  low instrumental resolution and the enormous amount of data flow generally inhibit any wanted high space-time mapping.  Nevertheless, over the last four decades of the unmanned space-flight age large amounts of observational data have been collected and accumulated which, already at the present time, allow drawing a preliminary though by far not ultimate picture of collisionless reconnection. 

\section{Fluid models of reconnection}
Reconnection, understood as the dynamics of magnetic fields in electrically conducting media is a problem of electrodynamics. Since, in fluids, the reorganization of magnetic fields naturally proceeds on diffusive time scales $\tau_\mathit{rec}\sim\tau_\sigma\gg\omega_e^{-1}$ much longer than the oscillation periods, $\omega_\mathit{em}^{-1}<\omega_e^{-1}$, of free electromagnetic waves, these processes are comparably slow, even for maximum diffusivity $D_\sigma\sim D_A$. Thus any relativistic effect on the fields can be neglected, and the equations reduce to the induction equation for the electric, $\mathbf{E}$, and magnetic, $\mathbf{B}$, fields, completed by Amp\`ere's law:
\begin{equation}
\frac{\partial \mathbf{B}}{\partial t} =- \nabla\times\mathbf{E}, \qquad \nabla\times\mathbf{B}=\mu_0\mathbf{J}(\mathbf{E}),
\end{equation}
where $\mathbf{J}(\mathbf{E})$ is the current density in the -- reconnecting --  current sheet in question. Its current density shall be determined from the dynamics of the differently charged components of the plasma as function of the self-consistent electric field under the boundary condition that the magnetic fields sufficiently far outside the current layer are directed antiparallel. From the induction equation, the magnetic field is trivially free of any magnetic charges, $\nabla\cdot\mathbf{B}=0$, which implies that magnetic field lines cannot be broken in no case and at no time; they are always continuous and must be closed either at finite distance or at infinity. At a later occasion we will return to this important point in relation to reconnection as, so far, it has never ever been properly addressed.

\subsection{Generalized Ohm's law}
All fluid physics is contained in Ohm's law $\mathbf{J}(\mathbf{E})$. Its MHD version is $\mathbf{J}=\sigma\left(\mathbf{E}+\mathbf{V}\times\mathbf{B}\right)$, with $\mathbf{V}$ the fluid velocity. In ideal MHD $\sigma\to\infty$, implying  $\mathbf{E}=-\mathbf{V\times B}$. The field is forever frozen to the fluid, and there is no reconnection. Moreover, the induction equation tells that an initially nonmagnetic plasma in the current sheet can by no means become magnetized unless magnetization currents flow at it boundaries. Reconnection thus requires that the current sheet itself is magnetized. Otherwise the magnetic field decays exponentially over one electron skin-depth (inertial length) $\lambda_e$ from its boundary towards the center of the current sheet thus keeping a current sheet of thickness $d>2\lambda_e$ free of magnetic field. In resistive MHD with $\sigma$ finite, the field becomes $\mathbf{E}=-\mathbf{V\times B}+\mathbf{J}/\sigma$, and the above two equations are combined into the induction-diffusion equation for the magnetic field
\begin{equation}\label{eq-ind}
\partial \mathbf{B}/\partial t =\nabla\times\left(\mathbf{V\times B}\right) +D_\sigma\nabla^2\mathbf{B}
\end{equation}
Applying its stationary version dimensionally to a plane current sheet under the condition of continuity of flow just reproduces the reconnection rate of the Sweet-Parker model due to \emph{resistive} diffusion of the magnetic field, a slow process. Moreover, MHD is valid only on scales larger than the ion gyroradius $\rho_i$ outside the ion inertial region $\lambda_i$  that is centered on the current sheet. Thus, even though one may speak about the diffusion of the field into the current sheet, the process of reconnection itself lies outside of MHD. MHD merely covers the post-reconnection effects of collisionless reconnection on scales $>\rho_i>\lambda_i$. 

The simplest generalized \emph{collisionless} Ohm's law is provided by two-fluid theory. It is essentially the electron equation of motion which, with $\mu=m_i/m_e$ the mass ratio, can be rewritten \citep[cf., e.g.,][]{krall1973}:
\begin{eqnarray}\label{eq-ohmslaw}
{\bf E}&\!\!\!+\!\!\!&{\bf V}\times{\bf B}\ =\ \mbox{$\mathbf{\sigma}^{-1}_a$}\cdot{\mathbf J}+\left(eN\right)^{-1}\Big[{\mathbf J}\times{\bf
B}- \nabla\cdot\Big(\textsf{{P}}_e+\mu\textsf{{P}}_i\Big)\Big]+ \cr &+& (\epsilon_0\omega_e^2)^{-1}\Big\{{\partial_t{\mathbf J}}+ 
+\nabla\cdot\Big[\mathbf{JV+VJ}-e\big(\Delta N\big)\mathbf{VV}\Big]\Big\} -\cr &-&(eN)^{-1}\Big[\Big(\mathbf{F}_e-\mu\mathbf{F}_i\Big)_\mathit{pmf}\Big]
\end{eqnarray}
as an expression for the electric field $\mathbf{E}$ in the plasma frame moving with velocity $\mathbf{V}\approx\mathbf{V}_i$ (the ion frame). The first term on the right accounts for a possible tensorial \emph{anomalous} conductivity $\mathbf{\sigma}_a$. We added a ponderomotive force term -- indexed \emph{pmf}. All terms on the right enter Faraday's law and produce a substantially more involved induction equation than Eq. (\ref{eq-ind}). Application of the curl operation makes all irrotational contributions vanish, leaving just the rotational terms contributing to the inductive electric and magnetic fields. Since $\mathbf{V\times B}$ freezes the plasma to the magnetic field, the surviving terms in Ohm's lay may violate the frozen-in concept contributing to diffusion of plasma against the magnetic field and causing reconnection in a current sheet. A reconnection theory can be based on each of the terms. 

It is instructive to consider each of the terms separately. In collisionless reconnection, the first term is non-zero only when the plasma develops an anomalous resistance. This requires a separate investigation which we delay to the end of this section.

\subsubsection{``Hall'' reconnection.}
The induction equation with only the Hall term included becomes
\begin{equation}\label{eq-hallind}
\frac{\partial{\bf B}}{\partial t}=\nabla\times\left({\mathbf v}\times{\mathbf B} -
\frac{1}{eN}{\mathbf J}\times{\mathbf B}\right).
\end{equation}

 \begin{figure*}[t!]
\centerline{\includegraphics[width=0.7\textwidth,clip=]{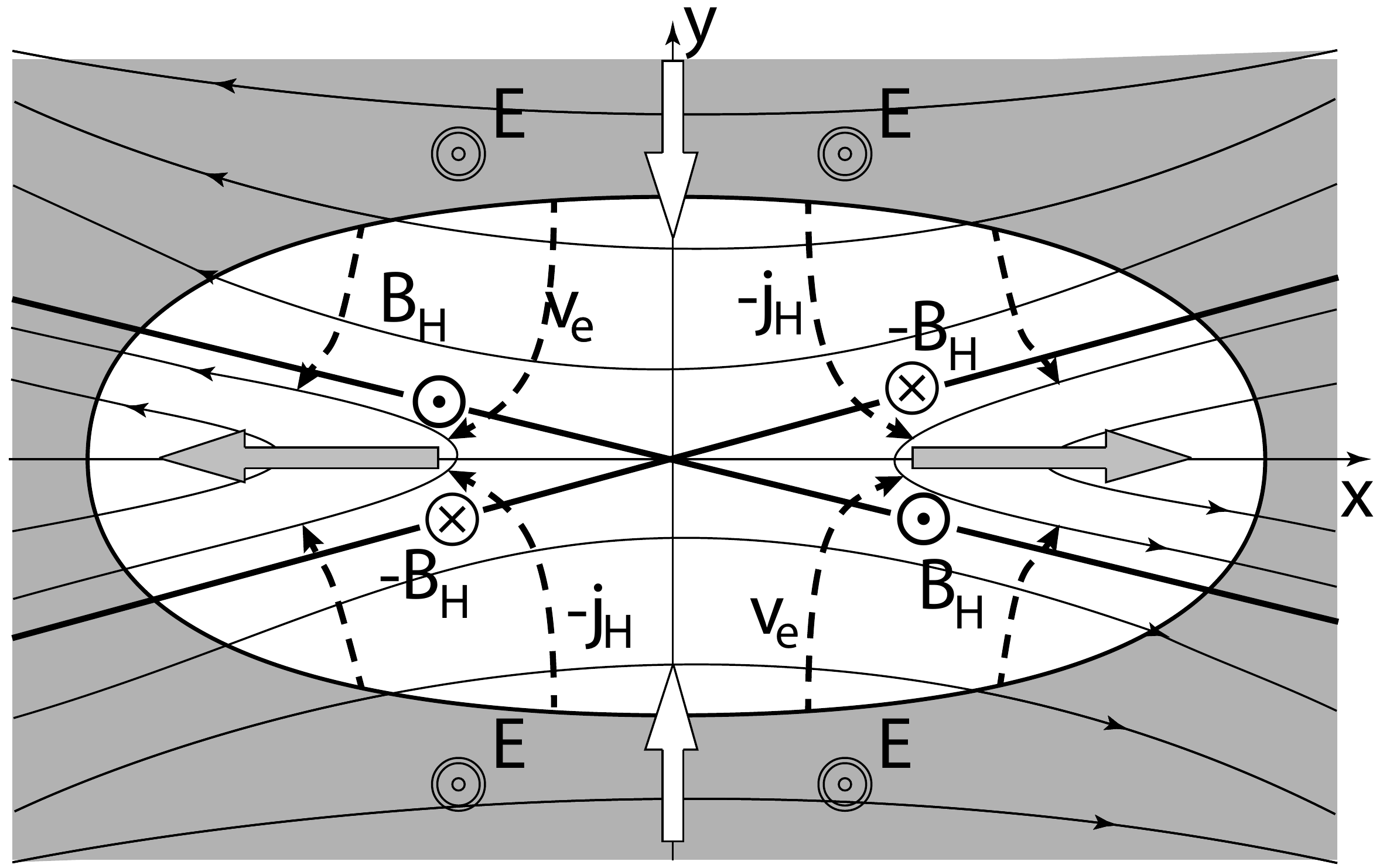} 
}
\caption[]
{Geometry of the Hall effect the X-line in collisionless reconnection. The white region is the domain $r\lesssim \lambda_i$ from the X-line accounting for a potential asymmetry between the inflow and outflow direction which elongates the region into the outflow direction.  Electron (dashed lines) performing an inward E$\times$B-drift. Hall currents $\mathbf{J}_H$ are centered on the separatrices forming four half-loops. They generate the quadrupolar Hall magnetic field $\mathbf{B}_H$. Open arrows show the inflow at low speed $V_\mathit{in}$ and the outflow at high speed $V_\mathit{out}$. The convection electric field points out of the plane. Ions are accelerated in this field out of the plane thereby contributing to the sheet current. This contribution effectively causes a broadening of the current layer.}\label{fig-hall}
\end{figure*}

Near the reconnecting current sheet electron and ion motions decouple almost completely on the inertial scale of the ions. Consider the ratio of ion gyro-radius $r_{ci}=v_{i\perp}/\Omega_{i}$ to ion inertial length $\lambda_i=c/\omega_{i}$. Inertial effects dominate the ion motion when $\beta_{i\perp}=r_{gi}^2/\lambda_i^2>1$. Ions decouple from the magnetic field. Electrons remain magnetized, perform gyrations and cross-electric field drifts, while being convected into the current layer with velocity $\mathbf{E\times B}/B^2$. They constitute the Hall current \citep{sonnerup1979}  
\begin{equation}
\mathbf{J}_{H}=-eN\mathbf{v}_E=-\left(eN/B^2\right)\mathbf{E\times B}.
\end{equation}
which flows in the ion inertial region \emph{outside} the electron inertial region and thus outside the current center, implying that it is primarily not involved into the reconnection process itself. Its maximum is located inside but close to the separatrices emanating from the reconnection {X} point, and it vanishes along the boundary of the electron diffusion region at distance $\lambda_e$ from the center of the current layer. 

The Hall conductivity tensor $\mathbf{J}_H=\mbox{$\mathbf{\sigma}_H$}\cdot\mathbf{E}$ has only two components and is antisymmetric, with element $\sigma_H= -eN/B$ corresponding to a Hall resistivity $\eta_H=-B/eN$. In the presence of an (isotropic) anomalous conductivity, the Hall conductivity becomes $\sigma_{H,a}=\sigma_H/\big(1+\sigma_H^2/\sigma_a^2\big)$. For large $\sigma_a\gg\sigma_H$ this reduces to $\sigma_H$. In the presence of large anomalous resistance, $\sigma_a\ll\sigma_H$, and the conductivity is isotropic with $\sigma_{H,a}\approx\sigma_a$. This Hall resistance is an inductive blind resistance and plays no role in the generation of heat. The Hall current evolves mainly in the absence of binary (Coulomb) collisions  and, assuming that anomalous resistance is negligible, has been taken as evidence for \emph{collisionless} reconnection in Earth's magnetotail \citep{fujimoto1997, nagai2001, oieroset2001}. It has been reclaimed also for the magnetopause \citep{mozer2002} of providing evidence for the existence of an electron diffusion layer there. Though an electron diffusion layer is reasonable and is supported by other observations, the claim of a Hall magnetic field and current system is to be taken with care because the Hall magnetic field in these particular observations is stronger than the main field almost everywhere which is inconsistent with the fact that Hall currents are secondary effects and, therefore, their field must necessarily be weak at any location. 

It may be noted that Sonnerup's (1979) realization of the Hall effect near a reconnection site has been repeated also by other groups \citep[e.g.,][]{uzdensky2006}.
Hall currents near a reconnection site generate a Hall magnetic field in the direction of the current flow perpendicular to the symmetry plane of the external magnetic field but parallel to the current and the convection electric field $\mathbf{E}$. This Hall field has quadrupolar structure and, though necessarily being weaker than the external magnetic field, introduces a guide field into reconnection \citep{baumjohann2009}, opening the possibility of accelerating electrons along the guide field in the direction anti-parallel to the convection electric field.  This is shown in Figure \ref{fig-hall}. 

Since Hall currents are a source of free energy with perpendicular scales comparable to the transverse scales of kinetic Alfv\'en waves, the reconnection site is a natural source region of kinetic Alfv\'en waves. In addition, Hall currents serve as sources of various plasma instabilities like the electrostatic modified-two stream instability or the electromagnetic Weibel instability \citep{weibel1959}.  Otherwise Hall currents do not \emph{directly} contribute to the excitation of reconnection. This is contrary to what has sometimes been claimed by the defenders of Hall reconnection. However, particle in cell simulations \citep{jaroschek2004a,jaroschek2004b} of electron-positron pair plasmas have convincingly demonstrated that fast reconnection takes violently place under non-Hall conditions; in pair plasmas Hall currents are (trivially) absent since $m_-=m_+\equiv m_e$ and, consequently, electron and positron diffusion regions coincide. These conclusions have been confirmed by subsequent similar simulations \citep{bessho2005} using substantially smaller particle numbers and less resolution but arriving at the same result {\citep[cf. also][anticipating these results in hybrid simulations when removing the $\mathbf{J\times B}$-Hall term, and considering the reconnection rate for both uniform and localized (anomalous) resistance; in both cases fast reconnection rates have been obtained]{karimabadi2004b,karimabadi2004c}}.

An important side product of the Hall currents is provided by the necessity of current closure. Such closure proceeds preferably along the separatrix magnetic field with the field-aligned current carried by electrons \citep[e.g.,][and others]{uzdensky2006,treumann2006}. The field-aligned currents also contribute to the guide field; this is, however, a  second-order effect. The main function of the field-aligned current is to couple the reconnection site to its external environment, possibly giving rise to important secondary effects far away from the reconnection region. In the magnetosphere such effects are observed as aurorae during substorms, particle precipitation, chains of electron holes, excitation of near-Earth plasma instabilities and electromagnetic radiation. In the solar atmosphere they serve as sources for solar radio emissions like, for instance, Type III bursts.

\subsubsection{Pressure-driven reconnection.}
Since no steep gradients neither in density nor temperature are expected near the reconnection site, the pressure term in a weakly inhomogeneous plasma contributes only when the pressure tensor is highly anisotropic or contains non-diagonal elements. If at all, contribution will come only from the electron pressure term $\textsf{P}_e$ which, in the center of the current layer, may develop inhomogeneity and possibly even electron-viscosity, generating non-diagonal elements. The latter, in particular, have been attributed to be important drivers of reconnection \citep[cf., e.g.,][and references to earlier papers of the same authors therein]{hesse2001}. 

In collisionless plasmas non-diagonal pressure tensor elements can occur as the result of finite-Larmor radius effects, argued to be preferentially caused on the meandering electron component in the current sheet that is trapped between the oppositely directed magnetic fields where performing so-called ``Speiser orbits'' {\citep[cf., e.g.][]{yin2004a}}. Non-diagonal pressure tensor elements have the character of viscosities in the electron fluid (similar effects hold also for ions on a much larger scale in the ``ion diffusion region'' but are of little interest or effect on the short electron scales in collisionless reconnection for,  inside the electron diffusion region of radius $\lesssim\lambda_e$ from the center of the reconnecting current sheet ions are clearly completely non-magnetic and finite-Larmor-radius effects are negligible here). It is, however, not completely clear whether non-diagonal pressure terms based on physical effects will indeed be generated at a reconnection site.  All wave-particle interaction-based nonlinear plasma effects that might generate viscosities here are probably too weak {though \citet{divin2010,divin2012a} proposed that electron acceleration near the reconnection site indeed forms non-diagonal electron pressure terms. The opinions are ambiguous on this matter. Numerical particle-in-cell simulations \citep{pritchett2005c} do indicate that it is indeed the pressure tensor effect which takes responsibility for collisionless reconnection. \citet{karimabadi2004a} demonstrated that electron temperature anisotropies have a strong destabilizing effect on the current layer and magnetic reconnection. } 

The case is a bit obscured by the imprecise definition of the main axes of the pressure tensor near a reconnection site. Pressure anisotropy $P_\perp,P_\|$ refers to the magnetic field which changes direction by $180^\circ$ across the current sheet. The \emph{natural} reference system is thus not that of the magnetic field but  that of the current sheet instead, centered at current maximum. In this frame of reference the collisionless pressure tensor has of course all nine components, \emph{provided it has maintained some  anisotropy} $P_\perp\neq P_\|$. Accounting for the three different non-diagonal elements in this \emph{geometric} representation of the tensors, pure \emph{anisotropic} pressure effects will naturally become important. The decision about the role of pressure needs clarification of which frame of reference has to be considered as basic in reconnection and how pressure anisotropy either is generated or maintained in the electron diffusion region. In fact, simulations seem to confirm that both is the case thus, apparently, supporting pressure-driven reconnection in collisionless plasma configurations to be a general, possibly even the canonical mechanism in collisionless non-forced reconnection. Support to this conclusion has been given by recent work \citep{zenitani2011} who derived an \emph{electron-frame}  ``dissipation measure'' to identify the location and extent of the electron diffusion region. This measure is the unrecoverably dissipated Joule energy, the scalar quantity 
\begin{equation}\label{eq-dissipmeasure}
W_{ed}=\Gamma_e\left[\mathbf{J}\cdot(\mathbf{E-B\times V}_e)-e(N_i-N_e)\mathbf{E\cdot V}_e\right],
\end{equation}
given here in Lorentz covariant form with $\Gamma_e=1/\sqrt{1-V_e^2/c^2}$ the bulk electron Lorentz factor. The last term is not necessarily zero because in the reconnection region electric fields arise partly due to charge separation. Nonrelativistically $\Gamma_e=1$. Since energy is frame dependent, transformation to the ion frame yields simply $N_eW_{ed}=N_iW_{id}$ as the relation between electron and ion dissipation measures. Similarly, transformation to the bulk MHD-frame defines the MHD-dissipation measure $W_\mathit{mhd}=W_{ed}(1+\mu)/(1+\mu N_i/N_e)$. These quantities are useful in mapping the dissipation region both in measurements as in simulations thus identifying the \emph{physical} reconnection site. Observations \citep{zenitani2011} support this measure in symmetric reconnection settings like in the magnetotail. In non-symmetric reconnection, however, the measure seems not as useful, as has been discussed by \citet{pritchett2013}.

\subsubsection{Inertial reconnection.}
The time dependent inertial term can be expressed as $\mathbf{E}_\mathit{inert}\approx\sigma^{-1}_\mathit{inert}\mathbf{J}$ through the ``inertial conductivity'' $\sigma_\mathit{inert}=\epsilon_0\omega_e^2\tau_\mathit{inert}$. The time $\tau_\mathit{inert}$ is the time of change of the current. This time is usually long compared with the period of plasma oscillations. In the magnetotail with density $N\sim 10^6$ m$^{-3}$ one has $\sigma_\mathit{inert}\approx 5\times10^{-3}\tau_\mathit{inert}$ S. Therefore, moderately fast current oscillations could well contribute to inertial resistivity providing sufficient inertial diffusivity for reconnection. Using it in the Sweet-Parker model one can show that current sheets of thickness of the ion inertial length $d\sim \lambda_i$ become of length $L\sim\sqrt{m_i/m_e}\lambda_i$. The reconnection rate is then found of order 
\begin{equation}
E_\|/E_\perp\sim\mathcal{M}_A^{-1}\sqrt{m_e/m_i}.
\end{equation}
Here $\mathcal{M}_A<1$ competes with the smallness of the mass ratio, but the inertial reconnection rate is nevertheless large compared with $\mathcal{M}_{SP}$. 

\subsubsection{Causes of the reconnection electric field.}
It is instructive to ask for the role of the reconnection electric field $E_\mathit{rec}$. This is most simply seen in two dimensional reconnection. From Ohm's law -- neglecting turbulence and ponderomotive terms -- one has (with $\nabla\equiv \nabla_x+\nabla_z$) for $E_\mathit{rec}\equiv E_y$:
\begin{eqnarray}\label{eq-erec}
 E_\mathit{rec}&=& -(\mathbf{V}_e\times\mathbf{B})_y-(eN_e)^{-1}\left(\nabla_xP_{e,xy}+\nabla_zP_{e,yz}\right)\cr
 &-& (m_e/e)\left(\partial_t+\mathbf{V}_{e}\cdot\nabla\right)V_{ey}
\end{eqnarray}
where we work in the X point frame. We explicitly wrote out the only two remaining non-diagonal electron pressure terms (neglecting ion pressure). These are combinations of $P_\perp,P_\|$. Otherwise, the non-diagonal elements may also have been produced by finite-gyro-radius effects causing non-gyrotropy {\citep[as argued by][]{yin2004a,karimabadi2004a,aunai2013}}. $E_\mathit{rec}$ is determined by the divergence of the electron pressure, i.e. the electron-viscous contribution, which \emph{in the X point frame} is non-zero in an extended region around the X point. Moreover, an inertial term contributes as well but is important only in the electron diffusion region where inertia comes into play, with the time derivative being of lesser importance. $E_\mathit{rec}(x,z)$ must depend on space such that the rotation $\mathbf{e}_y\times\nabla E_\mathit{rec}\neq 0$ contributes to the two magnetic components $B_x, B_z$.
 
\citet{pritchett2005b}, in an important paper, investigated the separate contributions of these terms in two-dimensional non-forced simulations, finding that electron pseudo-viscosity and inertia both  contribute to $E_\mathit{rec}$, compensating for the convection term. Their contributions are located in the electron diffusion region and have a pronounced spatial dependence. They are responsible for the breaking of the frozen-in state of electrons here. This gives support to the view that, in anisotropic plasma, collisionless reconnection is indeed made possible by electron  pseudo-viscosity, aided by nonlinear electron inertia (the nonlinear term in the inertial contribution to Eq. \ref{eq-erec}). Apparently (except for the ponderomotive term) all other terms in the induction equation provide minor corrections. One may even go further in concluding that any resistive reconnection is probably unrealistic. Resistive dissipation in a plasma will always be secondary to the simple electron pressure effects which generate electron pseudo-viscosity on the electron inertial scale. Reconnection proceeds, presumably, always on the micro-scale of the electron skin depth.    

\subsubsection{Flows, turbulence, ponderomotive effects.}
Ohm's law Eq. ({\ref{eq-ohmslaw}) contains the flow velocity, density, current and magnetic field. These quantities can be understood as composed of slowly variable mean values $\langle\mathbf{V}\rangle,\langle\mathbf{J}\rangle,\langle N\rangle,\langle\mathbf{B}\rangle$  and turbulent fluctuations v$\delta\mathbf{V},\delta\mathbf{J},\delta N,\delta\mathbf{B}$) changing on much faster time and shorter spatial scales. In a mean-field theory of reconnection, the nonlinear terms in Ohm's law then contribute additional correlation terms which, when violating the frozen-in condition, can also become sources of \emph{turbulent} reconnection. This is known from (collisional) dynamo theory where the correlations give rise to $\alpha\Omega$-dynamo action. Here they provide a collisionless fluid means of driving reconnection \citep[see, e.g.,][]{browning2013}. Similar effects are provided by the ponderomotive term.

Expressions for the ponderomotive force for different plasma models are found in the literature. For two-fluid plasmas \citet{leeparks1983,leeparks1988} and \citet{biglari1993}  derived some simplified forms. In contrast to the above fluid turbulence, ponderomotive forces account for the slow variability of plasma turbulence. They account for the role of kinetic effects on fluid scales. Reconnection theory has only recently begun implicitly including fluid and plasma turbulence in simulations {\citep[][]{daughton2011,roytershteyn2013,karimabadi2013a,karimabadi2013b}}.  

\subsection{Anomalous effects}
Of the above noted possible causes of reconnection of particularly interest are all those which may contribute by effects in plasma which are not covered by a simple two-fluid theory. These are effects due to heating and its reaction on reconnection as well as those which generate anomalous resistance, anomalous viscosity or other anomalous effects like current bifurcation {\citep[first observed by][]{runov2003}, broadening and ``disruption" \citep[recently reported to occur in laboratory experiments and simulations by][]{jain2013}. Since nonlinearity comes in at this place \citep[cf., e.g.,][ for an example referring to the nonlinear evolution of the lower-hybrid instability causing current bifurcation]{daughton2004},} these effects are more difficult to treat and, in principle, require a numerical investigation via particle or Vlasov simulations.

\subsubsection{Critical transition length scale.}
\citet{uzdensky2007} pointed out that Coulomb processes may come into play via heating the plasma. We know since \citet{sonnerup1979} that, {except for finite Larmor radius effects causing ion and electron viscosity on the respective relevant scales $\lambda_i,\lambda_e$} \citep{karimabadi1999,karimabadi2004a}, reconnection is basically collisionless whenever the current layer thickness $d<\lambda_i$ drops below the ion inertial length. In the Sweet-Parker model, with mean free path $\lambda_m=v_e/\nu$ of electrons ($v_e=\sqrt{2T_e/m_e}$ is the nonrelativistic thermal electron speed)  and $\beta_e=2\mu_0NT_e/B^2$, we have $d=\mathcal{M}_{SP,P} L<\lambda_i$. This yields an upper limit on $L<\lambda_i\mathcal{M}_A^{-1}=\lambda_m\sqrt{m_i/m_e\beta_e}\equiv L_\mathit{cl}$ for reconnection being collisionless. Thus, should it happen that the length $L$ of reconnection along the antiparallel fields drops below $L_\mathit{cl}$, reconnection went collisionless. Clearly, this depends on some process which is hidden in the electron mean free path and therefore depends on the mechanism which generates collisions. 

The collision frequency is proportional to the fluctuation level $W_w/NT_e$ of the scattering plasma waves. This allows expressing $L_\mathit{cl}$ quite generally through the Debye length $\lambda_D$ and $W_w$: 
\begin{equation}\label{eq-lcrit}
L_\mathit{cl}\simeq \lambda_D\left(NT_e/W_w\sqrt{m_e\beta_e/m_i}\right)
\end{equation}
Under pressure balance one has in the current layer $\beta_e+\beta_i\approx 2\beta_e=1$. The particular case of binary Coulomb collisions holding for the Sweet-Parker model gives $L_\mathit{SP,cl}\simeq\lambda_D(N\lambda_D^3\sqrt{m_i/m_e\beta_e})$. This latter expression scales as $L_\mathit{SP,cl}\propto B(T_e/N)^{3/2}$. One concludes that electron heating and dilution of plasma may increase the critical scale until reconnection becomes collisionless. \citet{uzdensky2007} has argued that in the solar corona $L_\mathit{SP,cl}\sim L$ about coincides with the lengths of loops providing evidence for a heating mechanism which keeps
the corona at marginal transition between collisionless and collisional reconnection. 

In contrast, in the magnetotail we have $\lambda_D\sim$ (6-10) m, $N\lambda_D^3\sim 10^8$, $\beta_e+\beta_i =1$, $3\beta_e\lesssim\beta_i< 10\beta_e$ \citep{artemyev2011}, thus $\beta_e\sim 1/6$. With these numbers we find $L_\mathit{SP,cl}\sim 10^{9}-10^{10}$ m. This indicates that, for the parameters in the magnetosphere, Sweet-Parker reconnection is indeed close to the claimed marginality. However, the scale where marginality would set on is much larger than any magnetospheric scale and, in particular, much larger than any observed longitudinal  scale of the reconnection region. This lets one doubt in the meaning of $L_\mathit{SP,cl}$ also in the solar corona.  

In order to bring the Sweet-Parker critical length in the magnetosphere down to observed reconnection scales of $<10$ R$_\mathrm{E}$ Earth radii,  the energy density of plasma waves should satisfy the condition
\begin{equation}\label{eq-lcritan}
3\times10^{-5} \lambda_D/\sqrt{N\beta_e}<W_w/NT_e\ll 1.
\end{equation}
Inserting for the tail then yields that the fluctuations must reach an energy density  of $W_w/ NT_e > 10^{-3}\sqrt{T_2/N_1}$. Here $T_2=T_e/(100$ eV), $N_1= N/(1$ cm$^{-3}$).  Such wave fluctuation levels require extraordinarily strong plasma wave excitation reaching far into the highly nonlinear regime. Wave intensities such high have not been observed, not even under extremely disturbed magnetospheric conditions. 

A condition for Petschek-like reconnection similar to Eq. (\ref{eq-lcrit}) reads $L_\mathit{P,cl}\lesssim 30 \lambda_i$. This condition is easier to satisfy, which indicates that Petschek-like reconnection based on some localized anomalous resistivity is not quite as unreasonable as Sweet-Parker reconnection under nearly collisionless conditions.

\subsubsection{Anomalous fluid effects.}
Lacking a resistivity under collisionless conditions in plasma, the importance of anomalous effects as a possible reason for anomalous resistance has early attracted attention. We already made use of Sagdeev's formula in Eq. (\ref{eq-lcritan}). Before returning to it we note that, formally, an anomalous conductivity can be defined for each of the terms in Eq. (\ref{eq-ohmslaw}) by simply defining some equivalent conductivities. This does not make sense for the Hall term, however. But the contribution of the pressure term, for instance, to the electric field could be written as $\mathbf{E}_P= -\sigma_P^{-1}\cdot\mathbf{J}=-(eN)^{-1}\nabla\cdot[\textsf{P}_e+(m_e/m_i)\textsf{P}_i]$. Multiplying from the left with the current density, one obtains a tensor equation $\mathbf{J}\cdot\sigma_P^{-1}\cdot\mathbf{J}=(eN)^{-1}\mathbf{J}\cdot\left\{\nabla\cdot[\textsf{P}_e+(m_e/m_i)\textsf{P}_i]\right\}$ which, for given pressure and with $\mathbf{J}=\nabla\times\mathbf{B}/\mu_0$, can formally be inverted to find the pressure induced contribution to conductivity. One immediately sees that the right-hand side is the projection of the pressure tensor divergence onto the direction of the current. Similar games can be played at the other terms in Ohm's law. This procedure brings the right-hand side of Ohm's law into the conventional form $\sigma^{-1}_\mathit{tot}\cdot\mathbf{J}$ with equivalent tensorial resistivity $\sigma^{-1}_\mathit{tot}= \sigma_a^{-1} +\sigma_H^{-1}-\sigma_P^{-1}+\sigma_\mathit{inert}^{-1}+\sigma_{(jV+Vj)}^{-1}-\sigma_\mathit{pmf}^{-1}$. Though this is a formal representation only, it shows that the different terms in Ohm's law may contribute to an \emph{equivalent resistance} each in its way. Except for $\sigma_a^{-1}$, which is a really anomalous resistance based on microscopic interactions, all other terms in $\sigma_\mathit{tot}^{-1}$ are macroscopic contributions to Ohm's law lacking the quality of anomalousness. They just refer to fluid conditions. This assertion applies also to the ponderomotive terms in Eq. (\ref{eq-ohmslaw}) for they contain just the slowly variable effects of waves in the momentum exchange between waves and the plasma fluid. The microscopic interactions which cause the ponderomotive forces are not included here.  

With this philosophy in mind, Ohm's law becomes an equation for the current density $\mathbf{J}$. Once it has been inverted, it enables one to estimate the dissipated energy $W_{ed}=\mathbf{J\cdot E}= \mathbf{E\cdot}\,\sigma_\mathit{tot}\cdot(\mathbf{E+V\times B})$ which is the energy that is available in reconnection for heating and acceleration in the dissipation region by all the processes contributing to the right-hand side of Ohm's law. Conversely, providing the current density and electric field can be unambiguously measured, $W_{ed}$ can directly be determined.

Another anomalous fluid effect arises if the velocity and current exhibit low frequency fluctuations according to $\mathbf{V}=\mathbf{V}_0+\delta\mathbf{V}, \mathbf{J}=\mathbf{J}_0+\delta\mathbf{J}$. The zero-order fields are mean fields $\mathbf{V}_0=\langle\mathbf{V}\rangle, \mathbf{J}_0=\langle\mathbf{J}\rangle$, and the averages  of the fluctuations vanish $\langle\delta\mathbf{V}\rangle=\langle\delta\mathbf{J}\rangle=0$. In this case the nonlinear terms contribute to Ohm's law in similar ways as known from dynamo theory. Ohm's law provides the link between dynamo theory and reconnection, with reconnection becoming driven by fluid and current turbulence \citep{browning2013} and interacting with dynamo effects providing merging in dynamo theory and enabling collisionless reconnection. Though it is clear that turbulence affects the dynamo, it is not clear whether it really causes reconnection. The scales of fluid turbulence are longer than any microscopic plasma scales. On those scales electrons should remain frozen to the magnetic field. Vice versa, onset of reconnection and the resulting tearing-mode turbulence will necessarily affect any ongoing dynamo by either speeding it up or braking it. In collisionless plasma one might suggest that reconnection-caused turbulence and anomalous resistivity will rather support the dynamo. On the other hand, in a recent mean field MHD model with the pmf in a turbulent Ohm's law \citep{higash2013} simulations seem formally to indicated that this kind of turbulence may drive reconnection, even explosively. Since this model is MHD it implies large scales; therefore its physical reality  remains unclear.

\subsubsection{Anomalous collisions.}
Microscopic nonlinear interaction can become a means of generating reduced electrical conductivities via the interaction of particles with self-consistently excited plasma waves. A spectrum of such waves may retard the electrons from their ballistic free flight motion due to scattering in the self-generated wave fields. The retardation is the cause of a fake friction exerted by the waves on the particles. It  results in real though anomalous (not based on binary interaction between particles) collision frequencies and contributes to an anomalous resistivity. Such a resistivity is necessarily anisotropic as it depends on the natural anisotropy of the plasma waves. 

The idea goes back to Sagdeev \citep[cf. his early reviews][]{sagdeev1966,sagdeev1979} who realized the equivalence between the scattering of electrons in thermal fluctuations of Langmuir waves in Coulomb interaction and scattering in unstably excited ion-acoustic wave spectra which we noted in the Introduction. \citet{sagdeev1966} favored ion-acoustic waves as the agents of anomalous collisions in non-magnetic, and electron-cyclotron waves in magnetized plasmas. However, in the magnetized reconnecting current layers like Earth's magnetotail or magnetopause, neither ion-acoustic waves nor Buneman modes \citep{buneman1958,buneman1959} don't grow, because $T_i>T_e$ there, and electron drifts $V_e<v_e$ are below electron thermal speeds. Also current-driven electron-cyclotron modes are stable in the extremely weak magnetic field in the current sheet. 

Some anomalous collision frequencies for various current driven plasma instabilities have been calculated by \citet{liewer1973} and \citet{davidson1977}. These papers favor the lower-hybrid drift instability for reconnection \citep{davidson1978,huba1981}. It grows in plasma density gradients of a magnetized plasma. Its source is the relative diamagnetic drift $V_\mathit{di}-V_\mathit{de}$ of charged particles, with $V_\mathit{dj}=|\nabla_\perp \ln N| (T_j/eB)$, giving rise to a net diamagnetic current {$J_\perp=|\nabla_\perp N|(T_i-T_e)/B$}. It has no threshold except $T_i=T_e$ and is thus universally present. It drives the lower-hybrid drift instability (as well as a relative of it, the modified-two-stream instability). 

Such density gradient exist at the boundary of the current sheet {and also at the boundary of the electron diffusion region while the magnetic field is weak and electrons are weakly magnetized}. The lower hybrid drift instability scatters the drifting particles reducing the relative perpendicular diamagnetic velocity between electrons and ions, stabilizing via heating the cooler plasma component. By Sagdeev's argument, it indeed yields high anomalous collision frequencies  of the order of the lower-hybrid frequency 
\begin{equation}
\nu_\mathit{a}\simeq\omega_\mathit{lh}=\sqrt{(\omega_i^2+\Omega_i^2)/(1+\omega_e^2/\Omega_e^2)}
\end{equation}
In weak magnetic fields with $\beta=\beta_e+\beta_i>1$ this collison frequency is the geometric mean between the electron and ion cyclotron frequencies $\omega_\mathit{lh}\simeq \sqrt{\Omega_e\Omega_i}=\sqrt{m_e/m_i}\,\Omega_e$. In stronger magnetic fields $\omega_\mathit{lh}\simeq\omega_i$, while in very strong fields $\omega_\mathit{lh}\simeq\Omega_i$. Unfortunately, due to depletion of the diamagnetic drift in weak magnetic fields, this instability is heavily damped in the high-$\beta$ conditions encountered in the center of the current layer and the ion diffusion region and should thus be of little value in driving reconnection, though it certainly affects the Hall and field aligned current system in the ion diffusion region causing ion heating and electron acceleration along the magnetic field here. Observations of wave spectra \citep{labelle1988,treumann1991} do not support its importance though measurements in magnetopause crossings \citep{treumann1990,bale2002} seem to provide evidence for its excitation even though lacking sufficiently high resolution for locating the waves to either the ion or electron diffusion regions. This argument may, however, be revised if a strong magnetic guide field is overlaid over the antiparallel fields and current layer. The physics then becomes quite different. Moreover, \citet{fujimoto2005} report modification of the electric convection field due to nonlinear evolution of the lower hybrid instability in the ion diffusion region. The additional inductive electric field component accelerates electrons to sustain a thin electron current layer which readily goes unstable with respect to tearing modes. It thus remains undecided whether or not the lower hybrid instability is directly involved into driving collisionless reconnection.

Anomalous collision frequencies require the excitation of some kind of plasma waves. The problem consists in the identification of the most effective instability and its saturation level. Most attempts rely on the lowest order interaction, i.e. quasilinear saturation, which is believed to provide the strongest quasi-stationary wave spectrum. However, wave-wave interactions and strong non-resonant wave-particle interactions may reduce the quasilinear level and deplete the expected anomalous collisions.

\section{Kinetic reconnection theory}
In view of the complexity (complicated geometry, all kinds of fluid aspects like flows, vortices, boundary layer effects etc., involvement of various modes of plasma instabilities and wave-particle interactions, coupling of the reconnection site along the magnetic field to other remote plasma conditions etc.) of the reconnection problem there is little chance for analytical results. {Pioneering work on the stability of a current layer concentrated on the tearing instability \citep{coppi1966,laval1966}.} Relevant analytical work terminated already in the seventies with papers by \citet{schindler1974} on the theory of substorms and fundamental work by \citet{galeev1975a,galeev1975b}, {followed by \citet{lembege1982}}. The former work \citep[see also][]{schindler1978,schindler1986} attributed the stability of the two-dimensional magnetotail current sheet solely to the ion-tearing mode, reducing the problem to the solution of a Grad-Shafranov equation for the only surviving component of the magnetic vector potential $\mathbf{A}=A\hat{y}$.  \citet{birn1977} extended this approach to three dimensions. 

In contrast to this quasi-fluid theory, \citet{galeev1975a} and \citet{galeev1975b} solved the full \emph{kinetic} stability problem  of a two-dimensional current layer with overlaid normal magnetic field including the difference between electron and ion dynamics. The assumption of a normal magnetic field component $B_z\neq0$ applies, for example, to the magnetotail, where the magnetic field is tied to the body of the Earth. It therefore remains to be weakly dipolar in spite of the extended current sheet. These authors provided a perturbation theoretical solution of the resulting Schr\"odinger equation for $A$ in the presence of the self-consistent non-vanishing electric potential. 

The key idea in this theory is to include the adiabatic oscillatory motion of nonmagnetized ions that are trapped in the current sheet and oscillate between the two oppositely directed magnetic fields thereby contributing to Landau damping while electrons remain magnetic due to the presence of the weak normal magnetic field component. From the electrodynamics wave equation for the vector potential $\mathbf{A}=A\hat{y}$ in the presence of a very low-frequency first order current perturbation $\delta\mathbf{J}=\sum_jq_j\int\mathrm{d}v^3\mathbf{v}\delta f_j(\mathbf{v})$ this yields an effective potential term in the equation for the only non-zero component $A$ of the perturbed vector potential 
\begin{eqnarray}\label{eq-schrd}
A''(z/d)&-&[(kd)^2+U(z/d)]A(z/d)=0, \cr U\equiv U_0&+&U_1=-\mu_0\sum\limits_{j=e,i}q_j\int\mathrm{d}^3v\, v_y\delta f_j(\mathbf{v}),
\end{eqnarray}
where $\delta f$ is the first order variation of the particle distribution function obtained from the linearized Vlasov equation, and $q_j=\pm e$. Through the Vlasov equation, $\delta f_j[A]$ is itself a functional of the vector potential $A$. Linearization reduces the functional dependence to linear proportionality $\delta f_j\propto A$. The Schr\"odinger-like equation (\ref{eq-schrd}) has been solved by perturbation technique \citep[e.g.,][]{galeev1975a,galeev1975b,galeev1979} assuming $U_1$ being small, and $U_0=-2\cosh^{-2}(z/d)$ forming a smooth potential well centered at $z=0$. To first perturbation order one obtains the condition
\begin{equation}
\int_{-\infty}^\infty \mathrm{d}(z/d) \left[U_{1e}+U_{1i}\right]=2kd\left[(kd)^{-2}-1\right].
\end{equation}
for the potential disturbance $U$, which yields a dispersion relation for the wave number $k$ tearing mode. The growing tearing mode is a long wavelength $k^2d^2<1$ mode. It causes pinching of the current into filaments forming `tears' (or plasmoids) containing closed field lines separated by magnetic X points with reconnected fields.

\subsection{Initial equilibrium}
Any kinetic theory, whether analytical or numerical simulations, requires the assumption of an initial equilibrium state. This, in a completely magnetized plasma containing an extended two-dimensional sheet current of width $d$ has been obtained by \citet{harris1962} under the assumption that the plasma is in thermal equilibrium obeying a Maxwell-Boltzmann distribution function. The equilibrium consists of a very simple shape $B=B_0\tanh(z/d)$ of the only available $x$-component of the magnetic field  $\mathbf{B}=B\mathbf{\hat{x}}$ which vanishes at the current center $z=0$ changing sign here when crossing from negative to positive $z$ values. The asumed pressure balance imposes an external $\beta=1$ everywhere. The sheet current maximizes in the current center: $\mathbf{J}=J_y\mathbf{\hat y}=\mu_0(B_0/{d})\cosh^{-2}(z/d)\mathbf{\hat y}$. With these definitions, index $j=e,i$, and $v_\perp^2=v_x^2+v_z^2$, the stationary particle distribution becomes
\begin{eqnarray}
F_j(z,\mathbf{v})&=& \left(\frac{m_j}{2\pi T_j}\right)^{\!\!\frac{3}{2}}\!\!\frac{N_0}{\cosh^2(z/d)}\cr
&& \exp\left\{-\frac{m_j}{2T_j}\bigg[v_\perp^2+(v_y-V_{j})^2\bigg]\right\}.
\end{eqnarray}
$N_0$ is the particle density {in the current sheet center}, and $V_{j}$ is the contribution of species $j$ to the current drift speed along $\mathbf{\hat y}$. It constitutes the current via $J_y(z)=eN(z)(V_i-V_e)$. As initial condition, this distribution function lies also at the base of most particle-in-cell and Vlasov particle simulations of collisionless reconnection. Whether it is experimentally justified, is an undecided question. Observations suggest \citep[e.g.,][]{christon1991} that the ion distributions belong to the type of $\kappa$-distributions exhibiting long nonthermal tails produced in collisionless wave-particle interactions \citep[see][for the microscopic derivation]{yoon2012}. Such distributions still satisfy the Vlasov equation but are typical for collisionless plasmas. They represent quasi-stationary states far away from thermal equilibrium \citep[e.g.,][]{treumann1999a,treumann1999b,treumann2008} which replace the Boltzmann distribution with parameter $\kappa[W_w]$ a functional of the wave energy density. They naturally account for an initial background of particles of higher energy \citep[for a recent review of the dynamics of particles in current sheets cf.][]{zelenyi2011}.

 \begin{figure*}[t!]
\centerline{\includegraphics[width=0.7\textwidth,clip=]{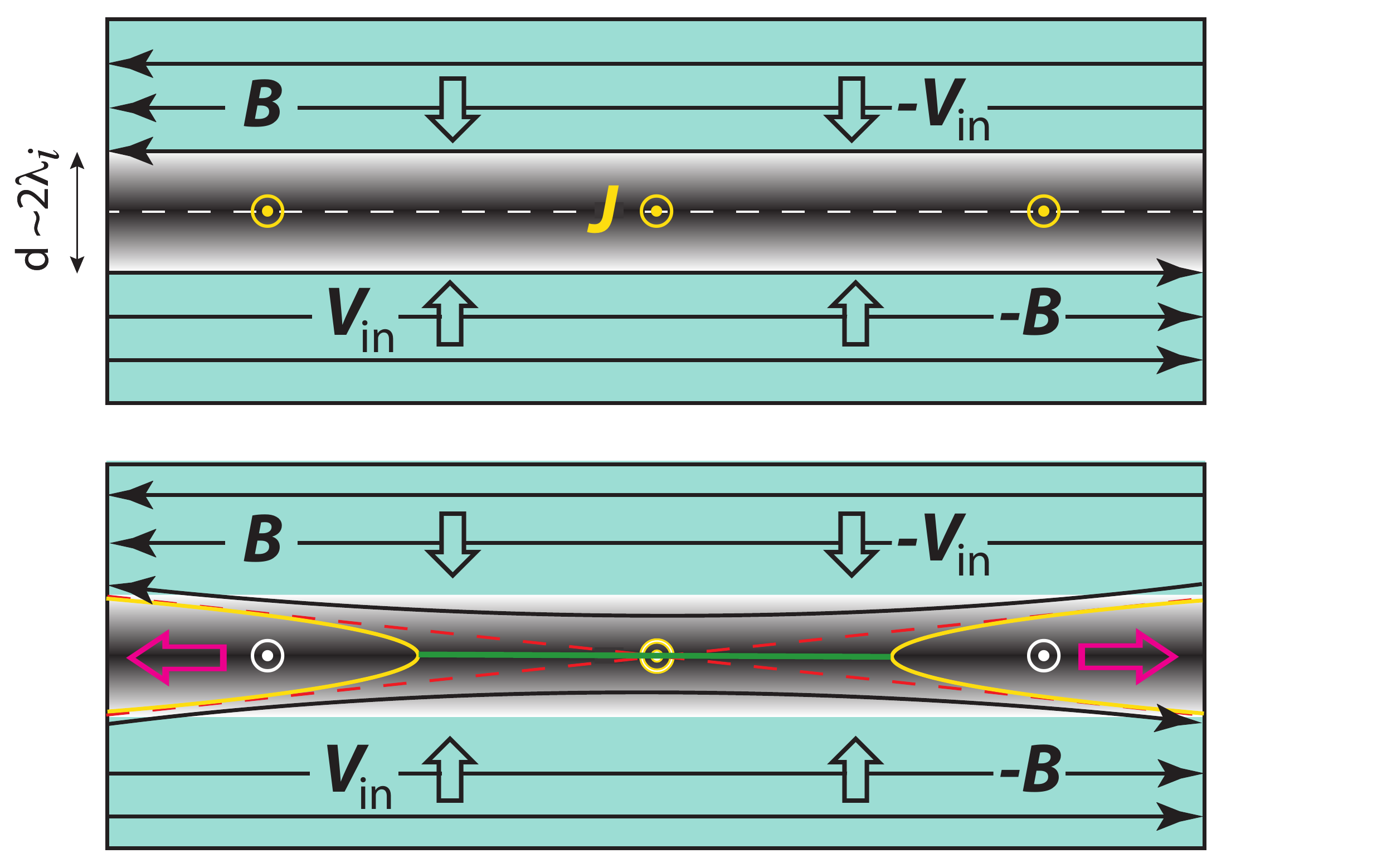} 
}
\caption[]
{\emph{Top}: Initial state of a two-dimensional thin current sheet $\mathbf{J}$ separating two antiparallel fields in a completely magnetized plasma. The width is of the current sheet is $d\sim2\lambda_i$. Open arrows show the plasma inflow from both sides before reconnection sets on. \emph{Bottom}: During reconnection the current sheet breaks off, and an X point is formed (indicated by the two dashed red separatrices). Connected field lines are shown in yellow forming plasmonds to both sides and producing two plasma jets (red arrows). The thin green line across the X point shows the extension of the \emph{electron diffusion region}. In the normal direction it is just two $\lambda_e$. along the current sheet its extension is much longer, possibly of the order of a few ion inertial lengths. In this region electrons are heated and jetted away from the X point to become two electron jets. These jets disappear at the end of the electron diffusion region where ions take part in the acceleration.}\label{fig-current}
\end{figure*}

One should note that these are two-dimensional equilibria assuming a flat infinitely extended current sheet of thickness $d$. Such sheets must be distorted in some way in order to undergo reconnection. In addition to the stable equilibrium current sheet distortions require a mechanism and an initial magnetic fluctuation level to start from. The distortion then selects some wave number range from this fluctuation level which becomes unstable and grows. Such magnetic fluctuation levels have only recently been calculated, for instance for the Weibel mode \citep{weibel1959}, assuming particular plasma response functions in an unmagnetized plasma \citep{yoon2007,treumann2012,schlickeiser2012,felten2013,felten2013a} which applies approximately to the very center of the current layer. \citet{felten2013a} recently reported a strongly damped isotropic mode which provides the highest level of magnetic thermal noise in an unmagnetized plasma. All these modes generate a fluctuating magnetic background from which other modes like the tearing may chose to grow. The spontaneous instability which is made responsible for reconnection is, however, the magnetized collisionless tearing instability \citep[for collision dominated plasmas first proposed by][]{furth1963}. Preliminary thermal magnetic fluctuation levels for this mode have been provided by \citet{kleva1982} in tokamak geometry, but  no full thermal fluctuation theory is available yet. 

\subsection{Collisionless tearing mode: Analytical theory}
Calculation of the contributions of electrons and ions in the above tearing dispersion relation can be done in either the absence \citep[e.g.,][]{galeev1979} or presence of residual Coulomb {\citep[e.g.,][]{zelenyi2013}} or anomalous collisions. The electron contribution is from oscillations of trapped electrons inside the electron inertial length, ions contribute nonmagnetically via Landau damping from the entire ion diffusion region. In the absence of collisions the trapped electron contribution, integrated over the cyclotron oscillations in the current sheet, is $\sim (c/V_{eA\perp})^2\sqrt{m_e/m_i}[1-(kv_e/\omega)^2]$, where $V_{eA\perp}$ is the electron-Alfv\' en speed in the weak magnetic field normal to the current sheet. The expression in the brackets contains the ratio of electron thermal speed to tearing mode phase velocity. Adding ion-Landau damping completes the  effective potential in Eq. (\ref{eq-schrd}). The total potential consists of a well with a hump in its center being formed by ion Landau damping and the additional rigidity of electrons in the normal magnetic field. The tearing, for becoming unstable, has to overcome both these humps, i.e. for being able to perform an oscillation in the main effective potential well it must manage to tunnel the central hump. In the absence of collisions and a normal magnetic field, the growth rate $\gamma_T=\mathrm{Im}\,\omega$ has been given by \cite{galeev1975a} for the electron tearing mode as $\gamma_{eT}=2kv_e\pi^{-1/2}(r_e/d)^{3/2}(1-k^2d^2)/kd$. Including both contributions from ions and electrons and a non-zero normal magnetic field {\citep[cf. also][]{galeev1975b,galeev1979,lembege1982,kuznetsova1991,pellat1991}}, the tearing instability needs to satisfy the following condition on the normal magnetic field $B_{z0}$:
\begin{equation}
\frac{k^3d^3}{(1-k^2d^2)(1+\theta)}<\frac{B_{z0}}{B_0}<\frac{(1-k^2d^2)(1+\theta)}{\sqrt{\pi \theta}}\left(\frac{r_i}{d}\right)^\frac{3}{2}
\end{equation}
where $r_i=v_i/\Omega_i$ is the ion gyroradius in the main field, $\theta\equiv T_i/T_e$, and $(kd)^2<d/r_i\lesssim1$ has been assumed {\cite[see also][]{schindler2006}}. Violation of this condition stabilizes the tearing mode. The upper limit is due to the residual magnetization of electrons. The stronger $B_{z0}$ the higher is the electron magnetization and consequently the hump in the potential well. When the size of the island reaches the current thickness, i.e. $kd\sim 1$, then the tearing mode stabilizes because electron magnetization becomes too strong. This is the theoretically maximum possible size of the tearing islands. On the other hand, this is also the size at which strong nonlinear effects are expected to come into play. {The combination of both limits implies a meta-stability of the tearing mode which has been applied to the stability of Earth's magnetotail during substorms \citep{galeev1975b,galeev1979,zelenyi2010b,zelenyi2011,zelenyi2013}. Recent extended three-dimensional simulations \citep{liu2013,leonardis2013,roytershteyn2012,roytershteyn2013,karimabadi2013a,karimabadi2013b} suggest that in three dimensions this inference for tearing stabilization may be invalid. The tearing mode in three dimensions is more easily destabilized due to oblique effects. Also, already in two dimensions, transient electron dynamics destabilizes the tearing mode at higher electron temperatures which may be  caused by energy dissipation and heating \citep{sitnov2002,divin2007,divin2012b} and/or} {embedding of the thin current sheet with a naturally present ion pressure anisotropy into a thick plasma sheet \citep{burkhart1992,zelenyi2008,zelenyi2010b}.} 

The lower limit is provided by ion Landau damping. It can be overcome by the presence of a sufficiently strong magnetic field $B_z$  normal to the current sheet. Therefore, the tearing mode will not grow spontaneously when the normal magnetic component vanishes. The current sheet is metastable awaiting the evolution of an increase in $B_z$ over the lower bound in the above equation in order to grow. This lower bound depends on the tearing mode wavelength. Preference is given to long-wavelength tearing modes. Moreover, in Earth's magnetotail we have {$3<T_i/ T_e<10$ \citep{artemyev2011,wang2012} with the most frequent values in the range of $T_1\sim (3\ \mathrm{to}\ 5)T_e$. Hence, the lower bound is the fraction $1/6$ of that in a plasma with $T_e\sim T_i$.} 

This linear theory of tearing mode growth does not yet account for a large number of additional effects such as weak collisions which contribute to a collision integral in the Vlasov equation and may both stabilize and destabilize the tearing mode further \citep{zelenyi2013}. In addition, the dynamics of electrons and ions in the plasmoids plays another essential role when the size of the plasmoids reaches the thickness of the current layer. These particles,oscillating in the plasmoids, are trapped in the plasmoids only for limited times and can be lost when passing the weak magnetic field in the X points, an effect which nonlinearly contributes to tearing growth because of heating ions and effectively decreasing the number of ions which contribute to Landau damping. If this happens, it has been shown \citep[cf., e.g.,][]{galeev1979} that the tearing mode might enter a regime of explosive growth. The estimated time for this effect is of the order of $\tau_\mathit{expl}\sim 10 \tau_\mathit{tear}$, with $\tau_\mathit{tear}$ the linear growth time of the collisionless tearing mode. 

Inclusion of collisions requires definition of the structure of the Boltzmann-collision integral in the Vlasov theory. In the simplest way this is done by assuming a BGK structure of the collision term. This leads to a modification of the Galeev-Zelenyi theory \citep{zelenyi2013}. Collisions, whether weak or strong, help breaking the frozen-in state of the electrons. So one expects that they diminish the stabilizing effect of magnetized electrons in the current sheet which is introduced by both normal magnetic components and also guide fields, which we have not yet discussed here. Since, however, the theory strongly depends on the assumptions of the structure of the collision term and the form of the anomalous resistance, i.e. the kind of waves which are responsible for generation of $\nu_a$, this is a difficult task which can barely be treated in sufficient generality by analytical means, in particular, because wave turbulence is genuinely anisotropic, the collision frequency will also become anisotropic. Moreover, anomalous collisions depend on the wave modes and through it on the distribution function which makes the problem nonlinear to a higher degree. The obvious way out is to switch to numerical simulations. This way has been gone since the early seventies. It is thus not surprising that, though the \citet{galeev1975a,galeev1975b} solution was intriguing, {until recently \citep{daughton2012} it has found less attention in the literature than it deserved. It was readily superseded by numerical simulation techniques. Most recent three-dimensional simulations \citep{daughton2012,liu2013,karimabadi2013b} do, however, surprisingly indicate that the idea of explosive reconnection and the evolution of multiply structured X points \citep{galeev1975b,galeev1979} may have been quite realistic. The three-dimensional investigation does, in addition, reveal that the above theory has to be corrected in two points: changes in topology due to the inclusion of the third dimension, and a different asymptotic limit for the growth rate which in three dimension has a pronounced dependence on the angle with respect to the magnetic field. The tearing mode in this case is oblique. The asymptotic tearing mode growth rate was given by \citet{daughton2012} as 
\begin{equation}
\frac{\gamma_\mathit{tear}}{\Omega_i}=\frac{1+\theta}{\sqrt{\pi\mu}}\left(\frac{r_i}{\theta d}\right)^3\left(1-k^2d^2\right)\frac{B_{y0}}{B_{x0}},
\end{equation}
with $B_{y0}$ the initial guide field. The growth rate decreases with increasing $\theta=T_i/T_e$ and $d/r_i$, and is limited to waves with $kd<1$. Companion three-dimensional simulations demonstrate, however, a much more complicated evolution of reconnection. This will be discussed below in relation to guide and normal field reconnection.} 

\subsection{Guide fields}
Another property of interacting plasmas in nature is their magnetization state. Frequently currents flow along a superimposed guide magnetic field which can be both weak and very strong. Such guide field reconnection has different properties. If the fields are weak, as is the case when self-generated quadrupolar Hall magnetic fields overlay over the external field in the ion inertial region, ions are accelerated in the ion inertial region and cause broadening of the current layer. If the electron inertial region at the center of the current layer remains free of such guide fields, reconnection is going on there in the sublayer of the gross current flow. 

If, on the other hand, the guide field points in the direction of the current and is externally applied -- as is the case at the magnetopause -- completely different effects set on. The guide field magnetizes the electrons to some degree thereby setting a threshold on the onset of reconnection. However, the presence of the guide field along the current, together with the external electric field which is also directed along the current, allows for acceleration of the electrons antiparallel to the electric field. This acceleration can become strong enough to excite the Buneman instability which has two effects: it causes anomalous resistance, heating of the bulk electron component, and it produces chains of electron holes along the guide field. Similar effects may already occur for Hall magnetic fields as well. 

Electron holes have a number of consequences. They structure the plasma in electric field-free regions and localized region of strong electric fields. They split the electron distribution into trapped and passing particles. They cause further heating of the plasma, generate cold electron beams and possibly radiation, transforming a strong current layer into a source of plasma waves and radiation, which might become of interest in astrophysical application. Strong guide fields, on the other hand, result mainly in these effects. Though reconnection takes also place in this case, it is restricted to the weak current-generated magnetic field and does not dissipate the energy of the strong field.  Below, guide field reconnection will be discussed in connection with numerical simulations of reconnection under various conditions.

\section{Simulation studies}

Since analytical theories of the complete complexity of reconnection are unmanageable, the proper and efficient approach is via numerical simulations. Simulations started in the mid-seventies where there were based mainly on MHD. Particle simulations were for a while inhibited by the necessity to include at least two spatial dimensions and thus large particle numbers. These remained intractable by the then available computer capacities. Two-dimensional particle-in-cell simulations became available to large extent in the nineties. Now, almost every decade, computational progress is made stepwise by increasing particle numbers, simulation boxes, more sophisticated asymmetric set ups, boundary conditions, flows, and including, recently, the highly desired three dimensions. 

\subsection{Preliminaries}
Contrary to early belief, reconnection is a \emph{microscopic} plasma process capable of breaking the frozen-in condition prescribed by the electrodynamic induction equation in a moving collisionless plasma. This breaking depends on the microscopic properties of the plasma in the current layer and its vicinity. It occurs in the so-called \emph{electron diffusion region} centered at the current sheet and being of width $d_e\sim\lambda_e$ (see Fig. \ref{fig-current}).  It is questionable whether any MHD theory could properly describe the reconnection process; MHD  applies only to the magnetic configuration and plasma flow on scales $>\lambda_i$ exceeding the ion inertial length \emph{after} reconnection has happened.  \citet{birn2004}, comparing kinetic and MHD simulations, interpreted the results as equivalent with the only difference of a lower MHD reconnection rate, i.e. longer time to reach the `same final state'. However, recent three-dimensional macro-to-microscopic scale simulations \citep[cf.][for a most recent review]{karimabadi2013b} completely vindicated these claims. Reconnection took place exclusively on the electron scale.  In this section we briefly review the recent achievements.

A general problem of any simulation is the initiation of reconnection. In most simulations this is artificially done imposing a disturbance in the center of the current sheet and box. It has turned out that the best approach is to impose an initial weak X point \citep{zeiler2000,zeiler2002}, an  effective approach in both two and three dimensions. Even a weak initial disturbance is, however, already a nonlinear disturbance and thus skips all the interesting physics which causes reconnection to start by itself. Such an approach makes sense if one is rather interested in applications, for the reconnection events encountered in nature are without exception well developed have for long left the initial growth phase. A shortcoming of this approach remains. Since the initial perturbation is necessarily large it skips any possible state where the reconnection process levels nonlinearly out at smaller amplitude than the initial disturbance. This approach therefore misses an entire class of fast but weak reconnection events. In those simulations without an initial disturbance \citep[e.g., three-dimensional simulations performed by][]{kagan2013} onset of reconnection is due to sufficiently large \emph{numerical} fluctuations which either fake an anomalous collision frequency or \citep[in really collisionless settings like those by][]{karimabadi2004a,pritchett2005b} start from scratch, or an initial as well as possible defined thermal level. Unfortunately numerical fluctuations are not localized, and thus those simulations belong rather to the class of a low but finite \emph{homogeneous} resistivity in the entire simulation box than to really collisionless conditions where anomalous resistances are caused self-consistently by collisionless wave-particle interactions. The two-dimensional simulations by \citet{pritchett2005b} without initial disturbance gave no indication of a saturation at low reconnection amplitudes. This provides confidence in the abbreviated approach of flux perturbation.

The most urgent challenges on the numerical simulation of magnetic reconnection by particle-in-cell codes in near-Earth space for the first decades of the 21. Century have been laid down in a widely referenced paper by \citet{birn2001}. They concern among others the identification of the reconnection process in the electron inertial layer, the structure of the electron diffusion region, asymmetric reconnection, forced reconnection, and as main challenges three-dimensional reconnection (including simulation-technical problems) and the transition to macroscopic effects. In the following we very briefly review the state of the art until about 2010 before referring to new evolutions that started in the past five years.

\subsection{Absence of guide fields}
This is the class of the overwhelming majority of published simulations in two- and, to a lesser extend, also in three-dimensional reconnection. The majority of simulations are of course two-dimensional in space, applying either periodic conditions, which implies the simulation of two antiparallel current sheets, i.e. doubling the simulation box. Otherwise open boundary conditions are used where plasma can readily escape and is thus lost. In the former case the simulation is valid only as long as the two sheets do not affect each other. In the latter case plasma flows freely out of the box and is either replaced or not. Limits are reached when the plasmoids generated reach the boundary, the current sheets start interacting (which is advantageous for investigating interacting current sheets, however) or when the plasma becomes locally too strongly diluted by jetting away from the X point. Any results depend highly on the settings and thus may differ considerably, in many cases they are incomparable which makes drawing conclusions about the general validity difficult. 

 \begin{figure*}[t!]
\centerline{\includegraphics[width=0.7\textwidth,clip=]{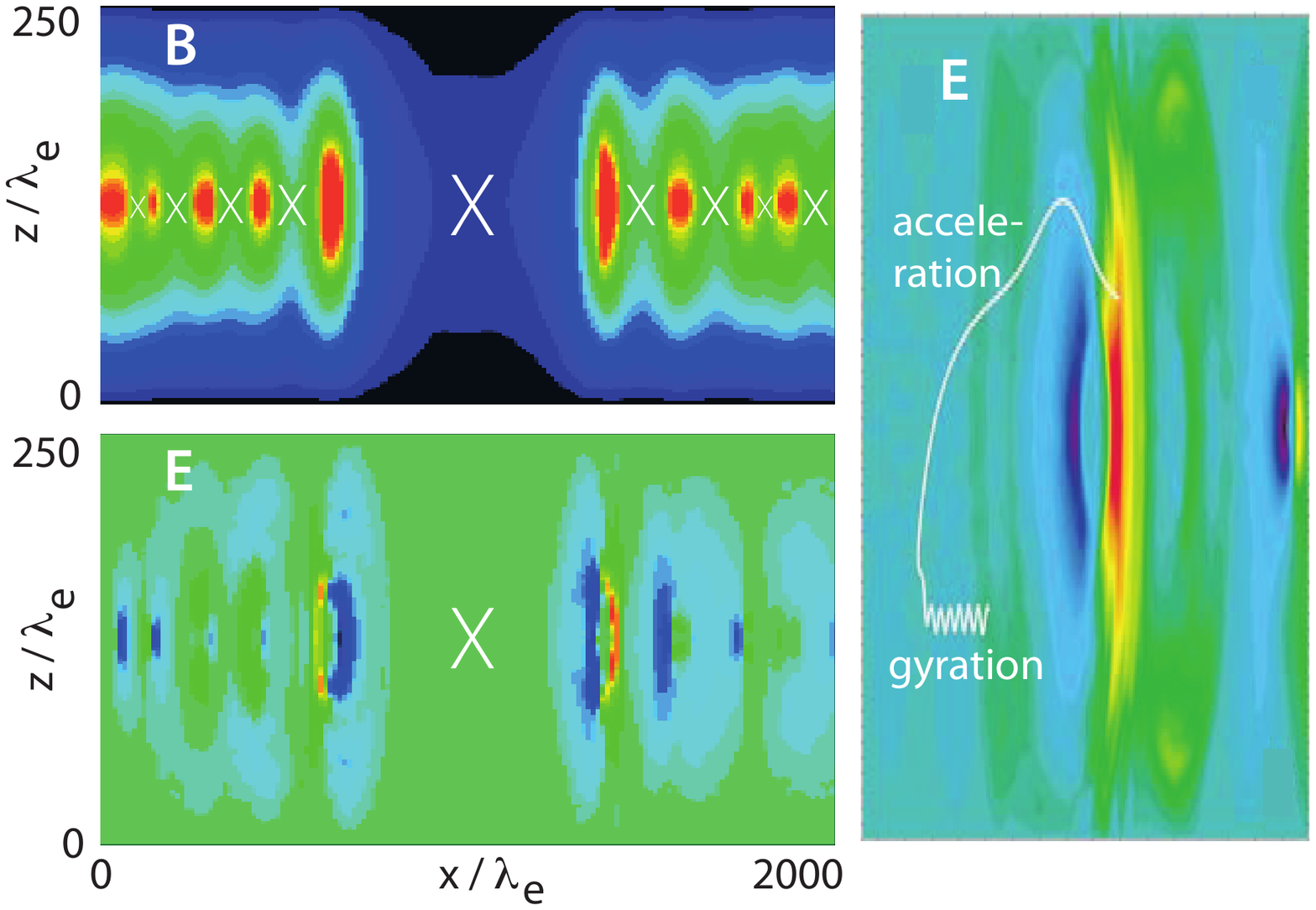} 
}
\caption[]
{Three-dimensional simulation of non-Hall reconnection in a pair plasma of high resolution in a large domain (simulation data taken from \citet{jaroschek2004a}.  \emph{Top}: Magnetic flux in color representation (only relative scales are given here). Contours indicate the magnetic topology. Several X point are generated separated by plasmoids (red blobs), indicating evolution of a tearing mode. The initial reconnection region dominates all secondary X points. These have their own dynamics. Note that the horizontal extension (length of the diffusion region) is several times its vertical extension indicating jetting. \emph{Bottom}: Electric fields (red is positive, blue negative). Strong fields evolve around the plasmoids causing electron-positron acceleration. Plasma jetting occurs around each X point thus leads to competition, jet braking and deviation. In this 3-dimensional simulation the X points have finite extension perpendicular to the plane shown. \emph{Right}: Orbit of one selected accelerated electron in the electric field near the first right plasmoid. Originally the electron performs a meandering oscillatory gyration in the current sheet between the two opposing magnetic fields until feeling the reconnection electric field, picking up energy and enlarging its gyroradius. At the end of the orbit shown it just enters the strong field on the backside of the plasmoid to become further accelerated.}\label{fig-halljar}
\end{figure*}

\subsubsection{Pair plasmas -- no Hall reconnection.}
Natural pair plasmas don't exist in near-Earth space. In reconnection theory they just serve for the most simple particle arrangement and simulation model. Because $m_e\equiv m_i$ in pair plasmas, the Hall effect is naturally absent, allowing for checking non-Hall reconnection. This was done first by \citet{jaroschek2004a,jaroschek2004b} in a \emph{three-dimensional fully relativistic} electromagnetic  particle-in-cell simulations with open boundary conditions in $x$ and periodic boundary conditions in $z$ and $y$. The simulation was repeated with the only emphasis on the non-Hall aspect in a strongly simplified \emph{two-dimensional} mini-set-up by \citet{bessho2005}. The relativistic treatment becomes necessary in a proper simulation in order to account for retardation, correct local charge neutralization, and to resolve the Debye scale. This is done by starting from the J\"uttner distribution function which properly accounts for the relativistic tail on the Boltzmann distribution (no bulk streaming is included in order to avoid non-reconnection shock formation and unwanted relativistic effects). Reconnection was ignited by imposing a weak initial disturbance of the current sheet.

A two-dimensional cut through the pair simulation box is shown on the left in Fig. \ref{fig-halljar} demonstrating that reconnection does indeed evolve. It, moreover, evolves on a very fast time scale and does not lead to Sweet-Parker reconnection. Several X points form rapidly and, most important, strong quasi-stationary inductive electric fields are generated at the plasmoid edges. Such electric fields have also been found recently in non-pair plasma simulations but were first observed here. They play a key role in particle acceleration in reconnection. 

The pair reconnection shown in Fig. \ref{fig-halljar} is dominated by one main X point that expands rapidly up to a final length scale roughly ten times the width of the initial current layer $\ell_{-}\sim 10 d_e$. This scale is determined not only by the expansion of the main X point but also by the reaction of the secondary X points and plasmoids. Each of them ejects quasi-symmetric pair plasma jets into both horizontal directions.  The jets interact with the neighboring X points, braking their the expansion and thereby shortening the extension of the jets. This is the reason for the diffusion layer of the main X point to terminate at a shorter scale, while all secondary diffusion regions are much narrower. 

Fig. \ref{fig-halljar} shows slices in the $x-z$ plane. The simulations were three-dimensional, however, permitting to infer about the extension of the reconnection site into $y$-direction. The finding was that the width of the reconnection site in $y$ was of the order of $\lambda_e$, the inertial scale in these simulations. Limitations in particle number and computer time did not allow to extend the box such that no information could be gathered about the further evolution of the reconnection site in $y$. However, some interesting inferences were drawn on wave excitation and particle acceleration (see the respective paragraphs below).

{\citet{daughton2007} performed similar pair plasma simulations applying various box sizes  between $75\times 75\lambda_e$ and $10^3\times10^3\lambda_e$ (the latter growing out of numerical noise, compared to the simulations of \citet{jaroschek2004a} reproduced in Fig. \ref{fig-halljar} being large-scale in the normal direction) also using open boundaries. Formation of many X points and plasmoids of different sizes was found. The focus was on the initial evolution of the tearing mode, determination of growth rates, interaction of plasmoids, and formation of the diffusion region. The main conclusion of this study  is that pair plasmas in these simulations never reached a final steady state reconnection regime, in agreement with the findings of \citet{jaroschek2004a}. The evolution of the tearing mode is highly dynamic with formation of an elongated unstable current layer which decays into a long series of plasmoids ejected along $x$ and mutually interacting.  Average reconnection rates remain fast and are insensitive to the size of the system. The authors conclude that pressure tensor effects cannot prevent the elongation of the current layer in pair plasmas thus inhibiting the final stationary state.}

\begin{figure*}[t!]
\centerline{\includegraphics[width=0.7\textwidth,clip=]{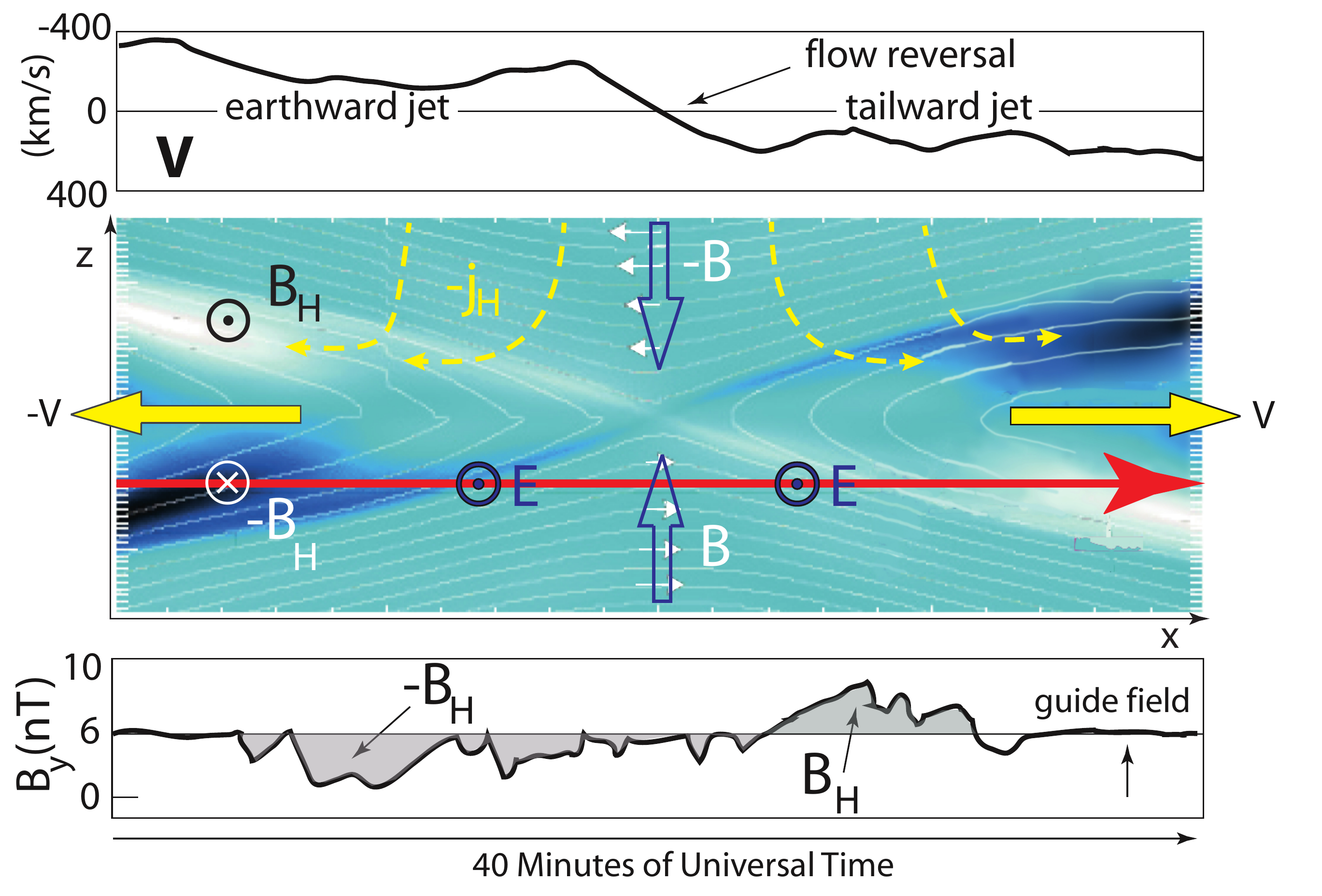} 
}
\caption[]
{Hall fields, observation and simulation. \emph{Central panel}: Two-dimensional simulation of  reconnection including different electron and ion dynamics \citep[data taken from][]{vaivads2004}. Shown is only the magnetic field with background field lines drawn in thin white.  Dark blue regions indicates positive Hall-$B_y$, white regions negative Hall-$B_y$. The Hall fields concentrate along and inside the separatrices and boundaries of the two plasmoids where the Hall electrons (dashed yellow lines) turn away from the X point. Hall fields twist the original magnetic field. They can be considered as self-generated magnetic guide fields in the ion diffusion region. Such guide fields embedded into the convection electric field $E_y$ (black circles) may accelerate electrons out of the plane thereby strengthening the current in the ion diffusion region. Open arrows show convective inflow, yellow arrows outward jetting of plasma. \emph{Top panel}: Plasma velocity measured along the spacecraft orbit (long red arrow in the central panel) in Earth magnetotail during a substorm reconnection event. Flow reversal is seen during passage near the X point indicating the two plasma jets emanating from X. \emph{Bottom panel}: Measured out-of-plane (Hall) magnetic field during the flow reversal event exhibiting the antisymmetric geometry of the Hall field \citep[data (schematized) after original measurements of][]{oieroset2001}. Note the presence of a weak (6 nT) magnetic guide field in $y$ direction.}\label{fig-vaivads}
\end{figure*}

\subsubsection{Electron-proton plasmas.}
Plasmas in Earth's environment consist of electrons and ions. Particle-in-cell simulation studies of collisionless reconnection in electron-proton plasmas are ubiquitous both in symmetric and non-symmetric current sheet configurations with or without guide fields. Depending on the set-up and the focus of the subsequent analysis they show a large variety of effects. Reviews of the achievements on various aspects (onset, Hall fields, current structure, field and current topology, bifurcation, dissipation, heating, acceleration, wave generation etc.) before 2010 is found, for instance, in \citet{birn2012}, \citet{buchner2007}, \citet{califano2007}, \citet{fujimoto2011},  \citet{karimabadi2013a}, \citet{mozer2011}, \cite{pritchett2009a}, \citet{yamada2010}, \citet{zelenyi2013}. Here we just summarize some of the most interesting insights before turning briefly to the most recent investigations.

\paragraph{Hall fields.} 
An important finding in particle-in-cell simulations was the confirmation of the difference between electron and ion dynamics near the current sheet center leading to the generation of Hall currents and fields, as predicted by \citet{sonnerup1979}. Figure \ref{fig-vaivads} shows an example of the fully developed magnetic field structure in a two-dimensional simulation of a thin current sheet with the background magnetic field exhibiting the typical X point-separatrix-plasmoid topology. The quadrupolar Hall field $B_y$ is along $y$ out of or into the plane. In the dark blue regions $B_y$ is positive, in the white regions negative. Hall fields maximize just inside but along the separatrices and at the outer plasmoid boundaries where the Hall electrons turn around. Added are confirmation measurements in the Earth's magnetotail \citep{oieroset2001}. The top panel shows the bulk velocity-field reversal typical for plasma jetting away from the X point. The bottom panel contains the magnetic out-of-plane field ($B_y$) exhibiting the plus-minus sequence expected of a Hall field near the X point. One may note that in these observations the Hall field undulates around a finite stationary guide field of 6 nT amplitude.    
\begin{figure*}[t!]
\centerline{\includegraphics[width=0.7\textwidth,clip=]{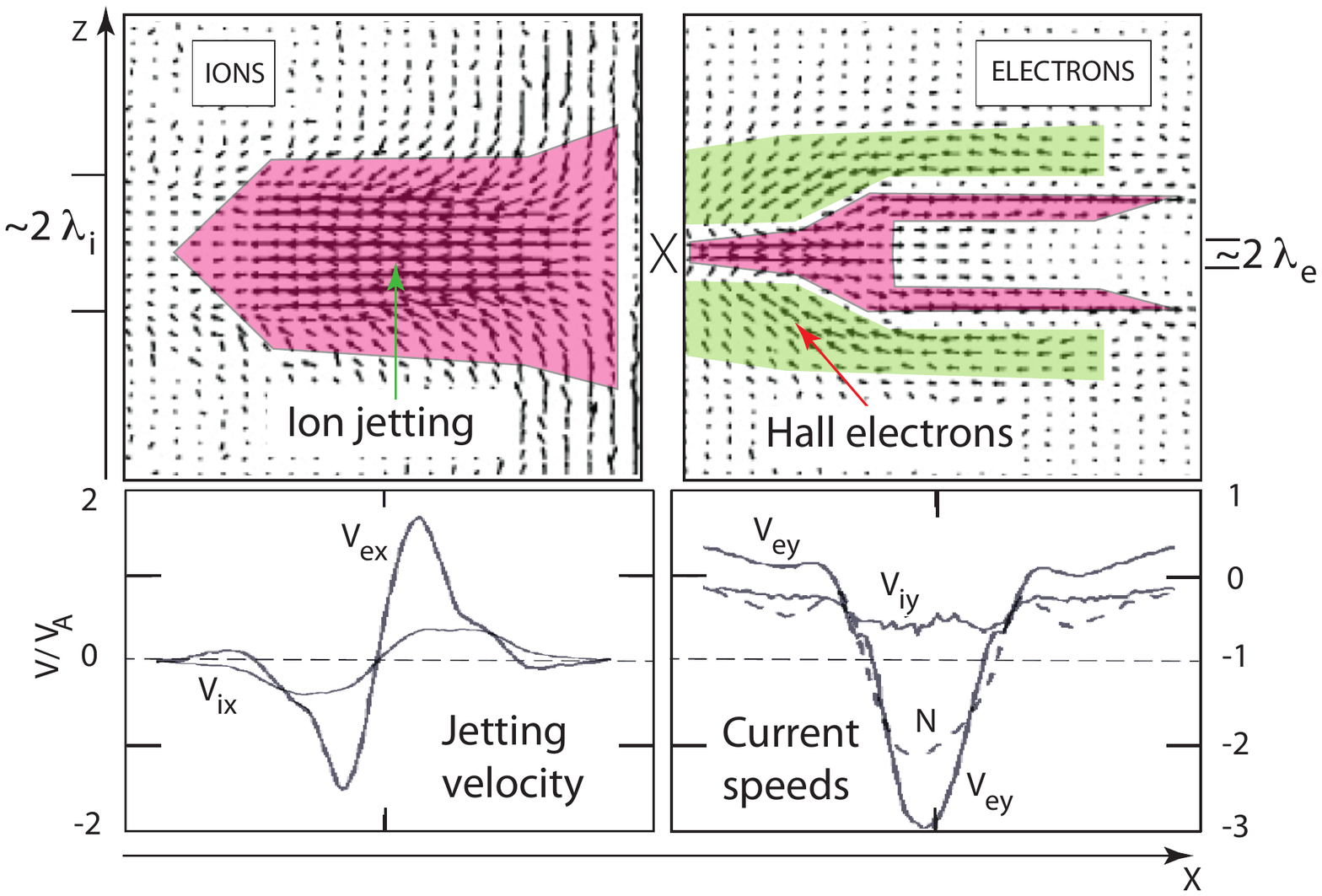} 
}
\caption[]
{Particle-in-cell simulations with small mass ratio $m_i/m_e=25$ \citep[simulation data  after][courtesy American Geophysical Union]{pritchett2001a}. \emph{Top panels}: Velocities near the X line. \emph{Left}: Left half (negative $x$) of space for ion velocity. Ions flow in from positive and negative $z$ before becoming diverted to negative $x$ by jetting from reconnection site in a region of width in $z$ proportional to $\lambda_i$. \emph{Right}: Right half (positive $x$) of space for electron velocity. Electron inflow continues across the ion inertial region until close to the center of the current sheet mapping the electron Hall current flow. In the current sheet center the electron velocity turns around by close to 180$^\circ$ to participate in the plasma jetting (in this half plane into positive $x$) in a region of width proportional to $\lambda_e$. \emph{Bottom}: Jetting and currents exhibiting different ion and electron dynamics. \emph{Left}: Electron and ion jetting velocities. Ion mass flow in jets barely reaches $0.5\, V_A$. Electron jetting comes up to $\sim 1.7\, V_A$. \emph{Right}: Current speeds indicate deceleration for the ions at the X point. Electron velocities are inflected such that the current in the center is carried about solely by electrons. }\label{fig-prit-1}
\end{figure*}

The pronounced differences in electron and ion flows in the ion inertial (diffusion) region has been nicely confirmed in particle-in-cell simulations by \citet{pritchett2001a} as shown in a summary plot in Fig. \ref{fig-prit-1}. The left part of this figure shows the usual ion jetting away from the X point, also clearly indicating that only a small fraction of the ion flow is passing the X point. At the boundaries of the ion jet discontinuities develop where the ion flow suddenly turns around. These are the separatrices.  Electron jetting sets on only inside the electron diffusion region which, in these simulation, has been artificially widened by assuming a small mass ratio such that $\lambda_i/\lambda_e=5$. The lower panels show the profiles of the jet and current velocities. Due to the large ion inertia, ion jet speeds barely reach half an Alfv\'en speed. Electron jetting in the narrow central region comes close to two Alfv\'en speeds thus carrying most of the jet current. Similarly, ion current velocities are decelerated in the X point region, while electron speed become completely deflected from positive to negative velocities, an effect of reconnection, such that the reconnection current in the X point region is carried almost solely by the electrons. \citet{pritchett2001b} also included the third dimension with open boundary conditions in order to investigate the extent of the electron diffusion layer and distortion of its two-dimensional symmetry. No such distortion was detected except for the evolution of an electric component  $E_z$ that is required by pressure balance. {These conclusions are valid in the absence of guide fields. When guide fields are included (see the corresponding section on guide field reconnection below) the Hall fields become distorted, asymmetric and compressed. This has been demonstrated by \citet{daughton2005} and \citet{karimabadi2005a}. A thorough comparison between different theory based simulations (hybrid, Hall-MDH and non-Hall hybrid, where the Hall term is removed) has been undertaken by \citet{karimabadi2004b,karimabadi2004c} focussing on ion-kinetics. Reconnection was found to be independent on the Hall effect even in these ion-kinetic simulations (including an anomalous resistivity) thereby anticipating strictly Hall-free simulation results \citep{jaroschek2004a}. Including the Hall effect reconnection turns becoming asymmetric, and the X line grows in the direction of electron drift with current carried by electrons. }

Hall fields were observationally inferred first by \citet{fujimoto1997}, followed in this sequence by \citet{nagai2001} and \citet{oieroset2001}, all in Earth's magnetotail. They represent \emph{self-generated guide fields}. Only along the separatrices they approach the electron diffusion region near the X point. Since the convection electric field $E_y$ is also along $y$, it acts as a guide-field-aligned electric field for the Hall electrons which are magnetized in the background field $\pm B_x$. For this reason electrons become accelerated in $E_y$ in the direction opposite to $E_y$. This acceleration amplifies the current in those domains where the Hall magnetic field is remarkable, an effect that leads to current bifurcation outside the reconnection site in the ion diffusion region. Bifurcation was observed first by \citet{runov2003} {and \citet{asano2003,asano2004}} in Earth's magnetotail current sheet. Though several different mechanisms have been proposed to produce current bifurcation \citep[cf., e.g.,][and others]{daughton2004,karimabadi2005b,ricci2005,zocco2009,jain2012}, one possibility is its ion-diffusion-region nature.

\begin{figure*}[t!]
\centerline{\includegraphics[width=0.7\textwidth,clip=]{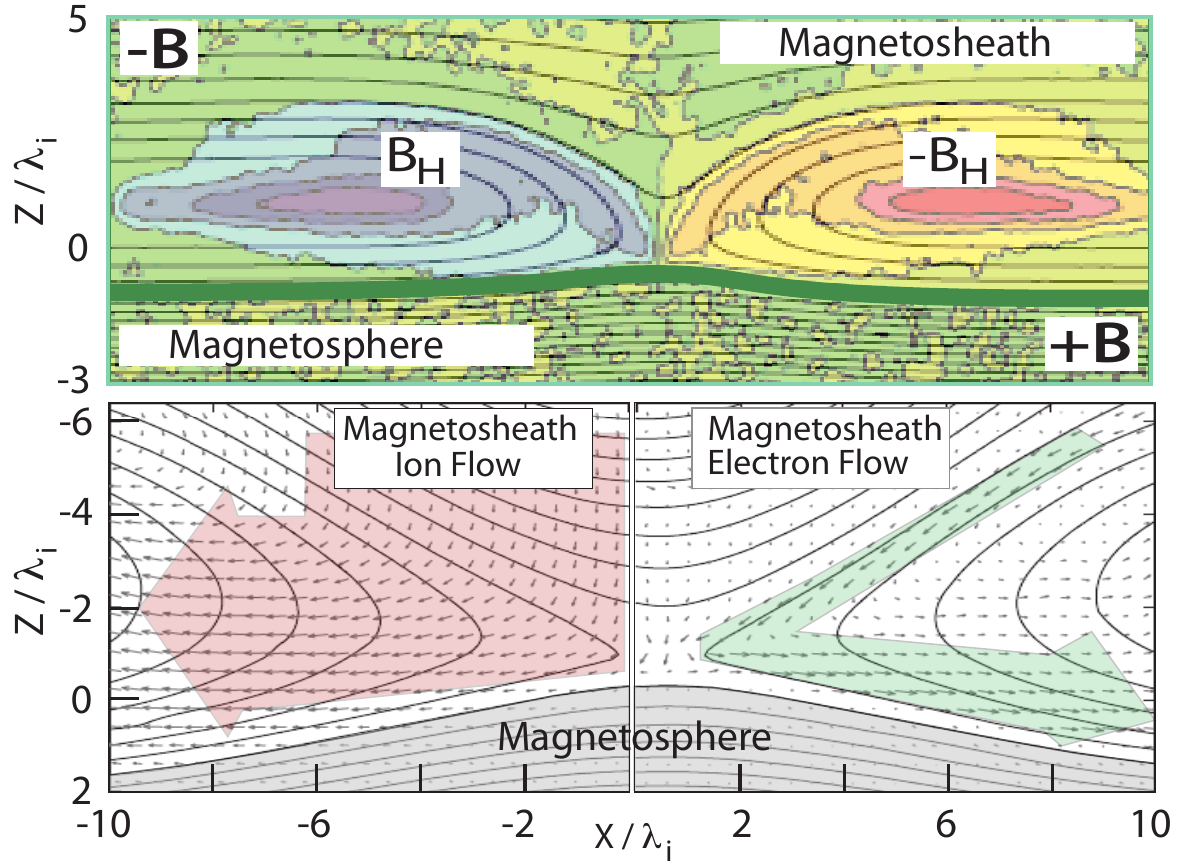} 
}
\caption[]
{Non-forced non-guide field particle-in-cell simulations of asymmetric magnetopause reconnection with(\emph{Top}) Hall field $B_H$ in color coding overlaid on background field  \citep[simulation data  after][courtesy American Geophysical Union]{pritchett2008}. The fat green line is the magnetopause which separates the magnetosheath from the undistorted magnetosphere. The magnetopause  remains intact except for having been pushed down from its original position at $z=0$ to roughly $-z=-1$ (in units of $\lambda_i$) and slightly distorted. The magnetopause coincides approximately with the inner separatrix. Only two large Hall field vortices develop on the magnetosheath side. \emph{Bottom panels}: Ion (left) and electron (right) bulk flows on the magnetosheath side develop (only half spaces are shown). Jetting of electrons is restricted to a narrow domain only along the magnetopause.}\label{fig-prit-mp1}
\end{figure*}

\paragraph{Asymmetric reconnection effects.}
Most cases of reconnection occur in interaction of plasmas with unlike properties. Such reconnection is non-symmetric. A famous example is reconnection between the interplanetary (solar wind) magnetic field and the geomagnetic at Earth's magnetopause. In contrast symmetric reconnection like that in the tail of the magnetosphere is a rare case. 

Theoretically one expects that asymmetric reconnection affects mainly the weaker magnetic field side than the high field side. This was demonstrated in asymmetric simulations under conditions prevalent at the magnetopause \citep{pritchett2008}.  Figure \ref{fig-prit-mp1} gives an impression on the \emph{non-forced asymmetric} case with \emph{no guide field} imposed. The most interesting effect is probably that the strong-field magnetosphere remains well separated from the distorted region by a slightly deformed but stable magnetopause which itself is adjacent to two legs of the separatrix system. The two newly formed plasmoids and the X point lie entirely on the weak field side. In contrast to the symmetric case, on the highly disturbed weak-field side just one \emph{dipolar} Hall current system $\pm B_H$ evolves along the outer edges of the two plasmoids (shown in violet and orange). The second (lower) Hall dipole is completely suppressed as there is no electron inflow from below. Electrons and ions flow in from the top and become diverted into jets along the magnetopause. This is shown in the lower part of Fig. \ref{fig-prit-mp1}. 

Partial support to these simulations has been given by magnetopause observations of reconnection events \citep{mozer2002,phan2004}. Observations should show jetting and Hall field signatures only outside though close to the magnetopause. The zero-field crossing should be asymmetrical if in agreement with the non-guide field simulation. In this point the observations disagree since the magnetic field is close to symmetry and one leg of the Hall field seems to lie inside the magnetosphere. This discrepancy with the simulations might possibly have been caused by not having recognized the presence of a finite guide field which should be subtracted from the data. {Recently \citet{malakit2013} found that asymmetry in reconnection is responsible for another way of contributing to reconnection electric field generation.} 

\subsubsection{Electron diffusion region.}
{Structure and extension of the electron diffusion region has attracted interest since this point had been made first by \citet{daughton2006} and \citet{fujimoto2006} and was investigated by simulations \citep{karimabadi2007,shay2007} who identified several scales in the diffusion region: an inner diffusion region and an electron outflow region or electron ``exhaust". Observations in space are difficult to perform. They have become available only recently \citep{scudder2012}.}

The width of the electron diffusion region equaling the electron skin depth applies to the direction \emph{perpendicular} to the current sheet. The length $\ell_-$ of the electron diffusion region is, however, a priori undefined along the current sheet and, in the absence of reconnection, can in principle be as long as the two-dimensional current sheet itself, for the electrons are in the absence of a guide field, by definition, unmagnetized in its center. Once reconnection has set on, however, a weak normal component of the magnetic field $B_z$ perpendicular to the current sheet is generated -- a process called ``(re-)dipolarization''. $B_z$ is strictly zero in the reconnection X point as well as in the center of each plasmoid of the X point-plasmoid (``magnetic island'') chain constituting the tearing mode. Outside, this component re-magnetizes the electrons. Hence, in plasmoids the length of the electron diffusion region is a fraction of the extension of the plasmoid along the current sheet, i.e. the wavelength of the tearing mode. The length $\ell_{-}$ of the diffusion region in the X point is less well defined. It is not identical to the extension of the Joule dissipation region defined earlier in Eq. (\ref{eq-dissipmeasure}) \citep{zenitani2011} which is the inner region of \citet{shay2007}, the region of unrecoverable Joule heating. It also includes the length of the electron jetting or exhaust which depends only weakly on the presence of the nonmagnetic ions but includes some energy transfer from electrons to ions resulting in ion heating. This is shown in observations of electrons at the geotail reconnection site and in accompanying simulations in Fig. \ref{fig-zenit-1}, both referring to \emph{symmetric} reconnection conditions (without or with only weak guide fields). 
\begin{figure*}[t!]
\centerline{\includegraphics[width=0.7\textwidth,clip=]{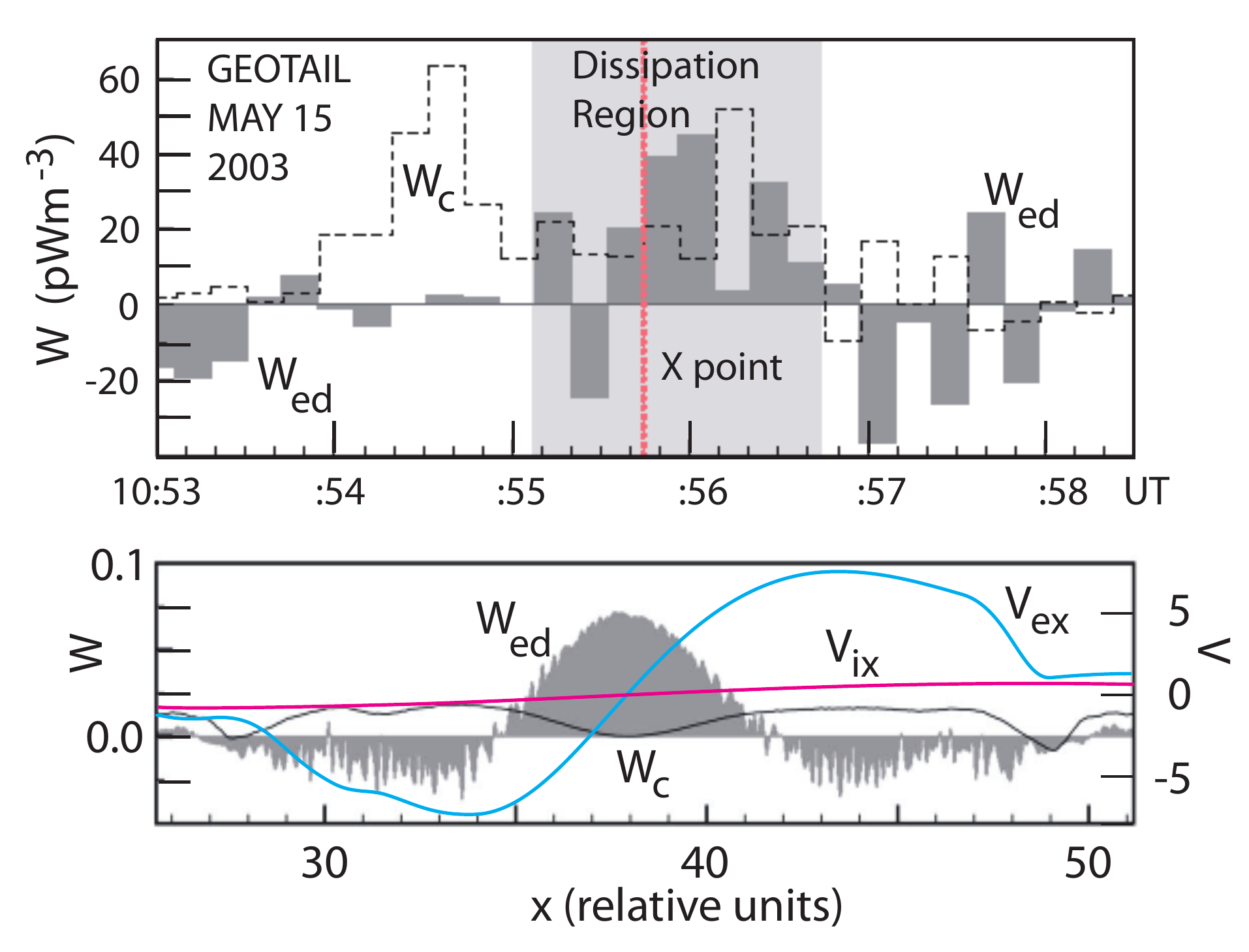} 
}
\caption[]
{Determination of the energy dissipated in the magnetospheric tail current sheet during ongoing reconnection from observations of Geotail \citep[observations and simulation data taken from][courtesy of  American Geophysical Union]{zenitani2012}. \emph{Top}: Geotail data of electron energy $W$ versus universal time UT. The grey histogram is the dissipated electron energy $W_{ed}$ determined from the measurements. $W_c$ is the convection energy. The vertical red line shows the time of crossing the location of the X point as estimated from the magnetic and plasma profiles when passing the current sheet. The electron dissipation region is indicated by grey shading. The dotted histogram shows for comparison the convective energy term in Ohm's law. \emph{Bottom}: Accompanying particle-in-cell simulations for the observational conditions. Shown are the mean electron (blue) and ion (red) velocities as well as the simulated dissipated and convection energies $W_{ed}$ (grey shading) and $W_c$ (almost unaffected black curve) along a cut through the X point in the center of the current sheet. The dissipated energy is positive in the dissipation region, where it maximizes thus indicating electron heating/acceleration. Acceleration of electrons into formation of two quasi-symmetric electron jets in opposite $x$-directions is indicated by the blue curve. Note that the electron jets extend substantially out from the dissipation region until becoming braked and assuming the same speed as the ions. Here the quasineutral plasma jets are born. Weak braking starts already soon after the electron jets have left the dissipation region indicating interaction with the ion fluid already here. Energy transfer from electrons to ions is indicated by negative dissipation energies $W_{ed}$. Since no effect is seen in the ion bulk flow the dissipated electron jet energy goes into ion heating. 
 }\label{fig-zenit-1}
\end{figure*}

The theoretical definition of $\ell_{-}$ can be based on the gyroradius of electrons in the normal field $B_z$ in the reconnection region: $r_e(B_z,x)\gtrsim \lambda_e(z)$. This, however, is a highly variable quantity, because $ B_z$ increases with distance from the X point until ultimately becoming comparable to the external field $B_0$. Moreover, electrons are strongly heated during reconnection. In addition, magnetic stresses accelerate them into a fast {electron jet} of super-electron-Alfv\'enic speed. Both effects increase the electron gyroradius in the $B_z$ field. The pure electron jet ceases once ions become involved. Then ion inertia retards the jetting electrons, and a quasi-neutral reconnection plasma jet forms. The length of the electron jet must be at least $\textstyle{\frac{1}{2}}\ell_{-}\gtrsim\lambda_i(0)\sim 43\,\lambda_e(0)$, where the inertial lengths are based on the density in the current center. This is then also the theoretical one-sided maximum length of the diffusion region. The total extension of the electron diffusion region therefore becomes roughly the order of $\sim 100\,\lambda_e$. This value can be larger since the electron density in the diffusion region decreases due to outflow. If the decrease is a factor of ten, then the total length of the exhaust can become several 100 $\lambda_e$ or several 10 $\lambda_i$.

{ \citet{shay2007} determined a scaling $\propto\mu^{-3/8}\lambda_i$ of the extension of the electron sheet current (inner dissipation region) along $x$,  indicating the shrinking of the length of the electron current layer to $\sim\lambda_i$ at a realistic mass ratio. Their estimate is based on the competition between the Lorentz force $v_xB_z$ of the electrons in the $B_z$ field and the reconnection electric field $E_\mathit{rec}$. At the point where these two become of equal magnitude, the electron current vanishes and reverses sign outside. \citet{karimabadi2013a} provided arguments for an extension of the order of $\sim2.5\lambda_i$ based on a comparison between the electron and ion based diffusion (reconnection) rates and the distinction between the fields $B_e, B_i$ upstream of the electron and electron dissipation regions, respectively.  The electron exhaust can be longer. \citet{karimabadi2007}  found elongations of up to several tens of $\lambda_i$ depending on convective plasma outflow and the nonideal divergence of the electron pressure tensor (thereby supporting the importance of the latter). Such lengths are in overall agreement with observations in the magnetosheath \citep{phan2007a}.  }

\citet{le2009} showed that trapping of electrons in the X point geometry allows to separate between trapped and passing particles. The trapping is mediated by the finite reconnection electric potential field along the field lines, not by magnetic mirroring. This potential, as was shown above, is the result of reconnection but is, at the same time, generated by the particle anisotropy which leads to non-diagonal terms in the pressure tensor. Preliminary confirmation of the pressure anisotropy  was provided via numerical simulations \citep{ohia2012} and made complete by the kinetic derivation of the pressure anisotropy \citep{egedal2013} accompanied by extended simulations. Earlier simulations by \citet{pritchett2005b} had already shown the various contributions to the reconnection electric field being localized in the electron diffusion region as a consequence of the pressure anisotropy produced by the reconnection process. The analytical asymptotic expressions for the pressure in the high-density low-field case applicable to reconnection are similar to the heat-flow suppressed CGL expressions
\begin{equation}
\frac{P_{e\|}}{P_{\|0}}\approx\frac{\pi}{6}\left(\frac{N}{N_0}\right)^3\left(\frac{B_0}{B}\right)^2, \qquad \frac{P_{e\perp}}{P_{\perp 0}}\approx\frac{NB}{N_0B_0}
\end{equation}
These expressions hold in any local frame along the magnetic field. Transforming them into the X point frame generates the wanted non-diagonal pressure tensor elements. Hence, in the electron diffusion layer both effects are closely related. Electron trapping in the reconnection electric field causes pressure anisotropy, and pressure anisotropy causes the non-diagonal elements responsible for reconnection. It should, however, be noted that the pressure anisotropy depends on the presence of a guide magnetic field \citep{egedal2013}. It disappears if the guide field become too weak. In the simulations of \citet{pritchett2005b} the pressure anisotropy was given. Hence non-guide field reconnection requires an initial anisotropy that is not generated by the reconnection process. In this case, however, an initial weak normal magnetic field will do the same service as a guide field. Moreover, the Hall field in the ion diffusion region which plays the role of a weak guide field \citep{baumjohann2009} may generate an anisotropy outside the reconnection site by similar processes; once this anisotropy is produced, it will be transported by the convective flow into the X point.
\begin{figure*}[t!]
\centerline{\includegraphics[width=0.7\textwidth,clip=]{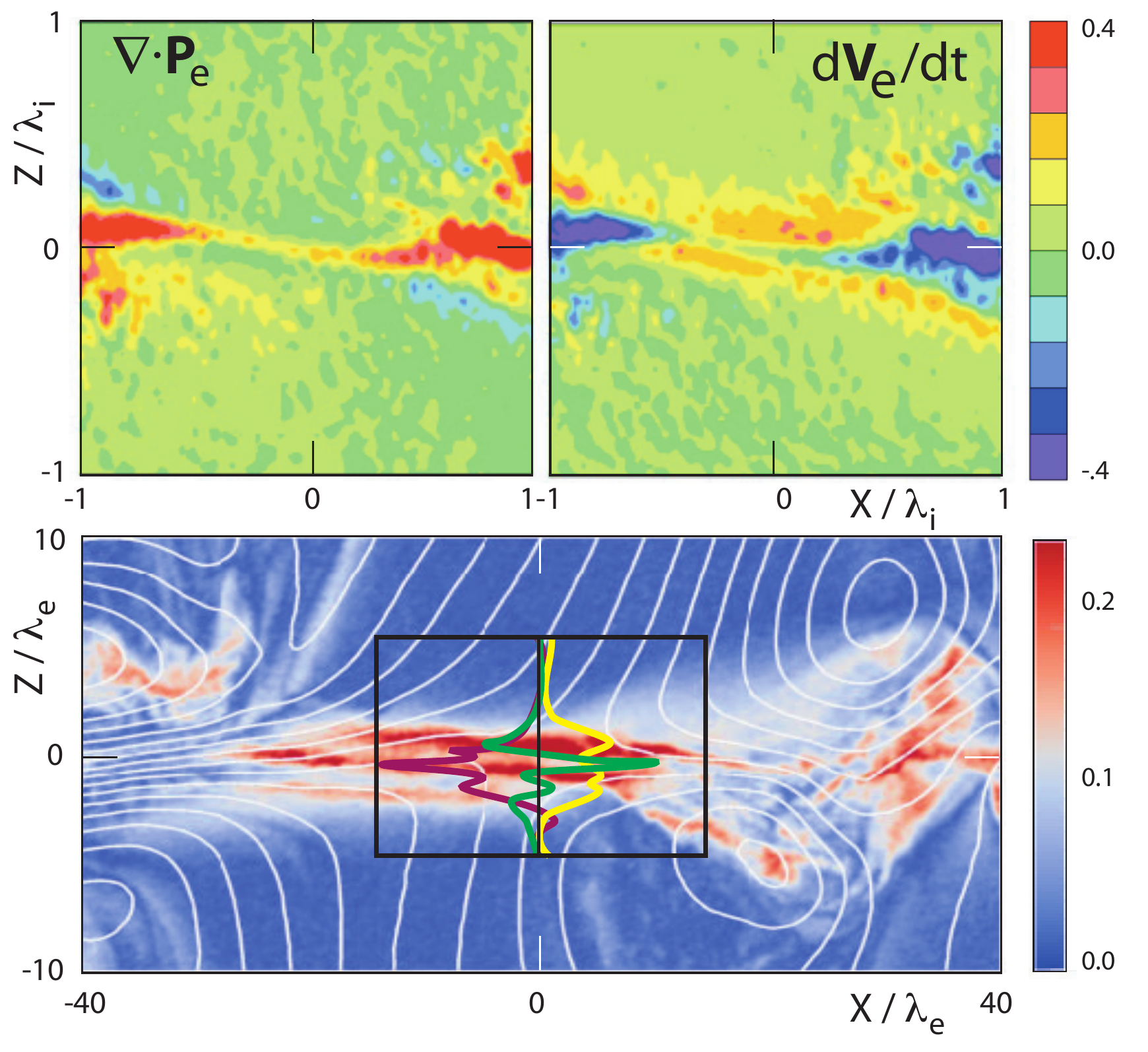} 
}
\caption[]
{Structure of the electron diffusion region. \emph{Top}: Contributions of the electron pressure tensor (\emph{left}) and inertial terms (\emph{right}) to the generation of the reconnection electric field and dissipation centered in the X point frame \citep[after][courtesy American Geophysical Union]{pritchett2005b}. The pressure tensor elements contribute mostly away from the X point near the plasmoids. Note the spiral form of the region. The nonlinear inertial term contributes near the X point above and below the symmetry line (all relative units given in the colour bar). \emph{Bottom}: Three-dimensional simulation \citep[data from][]{liu2013} showing the complicated spatial distribution of the electron current (from white to red; relative units in color bar) splitting into three distinct layers and twisted layers. Overlaid (transparent black box) are vertical profiles through the center (along the vertical black line) of the parallel electric field $E_\|$ (\emph{yellow}) and its contributions (nonlinear inertial term \emph{green} and divergence of pressure tensor \emph{dark red}), all in relative units with amplitudes in the interval [-5\%,5\%]. White lines are three-dimensional magnetic field contours in two-dimensional cut exhibiting the complicated flux rope structure generated in three-dimensional reconnection.  }\label{fig-liu}
\end{figure*}

A recent detailed analysis of three-dimensional reconnection \citep{pritchett2013} in an \emph{asymmetric} setting with weak field on one side and strong field on the other side of the current layer, a scenario that applies to the conditions at the magnetopause, partially disqualifies the generality of the dissipation measure of \citet{zenitani2011}. Strong electron current layers and intense dissipation are found widely distributed and not to coincide with the region of largest dissipation measures around the X line in this case for the reason that in three dimensions a turbulent drag term $-\langle \delta N\delta E\rangle$ caused by wave fluctuations makes a large and not negligible contribution to the electric field and thus to dissipation. It is of interest at the separatrices where wave fluctuations maximize. This conclusion is also strongly supported by the recent symmetric three-dimensional simulations of \citet{liu2013} and \citet{roytershteyn2013}.

{Recent high-resolution simulations in three spatial dimensions by \citet{liu2013} following those of \citet{karimabadi2007} and \citet{shay2007} have been used to further investigate the structure of the electron diffusion region. The interesting new feature found was that the current $J_y$ in the electron diffusion region, where it is carried mainly by electrons, splits into two or more thin tilted current layers. This splitting is different from the Hall induced one proposed above as it restricts to the electron diffusion region. Its provisional explanation is that oblique tearing modes evolve in three dimensions which may form chains as predicted by \citet{galeev1979} in the kinetic case. Such narrow layers indicate the evolution of turbulent reconnection structures. Figure \ref{fig-liu} shows a combination of the two- \citep{pritchett2005b} and three-dimensional \citep{liu2013} simulation results on the electron diffusion layer. In both cases the dissipation and generation of $E_\|$ takes place here. The three-dimensional case shows in addition the splitting of the single electron current layer into three distinct layers.}

The Polar spacecraft observations of the electron diffusion region near the magnetopause \citep[reported by][]{scudder2012} were taken at relatively large Mach numbers of $>1.5$, finding the electron diffusion region located asymmetrically shifted to the low density high flow side. They identified a large pressure anisotropy $>7$, according to our knowledge favorable for reconnection driven by pseudo-viscous pressure effects, and non-gyrotropy of electrons and indicating that the electrons were at least partially demagnetized. These observations are not in contradiction to the simulational results \citep{pritchett2008} of asymmetric reconnection. 

\subsubsection{Waves.}
An important question is which plasma waves do and to what degree do they participate in reconnection. Reconnection is an instability, whether forced or not; it grows out of some initial state and thus is accompanied by waves. In the non-guide field, reconnection has been brought into relation to the generation of whistlers that are localized at the reconnection site. In the guide-field case, it is believed that kinetic Alfv\'en waves take over instead, at least simulations seem to suggest this difference \citep{rogers2001,pritchett2005b}. 
\begin{figure}[t!]
\centerline{\includegraphics[width=0.5\textwidth,clip=]{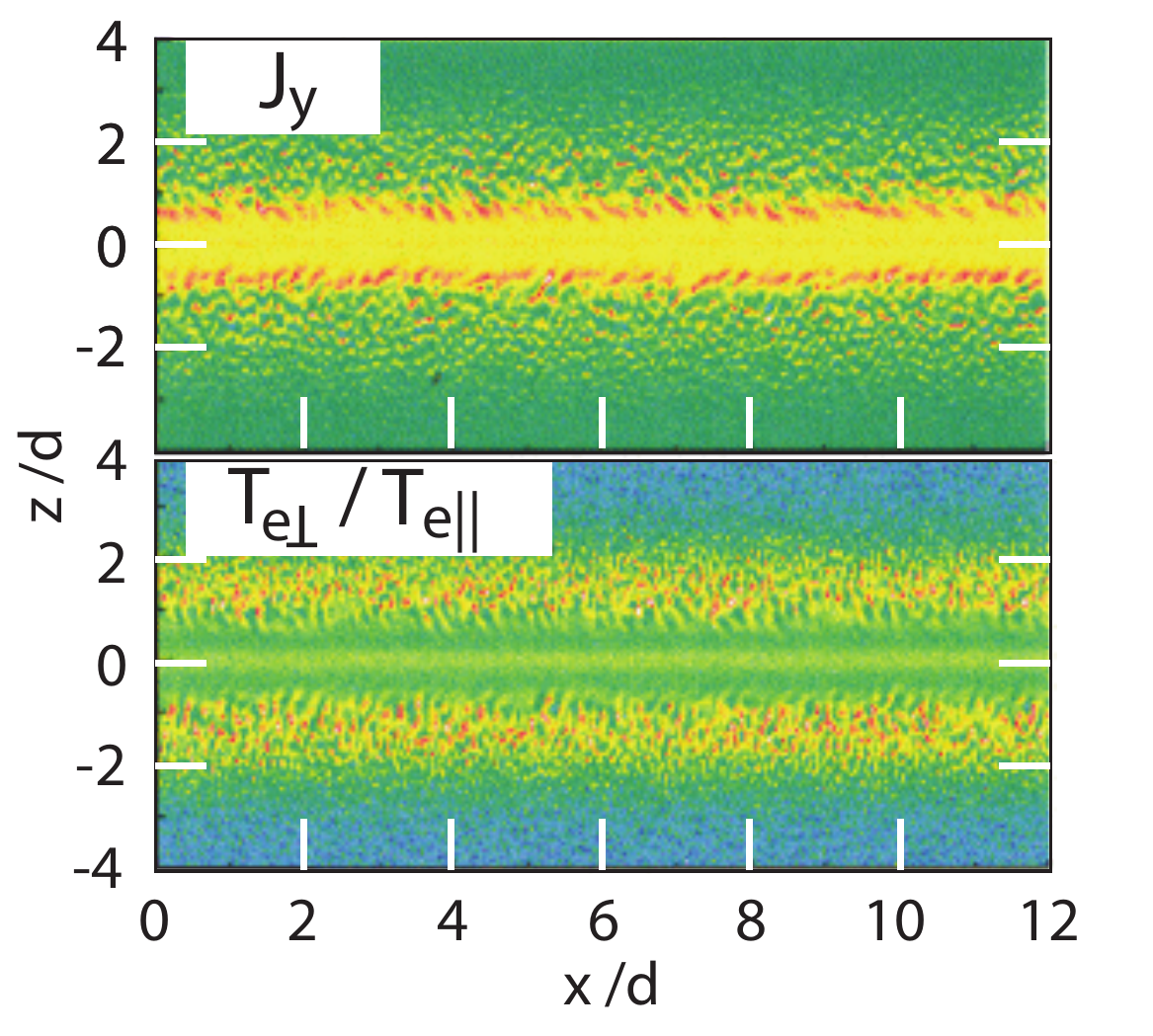} 
}
\caption[]{{Bifurcation of a thin current layer caused by the lower-hybrid-drift instability \citep[simulation data taken from][]{daughton2004}. \emph{Top}: Sheet current strength color coded (highest intensity red, lowest green). The initial current was concentrated at $z/d=0$. After evolution of the lower-hybrid drift instability it has split into two current layers separated by about one initial layer half-width $d$ indicating lower-hybri-ddrift instability mediated current bifurcation. The bifurcated current is structured in $x$ by the lower-hybrid wavelength. \emph{Bottom}: Electron temperature anisotropy $T_{e\perp}/T_{e\|}$ caused in current bifurcation. The anisotropy is positive and concentrated mainly on the bifurcated current. A weak anisotropy is attributed to the original current center. Lengths are measured in halfwidths $d$.} }\label{fig-lh-bif}
\end{figure}

There is a simple reason for this difference which is barely mentioned in the literature. In no-guide field simulations electrons are magnetized in the non-magnetic ion diffusion region. Hence, under conditions when electron anisotropies are produced, whistler waves can be excited on the ion diffusion region scale and below, propagating at speed close to the electron Alfv\'en velocity. These may form large amplitude quasi-localized waves which structure the electron outflow region and also radiate along the separatrices. In a sufficiently strong guide field \citep[as discussed by][]{rogers2001} ions become involved by re-magnetization. Since the transverse scale of the ion diffusion region  is still of the order of $\lambda_i$, this scale is just the typical transverse scale of kinetic Alfv\'en waves with $k_\perp\lambda_i\sim$ O(1). Therefore, a spectrum of short transverse wavelength kinetic Alfv\'en waves takes over in structuring the exhaust region if only the guide field is not too weak, even though there is no principle reason for that the waves should not be excited also in the absence of guide fields since the active inertial ion scale is identical to the transverse wavelength and could thus be in resonance with kinetic Alfv\'en waves. Transferring energy from the drifting electrons is the source feeding the waves.  

Kinetic Alfv\'en waves propagate obliquely to the magnetic field. Their dispersion relation is $\omega_\mathit{kAw}=\pm(k_\perp v_i/\Omega_i)(k_\|V_A)\alpha$, with ion cyclotron frequency $\Omega_i=eB_0/m_i$, $\theta\equiv T_i/T_e$, and $\alpha=\sqrt{(1+\theta)/[2\theta+\beta_i(1+\theta)]}$. The transverse wave-electric field can, in the ion diffusion region, stochastically heat ions in the perpendicular direction.  The parallel wave-electric field component accelerates or heats the electrons parallel to the field. The latter process is of particular interest. The parallel electric field is expressed through the potential $\Phi$ and $\delta B_x$, the wave magnetic field, as
\begin{equation}
E_\|(\mathbf{k})=-i\frac{k_\|\Phi}{\theta}=-\frac{2ik_\|k_\perp}{k_y}\frac{\alpha\theta T_e\delta B_x}{(1+\theta)^2e\sqrt{\beta_i}}
\end{equation}
The normalized spectral energy density $\mathcal{E}_E=\frac{1}{2}\epsilon_0\langle \left| E_\|(\mathbf{k})\right|^2\rangle/NT_e$ that is available per electron for heating is obtained by integration with respect to time over the wave frequency
\begin{equation}
\mathcal{E}_E({\bf k})=\left[\frac{\alpha}{\theta(1+\theta)}\right]^2\frac{k_\perp^2}{k_y^2}\frac{V_A^2}{c^2}(k_\|\lambda_i)^2\mathcal{E}_{\delta B}(\mathbf{k}).\end{equation}
It is expressed in terms of the normalized magnetic spectral wave-energy density $\mathcal{E}_{\delta B}=\langle |\delta B_x|^2\rangle/B_0^2$. Under conditions in the magnetotail during substorms kinetic Alfv\'en waves, by this parallel wave power, are probably responsible for the acceleration of electrons along the separatrices for closing the Hall current system in the ion diffusion region. There, it is the kinetic Alfv\'en wave spectrum which transports the field-aligned currents from the tail reconnection site into the ionosphere. {Recently, kinetic Alfv\'en waves have been revived as also contributing to plasma mixing during reconnection \citep{izutsu2013}.}

The three-dimensional pair-plasma non-Hall simulations with open boundaries in $x$ performed by \citet{jaroschek2004a} showed strong electrostatic wave activity  \emph{outside} the X point diffusion region in the $x-y$ plane including the third dimension. These waves propagate in oblique direction causing density ripples above and below the X point. In the pair plasma, the waves are the equivalent of lower-hybrid-drift waves in the current density gradient in electron-proton plasmas and have a similar effect in scattering the plasma particles. Nonlinearly they evolve to large amplitudes and cause structuring of the current, generating of anomalous resistance, and accelerating particles along the magnetic field into field-aligned beams. These beams are similar to the electron fluxes which close the Hall current system along the separatrices in electron-proton reconnection, in pair plasma they are of different nature, i.e. contributing to tail formation on the pair distribution functions. The waves do not necessarily cause field-aligned current flows as no Hall current system requires any closure. Moreover, though wave activity was remarkable, its visible heating effect on the distribution was small and did not have any profound effect on the reconnection process itself.

Lower-hybrid (drift) waves are the main mode which is known to be generated in the ion diffusion region plasma gradient. At short wavelength it is purely electrostatic and related to whistlers, at long wavelength it is electromagnetic. These waves have been predicted long time ago to be of importance in reconnection {\citep{huba1977,huba1981}}. However, as has been discussed earlier, observations show \citep{bale2002,treumann1990} that they are of very weak amplitude in the vicinity of reconnection sites and in simulations do barely appear in the electron diffusion region, in spite of the relatively strong density gradients at its boundaries. Presumably they play little role in reconnection other than in accelerating electrons outside the diffusion region near the separatrices.  {However, importance has been attributed to their nonlinear evolution of being capable of causing current bifurcation \citep{daughton2004,karimabadi2005b,roytershteyn2013} and, by modulating the electric field and flow, affecting the reconnection rate. In the realistic mass-ratio ($\mu=1836$) simulations of lower-hybrid waves by \citet{daughton2004}, no anomalous resistance effect was observed.} 

{\citet{roytershteyn2012} investigated excitation of lower-hydrid drift waves in the high and low $\theta=T_i/T_e$ regimes in three-dimensional reconnection, finding that, in support of the earlier findings \citep{daughton2004}, in the high regime long wavelength lower hybrid waves are substantial near the X point, strong enough to have an effect on the electric and flow fields and causing dissipation and, most important, break-off of the electron current layer. By the above equation, $\mathcal{E}_E\propto \theta^{-4}$ in this case, suggesting, however, that the effect in the magnetotail is small. This case is, anyway, not realized at the magnetopause and also not in the tail. In both cases $T_i>T_e$, conditions to which the \emph{asymmetric} three-dimensional simulations by \citep{pritchett2013} apply who found strong excitation of lower-hybrid waves in the steep density gradient on the low density magnetospheric side (see, e.g., Fig. \ref{fig-prit-mp1}). The waves had the expected lower-hybrid polarization with electric field highly oblique to the magnetic field. They formed very-large amplitude localized patches along the separatrix.}

Lower-hybrid waves in magnetopause crossings related to reconnection had been occasionally observed at their high frequency whistler tail \citep{treumann1990} and on the low-frequency branch \citep{bale2002}. \citet{mozer2011a} reported a large statistics of THEMIS observations of lower-hybrid drift waves in magnetopause reconnection having large amplitudes but otherwise no effect in causing drag. The latter conclusion is true since the waves do not occur in the electron diffusion region near the X point as confirmed by the simulations. At the separatrix the amplitudes and the effect of waves may be underestimated due to spacecraft resolution and smearing out the localized wave patches. Much earlier investigations \citep{labelle1988,treumann1991} had already ruled out their importance in generating anomalous diffusion at the magnetopause, even when localized. 

Finally, the Weibel instability \citep{weibel1959} is another mode that could be of importance in contributing to the spectrum of waves near the reconnection site. {This was mentioned first by \citet{karimabadi2005a}. }It is a purely magnetic quasi-stationary $\omega\approx 0$ mode. If it can be excited, it contributes to the formation of seed X points in the current layer which serve as initial disturbances in causing reconnection. A possible scenario refers to the acceleration of electrons along a guide field  \citep{treumann2010} producing the required temperature anisotropy of the electrons for excitation {of} the Weibel mode in the current sheet.

\subsection{Guide field reconnection}
Guide fields do arise naturally in interaction of asymmetric plasmas {\citep[as suggested by][]{karimabadi1999}} for only the exactly anti-parallel field components do merge and annihilate each other. Tilted magnetic fields therefore carry guide fields. \citet{eastwood2013} recently summarized their effect on the diffusion region.
\begin{figure*}[t!]
\centerline{\includegraphics[width=0.7\textwidth,clip=]{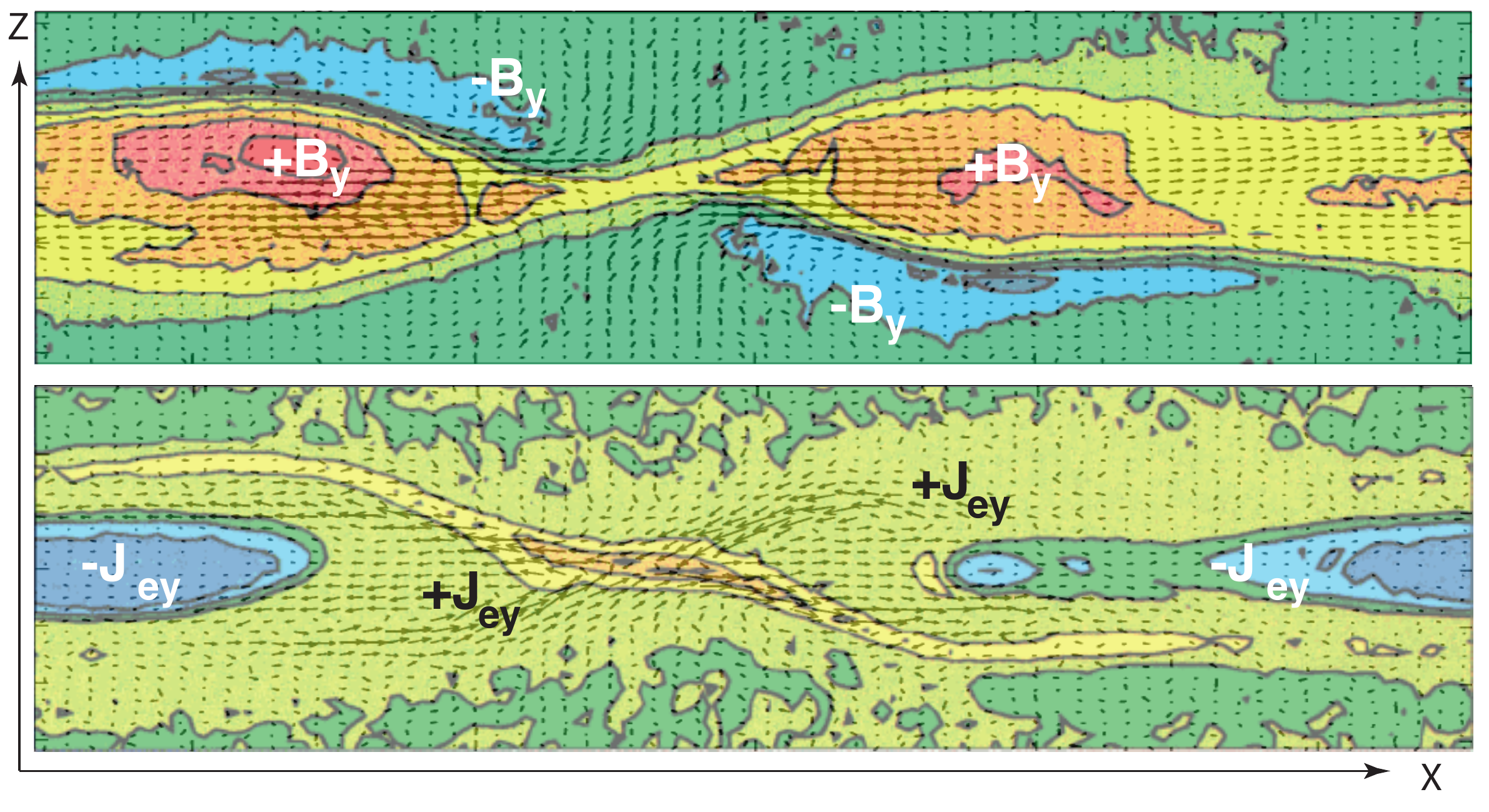} 
}
\caption[]
{Guide field reconnection. \emph{Top}: Ion velocity (arrows) overlaid by the out-of-plane magnetic component of the magnetic field \citep[simulation data  after][courtesy American Geophysical Union]{pritchett2001a}. This field includes the guide field  $B_{yg}$ and the self-consistent Hall field. The guide field causes an asymmetry in the field and ion flows. The diffusion region becomes tilted. \emph{Bottom}:  The highly asymmetric electron flow pattern  overlaid by the out-of-plane electron current $J_{ey}$ which is opposite to the out-of-plane electron flow. Yellow-to-red colors indicate positive, blue-to green colors negative values. The electron current is mainly in $+y$ direction along the induction electric field. It is highest along the yellow-red diagonal, indicating that it has been accelerated along the reconnection amplified convection electric field.  }\label{fig-prit-mp2}
\end{figure*}

{Including a guide field leads to profound changes of the reconnection process. A guide magnetic field parallel or anti-parallel to the sheet current acts stabilizing on reconnection once the guide field becomes strong. This is a consequence of the magnetization of the electrons by the guide field which enforces the frozen-in condition. In addition, guide fields cause a tilt of the X line against the $z$ axis, i.e. the normal to the Harris layer. The upright X becomes an italic $X$. They also tilt the Hall field geometry and favor one leg of the Hall current. This was demonstrated first by \citet{karimabadi1999} and confirmed by \citet{pritchett2001b}, \citet{daughton2005}, and \citet{daughton2006} in open system simulations.} An example is shown in Fig. \ref{fig-prit-mp2}.

{Evolution of reconnection depends strongly on the strength of the imposed guide field. This has been investigated by \citet{daughton2005} and \citet{karimabadi2005a} who indentified three different guide field regimes, depending on a critical magnetic field strength $B_c/B_0=\sqrt{r_i/2d}(\theta/\mu)^{1/4}$. For weak and strong guide fields $B_y<B_c$ respectively $B_y\gtrsim 3B_c$, the intermediate regime is $B_c<B_y\lesssim3B_c$. In the strong regime, the tearing mode growth rate decays as $\gamma_\mathit{tear}\propto (B_y/B_0)^{-2/3}$ indicating the stabilizing effect of strong guide fields. In the weak regime the effect of the guide field is small. In the intermediate regime,  the fastest growing tearing mode grows in oblique direction (within an angle of $6^\circ$ to $10^\circ$). In this regime the tearing mode becomes the drift-tearing mode being modified by the diamagnetic drift of electrons magnetized in the guide field. This mode has finite frequency comparable to the electron drift frequency. Two unstable modes $\pm k_x$ exist, localized on the resonant surfaces on opposite sides of the current sheet, which indicates that a full theory requires a three-dimensional treatment. Moreover, the presence of a guide field has an important effect on the quadrupolar structure of the Hall magnetic field \citep{karimabadi2004c,daughton2005}. It not only distorts the Hall field, introducing an asymmetry, but compresses the spatial range of the Hall field down to the electron gyro-scale. This Hall field structure survives into the nonlinear regime of the tearing mode. \citet{karimabadi2005a}, for realistic mass ratio $\mu$, found that nonlinear saturation of the tearing mode in the presence of guide fields is determined by two effects: electron non-gyrotropy and development of electron pressure anisotropy. In absence of guide fields (anti-parallel case) electrons become non-gyrotropic and anisotropic. Parallel heating of electrons causes a tearing saturation amplitude (tearing island half-width in terms of electron gyro-radii $r_e$) of $\Delta_\mathit{tear}/r_e\sim 3$. This process dominates over saturation at singular layer thickness due to electron trapping. The latter becomes important only when Weibel-modes or turbulence are excited causing pitch angle scattering. Guide fields allow only for the electron trapping mechanism leading to saturation at  very low amplitudes $\Delta_\mathit{tear}\sim 1.8 r_e$ in the strong regimes, and $\Delta_\mathit{tear}\sim (2r_ed)^{1/2}$ in the intermediate regime, where $r_e$ is the electron gyroradius in the guide field.}

Guide fields allow for a number of other important effects: due to the magnetization of electrons in guide fields, they introduce the whole spectrum of magnetized electron plasma modes, if only conditions can be generated which drive one or the other mode unstable.  Guide fields thus open up the gate to excitation of electron whistlers and electrostatic electron cyclotron waves if the electron pressure becomes anisotropic, a situation which is quite realistic under various conditions like forcing of the current sheet and has been observed in simulations (see below). The density gradient also causes excitation of electron drift waves. Some of these waves are known to produce anomalous resistivity and may thus contribute to plasma heating, dissipation weakly supporting reconnection. If electrons are accelerated along the guide fields, which is expected in the presence of convection electric fields, the Buneman instability \citep{buneman1958} can be excited either leading to anomalous resistivity and plasma heating, or allowing for the growth of chains of electron holes found in simulations \citep[cf., e.g.,][]{drake2003,che2010,che2011}. {These simulations inferred the presence of electron holes near the reconnection site finding two kinds of electron holes: genuine ``slow'' Buneman instability-driven holes and ``fast'' holes attributed to lower-hybrid-drift waves. The latter family of holes indicates hole generation away from the electron diffusion region near  the separatrices. It may be related to the nonlinear lower-hybrid wave evolution reported by \citet{daughton2004}.} Without proof \citet{che2011} suggested the holes of being caused by electron trapping in the parallel lower-hybrid wave-electric field and afterwards evolving nonlinearly into holes and transported away with the fast parallel phase speed of the lower-hybrid waves. 

Electron holes are local concentrations of electric potential drops along the guide magnetic field due to charge separation between trapped and passing electrons. Their occurrence in relation to substorms and reconnection was supported by observations in the magnetosphere \citep{cattell2005}, in the laboratory \citep{fox2012}, and in one-dimensional Vlasov simulations \citep{goldman2008} of electron hole formation based on electron distribution obtained from the symmetric numerical simulations of reconnection in presence of a magnetic guide field by \citet{pritchett2005b}. The latter simulation demonstrated that the simulated electron distribution functions inside the electron diffusion layer and exhaust region are consistent with electron hole formation during reconnection. They thereby confirm the assumption made earlier that electron acceleration by the reconnection electric field in the electron diffusion region is strong enough to exceed the Buneman threshold for instability and nonlinear evolution of the Buneman instability which results in the formation of electron holes. Electron holes give rise to further violent local acceleration; by cooling the passing electron component, electron holes may generate bursts of fast cool electron beams which result in the excitation of Langmuir waves  and non-fluid plasma turbulence. For all these reasons, recent simulation research favors guide-field reconnection in two or three dimensions.

\subsubsection{Forced reconnection}
So far we considered the case of spontaneous reconnection when a sufficiently thin current sheet spontaneously pinches and decays into a number of current braids located in the centers of a chain of plasmoids separated by magnetic X points of weak magnetic fields. We noted that in this case the Parker idea applies that the plasma is sucked in into the X point reconnection site. 

The philosophy changes when reconnection is driven by a continuous inflow of plasma into the current layer. This case is analytically almost intractable and requires a numerical treatment which will be discussed below. One, however, immediately realizes that the simple definition of reconnection rates being identified with the Mach numbers breaks down in this case, simply because the inflow and Alfv\' en speeds are both prescribed under these conditions. Continuous plasma inflow forces the current sheet to digest the excess plasma and magnetic fields. This can happen only by violent reconnection and plasma ejection from the X points.

One of the first full-particle kinetic investigation of forced reconnection was reported by \citet{birn2004} who exposed the Harris current sheet to a temporary inflow of plasma from both sides. Such an inflow is produced by imposing a cross-magnetic electric field (here for a limited time). It results in a compression of the initial current sheet causing current thinning. This study, called ``Newton Challenge'', did however not lead to any profound insight. It seemed to demonstrate that all kinds of simulations would lead to about the same reconnection rates, except for the MHD codes where the rates were much less. But the amount of reconnected flux at the final state was about the same. 

This has by now changed profoundly having been superseded by more sophisticated two-dimensional forced simulation \citep{pritchett2005c} in which an external cross-magnetic electric field was imposed at the boundaries above and below a symmetric electron-proton Harris current sheet for a temporally limited time. This field is modeled as $E_y(x,t)=E_{0y}f(t)[\cosh(x/\Delta x_e)]^{-2}$, with $f(t)$ a free function of time and $\Delta x_E$ the range of the electric disturbance in $x$. It causes a spatially varying inflow of magnetic flux that compresses the initially assumed thick Harris current sheet. The simulations assume low mass ratios $m_i/m_e=25$ and 100. Initiating of reconnection is not done artificially. Instead it is waited until it sets on by itself which happens due to both thinning of the current layer due to inflow and numerical noise. This causes significant delay of reconnection, however. The main finding is that an electron temperature anisotropy $T_\perp\sim 1.1T_\|$ develops readily, and an electron current layer evolves below the ion inertial scale. These effects lead to reconnection based on non-diagonal pressure tensor elements in thin current layers, significantly different from the case of an isotropic Harris layer. Of course, quadrupolar Hall-$B_y$ are as well generated. The reconnection rate is $E_y\sim 0.1 V_AB_0$ indicating fast reconnection which persists until the flows interact with their neighboring images, independent of the two low mass ratios. Mass ratio effects come into play at the stage when plasmoid formation takes place slowing down further reconnection. Thus forced reconnection proceeds through two stages: fast initial and slow plasmoid formation. 
\begin{figure*}[t!]
\centerline{\includegraphics[width=0.7\textwidth,clip=]{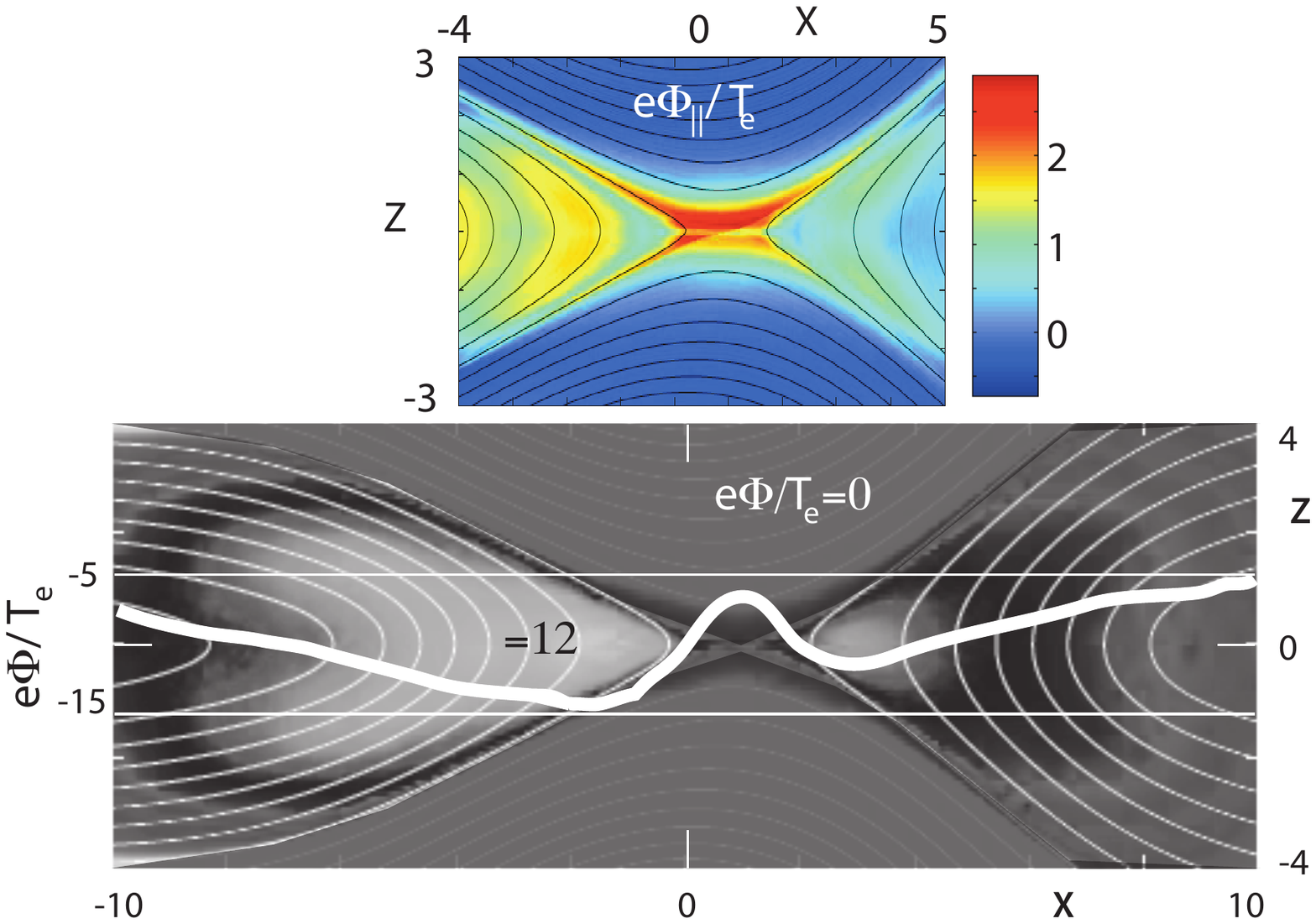} 
}
\caption[]
{Finite normal magnetic field component. \emph{Bottom}: Grey scale image of the total electric potential in a real mass ratio $m_i/m_e=1600$ two-dimensional simulation of forced reconnection \citep[after][courtesy American Geophysical Union]{pritchett2010}. The potential is normalized to thermal energy.  It varies between the values 0 and -15. The fat white line is the potential along $z=0$ showing a minimum of about $-7$ in the X point and maximizing at around $-15$ at the earthward plasmoid boundary. \emph{Top}:  Normalized parallel potential in the inner region around X point in color coding. The potential is asymmetric in $z$ and in $x$ maximizing at the separatrices and in the boundary of the earthward plasmoid. This potential is responsible for generation of field aligned currents and acceleration of electrons along the field. Asymmetry is caused by the finite $B_z$.}\label{fig-ebz}
\end{figure*}

\subsubsection{Finite normal fields and dipolarization.}
The early kinetic theory of \citet{galeev1975b,galeev1979a} stressed the importance of the presence of a residual normal magnetic field component $B_z$ in reconnection. Such a component is natural in magnetospheres where the magnetic fields are anchored in the central object, a planet like Earth or Jupiter or a Star. Simulations of reconnection in presence of finite $B_z$ have recently been performed under forced reconnection conditions \citep[a first attempt was published by][as an extension of the forced simulations]{pritchett2005a} and, most importantly, for an about realistic ion-to-electron mass ratio $m_i/m_e=1600$ \citep{pritchett2010}. Open boundary conditions were imposed in $x$.  Forcing of reconnection was achieved by switching on a uniform external cross-$B$ electric field for a brief period of the order of one ion-gyro period to both sides of the Harris current layer. This causes continuous thinning (compression) of the current layer until reconnection initiates. The initial $B_z>0$ was organically included into the simulation from beginning in modeling the residual geomagnetic  northward dipole field component similar to analytic theory \citep{galeev1975b}.

There are several stages in this simulation which deserve mentioning. As in the above non-$B_z$ simulation of forced reconnection \citep{pritchett2005c} reconnection was not ignited. This caused again an about mass-ratio independent comparably long waiting time of  $\Omega_it\sim 25$ until onset of reconnection, the time of propagation of the electric disturbance from the boundary into the current sheet and formation of the $\sim2\lambda_e$-thick electron current sheet and density increase by 20\%  in the current sheet center. Afterwards the reconnection rate increased steeply. The rate in this driven case is significantly higher than in non-forced reconnection, where the canonical value always obtained is $\sim 0.1$, here being $\sim 0.2$. {\citet{sitnov2011}, in a larger particle-in-cell simulation set up, recently confirmed the results of \citet{pritchett2005c}  on the formation of dipolarization fronts. They further investigated their dynamics and dependence on forcing, finding that, after displacement of the fronts from the X line, several new X lines may form in the current sheet behind the front when the forcing is not switched off but continues. This may lead to steady production an ejection of dipolarization fronts from a current layer.}

The high-mass ratio runs show the initial evolution of a distortion of the normal magnetic component during the thinning phase with $B_z$ evolving a sinusoidal shape with negative $B_z$ in the distorted $x$ range. The final $B_z$ shape during fast reconnection is quite regular. In the negative $B_z$ region electrons lose magnetization, and thus reconnection is augmented by demagnetization here. This causes steeping of the $B_z$ profile which resembles experimentally observed so-called dipolarization fronts during reconnection in the magnetospheric tail \citep{runov2011,runov2012,runov2013,oka2011}. Figure \ref{fig-dipole} shows the evolution of $B_z$ in the simulation with realistic mass ratio for four times. The final quasistationary profile of $B_z$ as function of $x$ forms a steep ramp of nearly symmetric positive and negative amplitudes being localized in space. This profile moves into left direction as a dipolarization front implying that locally at its location the magnetic field has become more dipole-like than it was initially. This is the result of the forced reconnection to which the initial current layer was exposed. 
\begin{figure*}[t!]
\centerline{\includegraphics[width=0.7\textwidth,clip=]{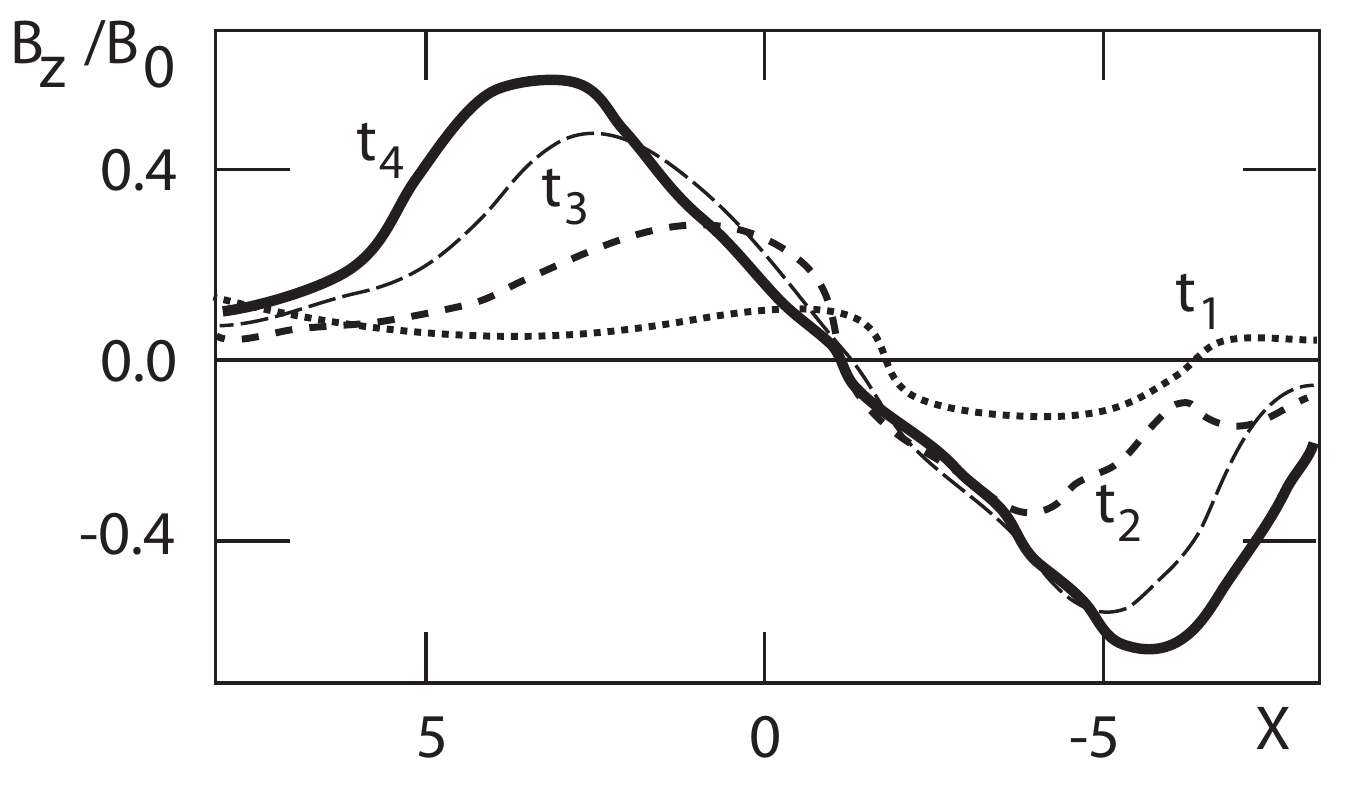} 
}
\caption[]
{Evolution of `dipolarization fronts' in forced reconnection simulation with finite $B_z>0$ shown for four consecutive times. These fronts move to the left in the direction of Earth \citep[schematized, simulation data after][courtesy American Geophysical Union]{pritchett2010}}\label{fig-dipole}
\end{figure*}
Another important observation is that the electron outflow speed substantially exceeds those of outflows in non-forced non-$B_z$ simulations. In the realistic mass-ratio simulation they reach flow velocites of $V_\mathit{out}\sim 18\,V_A$ thus forming fast electron jets which survive until distances of the order $\sim 5\lambda_i$ (in this simulation) away from the X line where the electron jets merge with the ion outflow. This distance may become even longer in larger simulation set-ups.

The reconnection electric field $E_\|$ near the X line evolves a quadrupolar structure at normal distance $|z|\sim 0.8 \lambda_e$ and extends over a parallel distance $|x|\sim \lambda_i$ to both sides of the  X line.  More informative is the structure of the field components $E_x, E_z$ which both maximize along the separatrices and have dipolar structure. Hence the separatrix regions are the strongest carriers of electric fields. Here the acceleration of particles is strongest. Calculation of the parallel electric potential can be done along the field lines from a given $x$ to the end of the simulation box. Figure \ref{fig-ebz} gives an impression of the field aligned potential in color code for the inner part of the simulation region. The parallel potential maximizes along the separatrices but exhibits a certain asymmetry which is due to the presence of the finite $B_z$ field and its evolution. This potential field is responsible for electron acceleration. Below is the total electric potential shown. It exhibits a quasi-sinusoidal structure around the X point but is clearly asymmetric in $x$ caused by the positive initial $B_z$.  

\subsubsection{Large three-dimensional systems, flux ropes, turbulence.}
{This final section reviews some most recent simulation results obtained with extremely large simulation set ups allowing for following the evolution of tearing modes in three dimensions up to mesoscopic scales. It had already been suggested \citep{daughton2005,daughton2007,karimabadi2004b,karimabadi2007} that oblique evolution of the tearing mode in guide field systems would require a full three-dimensional treatment of reconnection under collisionless conditions. Indications of the importance of the third dimension were also seen in pair plasma simulations including a narrow three-dimensional extension of the simulation box \citep{jaroschek2004a}. However, extensions to three dimensions and larger scale have become available only recently by supercomputer technology. Such simulations have been reported by \citet{daughton2011}, \citet{roytershteyn2013} and were summarized by  \citet{karimabadi2013a,karimabadi2013b}. The extended three-dimensional meso-scale simulations \citep{daughton2011} with realistic mass ratio $\mu=1836$ confirm the importance of formation of electron layers in the center of the plasma sheet during reconnection. These layers readily form, as described above for two-dimensional simulations, during the initial phase and are responsible for field line merging/reconnection in the electron diffusion layer.  As before they are caused by the divergence of non-diagonal pressure tensor elements and electron inertia also in three-dimensional simulations with the remarkable difference that in extended three dimensions with guide fields the electron layers are warped and much shorter than in two-dimensional simulations. Helical magnetic structures are formed and organize in flux ropes rather than the two-dimensional chains of new magnetic islands. Thus three-dimensional reconnection is dramatically different from two-dimensional reconnection in thin current sheets. The cause of the destruction of long electron layers is the magnetic shear that is produced by instabilities which evolve due to the available freedom in the third dimension which allows for the organization of the magnetic field into twisted ropes. The initial Harris layer breaks completely off at the best consisting subsequently of short skewed pieces only after reconnection has structured the field into ropes (as can be seen in Figures 3 and 4 of \citet{daughton2011}).}

{In general, when complicated magnetic structures evolve due to some unspecified reason (interacting flows, fluid instability, turbulent cascades etc.), it is clear that thin current layers are produced. This will always happen when the field is not a priori forcefree. If such thin current layers evolve, tearing modes will become excited spontaneously, and collisionless reconnection sets on. This is the case in pre-existing turbulence resulting, for instance, from Kelvin-Helmholtz instability in shear flows at the contact of two interacting plasma streams \citep[as a possible site of reconnection predicted thirty years ago by][when the formation of narrow flow, current and magnetic vortices had been observed in the nonlinear evolution of the Kelvin-Helmholtz instability]{miura1982a,miura1982b}. Another possibility is that field lines displace chaotically \citep[cf., e.g.,][]{boozer2012} which, presumably, is the normal case in three dimensions.} 

On the other hand, in three dimensions current layers have sufficient freedom to themselves evolve into narrow warped structures during reconnection resembling turbulence as found in simulations by \citet{daughton2011} where the reconnection layer decayed into many ``turbulent'' flux ropes forming short and chaotically distributed electron diffusion regions. This suggests that between reconnection and turbulence obviously exists a close relation which in retrospect is not surprising. If narrow current layers evolve in turbulence, they will undergo current and magnetic field rearrangement, which is necessarily accompanied by reconnection, thereby dissipating the stored energy and contributing to the turbulent cascades at reconnection scales $\sim\lambda_e$ typical for the electron diffusion region. The electron inertial length becomes the relevant turbulent dissipation scale, which is far less than any of the meso-scale fluid scales but still much longer than the microscopic atomic dissipation scales. \citet{karimabadi2013b} recently reviewed three-dimensional simulations of turbulence including reconnection spanning the regime from mesoscales of fluid turbulence to the scales of the electron diffusion region. Their finding is that the turbulent cascade indeed generates current sheets in plasma of scales down to the electron skin depth which, through reconnection, causes the expected  plasma heating and reorganization of current filaments and magnetic fields. In a series of impressive figures zooming in from mesoscales to microscales they demonstrate the warping of a current sheet at the edge of a turbulent vortex until it folds and cascades down to the electron scale showing the generation of tearing mode structures and rearranging magnetic fields. The observed turbulent heating is primarily due to parallel electric fields associated with reconnection in the thin current sheets, indicating that reconnection is the dominant dissipation and for the first time identifying the dissipation process in truly collisionless plasma turbulence. In addition generation of various plasma modes was found, kinetic Alfv\'en waves and several types of coherent structures like the mentioned flux ropes on larger scales and electron holes on shorter scales. \citet{leonardis2013} analyzed such simulations to infer about the presence of intermittent multifractal structures as indication of self-organization of the turbulence. {Recently, \citet{karimabadi2013c} reviewed the consequences of these findings in view of their application to astrophysical turbulence and the driving of reconnection by external turbulence.}  Spacecraft observations suggest that reconnection and turbulence are indeed closely related. Reconnection in the turbulent magnetosheath was reported in thin current layers observed by \citet{retino2007} and suggested triggering reconnection onset by \citet{phan2007b}. It is also expected to take place in the turbulent heliosheath plasma, the region between the termination shock and heliopause.

\subsection{Acceleration of charged particles}
{Acceleration of charged particles during collisionless reconnection has become a hot topic in recent years. It has been known for long that reconnection ejects about symmetric quasi-neutral plasma jets which since the first in-situ identification of reconnection by  \citet{paschmann1979} have been taken as the unambiguous canonical signature of ongoing reconnection. These jets are but slow quasi-neutral plasma streams confined to the narrow plasma outflow region from the X point outside the electron diffusion region that asymmetrically surrounds the X point. We already mentioned that electrons are also ejected from the X point region to both sides forming extended electron jets in the inner diffusion region and giving rise to ``electron exhausts'' of narrow opening angle and extension of several $\lambda_i$ along the current layer, much longer than the electron skin depth, until matching the ions and ultimately slow down to become members of the above mentioned canonical slow neutral plasma reconnection flows. These jet electrons are much faster than the quasi-neutral flow, being of the order of or faster than the electron Alfv\'en speed. Their existence implies that electrons and ions at those distances from the X point couple together in their motion while inside the electron diffusion region the dynamics is mainly that of the electrons. Particle acceleration thus refers separately to electrons and ions. In addition, electrons become accelerated by the reconnection-caused electric fields and also by lower-hybrid waves along the separatrices where they may contribute to the formation of electron holes and where they may cause further acceleration and beam formation.}

\subsubsection{Electron acceleration.}
{The realization that collisionless reconnection is basically determined by electron dynamics has recently lead to  interest in acceleration of electrons up to high energies in  reconnecting current sheets}.  Observations of high energy electrons were first reported in evident relation to reconnection in the geomagnetic tail by \citet{oieroset2002}.  There, electron energies reach up to $\sim300$ keV, far above any ambient tail-electron temperature which is of the order of mere $0.3\lesssim T_e\lesssim 1$ keV. These energetic electrons form high energy tails on the electron distribution function. The 1000 times higher than thermal energy make them of considerable interest for astrophysical systems. For obvious reasons it is unlikely that these energies can be obtained by acceleration in a single X point in reconnection in the geomagnetic tail. Direct acceleration would require  unreasonably strong electric fields that can hardly be sustained in the comparably weak reconnecting tail-magnetic fields of typical strength of several nT only and reconnection scales of the order of the electron skin depth. Reconnection-produced chains of electron holes, as have been observed in the mentioned particle-in-cell simulations \citep{drake2003} and in space \citep{cattell2005}, can probably be ruled out of being responsible for the high observed energies, though they have also been invoked as a viable mechanism \citep[cf. discussion by][]{drake2003}. Sophisticated simulations of electron holes performed by \citet{newman2001} confirm electron heating in electron hole formation by a factor of less than ten, insufficient to explain the high observed energies. More interesting is the barely mentioned formation of the residual high energy electron beams (see the discussion in the electron hole section above) which result from selective cooling of the original hole-passing current electrons by the chain of holes. The reported beam energies may reach a factor of 100 times the original electron temperature. This is still below the observed upper limits, though further acceleration of some selected electrons by very long hole chains would not be unreasonable. 
\begin{figure*}[t!]
\centerline{\includegraphics[width=0.7\textwidth,clip=]{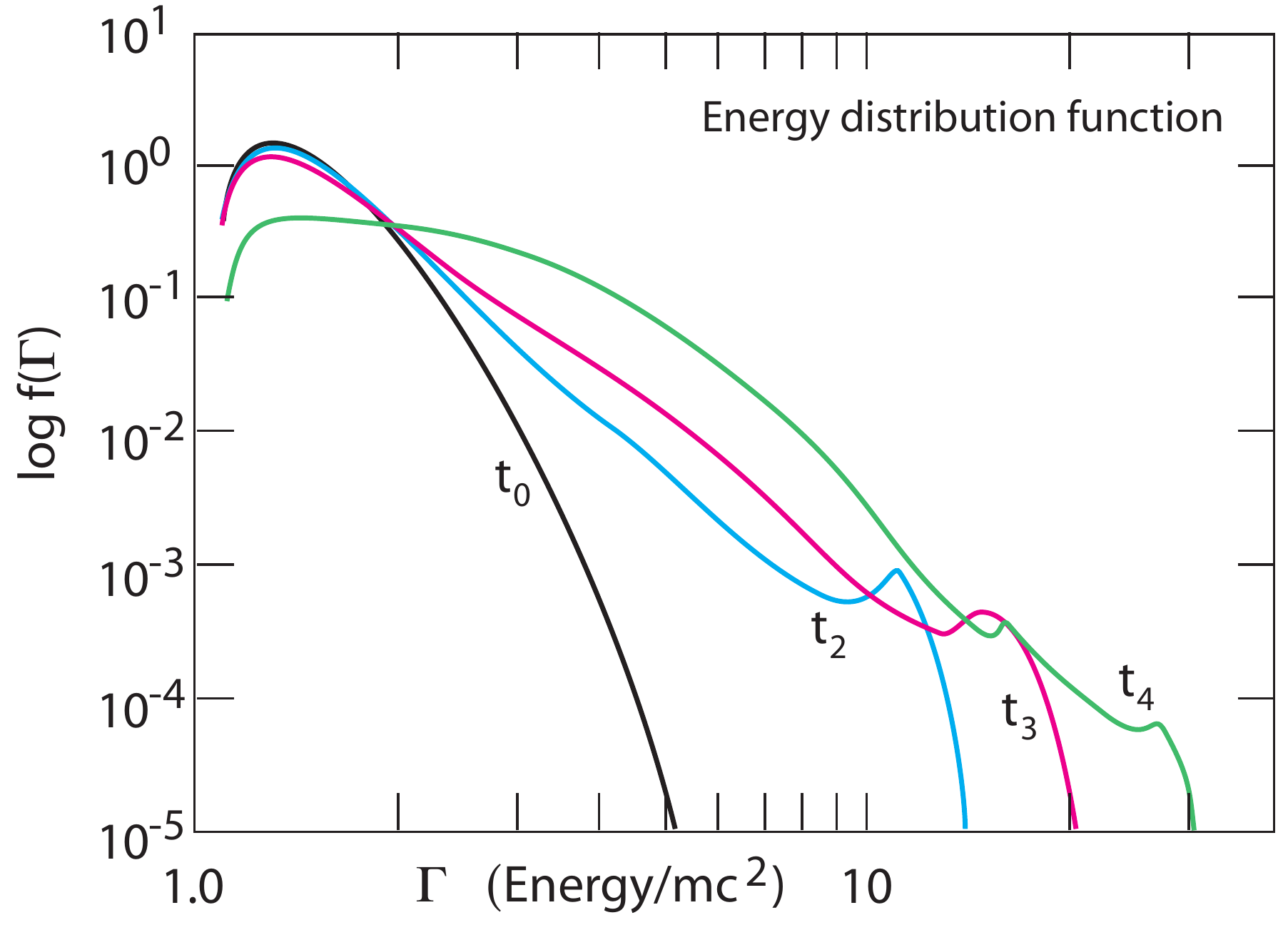} 
}
\caption[]
{Evolution of a high energy tail on the initial particle distribution function in pair plasma simulations of three-dimensional reconnection \citep[schematized, simulation data after][]{jaroschek2004a}. Three of four equally space consecutive simulation time snapshots of the relativistic distribution function $f(\Gamma)$ are shown. The black distribution at time $t_0$ is the relativistic J\"uttner equilibrium distribution. The gradual generation of a high energy tail by interaction of the charged particles with the reconnection electric field is seen during the evolution of the distribution from time $t_0$ to $t_4$. {The final distribution (not shown) at time when the particles reach the end of the box has evolved up to $\Gamma\sim 80$, indicating strong particle acceleration. The distributions exhibit an intermediate  region of quasi-power law $f(\Gamma)\sim\Gamma^{-s}$, with index $3<s\lesssim4$, followed by a steep high energy cut-off. Note the formation of a (weak though statistically significant) high energy bump.}}\label{fig-acc-jaro}
\end{figure*}

Current interest in acceleration has instead focussed on two directions. \citet{jaroschek2004a,jaroschek2004b} demonstrated (in two widely ignored papers) the formation of high energy tails on the particle distribution function in three-dimensional simulations of pair plasma reconnection. A typical trajectory of an accelerated high energy tail particle  projected onto the $x-z$ plane in these simulations is shown on the right in Fig. \ref{fig-halljar}. Initially the particle located in the current sheet performs its meandering Speiser orbit, gyro-oscillating in the space between the antiparallel magnetic fields until becoming picked up by the reconnection electric field and gaining energy. This field accelerates the particle (electron, positron) up to a gyroradius of the size of the entire width of the current layer and plasmoid. Passing the plasmoid, it becomes just weakly decelerated, but when re-entering the rear side of the plasmoid further acceleration in the electric field heats the particle up again, lifting it up and augmenting its energy. During the long time the particle needs to jump ballistically from plasmoid to plasmoid until reaching the boundaries of the box it is  in this way accelerated non-stochastically. All particles of similar kind finally add up to an extended tail on the distribution function (Fig. \ref{fig-acc-jaro}). This tail obeys a substantial section exhibiting a quasi-power law shape which is exponentially cut-off at its high energy end where the most energetic particles accumulate to form a beam-like bump. The acceleration process is limited by the geometric size of the box with its open ends in $x$ direction. These simulations were preformed with one current sheet only carefully avoiding interaction between the two antiparallel currents in the symmetric simulation set-up in $z$. 

The second kind of simulations relies on inclusion of a multitude of parallel-antiparallel current sheets. In periodic settings one already has two current sheets available for studying the long time behavior until the reconnection-produced plasmoids start interacting. Usually at this time the simulation is stopped. However, in view of investigation of acceleration, plasmoid interaction attracts considerable interest, probably coming close to what happens in real nature in systems where many current sheets are generated. Such systems form behind shocks or also in turbulent plasmas. {This kind of acceleration with two antiparallel current layers has been intensely investigated by \citet{drake2006}. In the plasmoid interaction scenario \citep[see also][]{oka2010}, the electrons in the two closely adjacent current sheets, when moving along the magnetic field enclosing a plasmoid, start feeling the mirror field of the adjacent plasmoid which is produced in reconnection in the mirror current sheet. The resulting mirror field configuration causes trapping of electrons between the plasmoids in the two adjacent reconnection sites and strong particle acceleration in the combination of the two reconnection electric fields and also by the relative motions of the plasmoids, a process much simpler than acceleration in the chain of plasmoids of one single reconnection line and obviously leading to even stronger acceleration.}

A similar scenario was applied by  \citep{jaroschek2008} who extended the \citet{jaroschek2004a} simulations to the inclusion of up to 10 adjacent, fully relativistic current sheets \citep[][including also radiative losses]{jaroschek2009}. Their findings indicate that well defined power law ranges are produced in the acceleration of the electrons when interacting with the many current sheet reconnection sites. The distributions exhibit sections of power law shape  $f(\Gamma)\sim \Gamma^{-s}$ with power law index in the range of $3<s\lesssim4$ resulting from the two-dimensional simulations {(cf. Fig. \ref{fig-acc-jaro})}. The ultimate spectral cut-off is generated mainly by the particle losses from the simulation box. 

{In view of astrophysical application and generation of synchrotron emission, \citet{cerutti2013a,cerutti2013b} refer to electrons performing Speiser orbits across the reconnecting current layer in their two-dimensional simulations of ultra-relativistic pair plasma reconnection. This problem closely resembles that of \citet{jaroschek2004a,jaroschek2004b} where the cross current layer motion of the particles can also be seen as Speiser-like and comprehensive particle distributions have been used to determine the synchrotron emission spectrum with focus on gamma-ray afterglow. \citet{cerutti2013a,cerutti2013b} include, in addition, radiation damping through the radiation damping force in the equation of motion. Damping at high energies quenches the emission of synchrotron radiation.}

\subsubsection{Accleration of ions.}
{Proton acceleration comparable to the acceleration of electrons has not been observed during reconnection events in the magnetotail, as was explicitly stated by \citet{oieroset2002}. This lack of observations raises the question whether ions can become \emph{substantially} accelerated during reconnection. The problem of ion acceleration in current sheets and reconnection  has been addressed first by \citet{bulanov1976} and further explored by \citet{zelenyi1984}, \citet{burkhart1990}, \citet{divin2010} and others, who attributed acceleration to direct electric field acceleration along the X line in combination with quasi-adiabatic acceleration in the magnetic mirror geometry during reconnection under conservation of a (quasi-adiabatic) invariant in $B_z/B_x$ field geometry approximately conserved for particles which already possess a sufficiently high energy. \citet{zelenyi1984} argued that the ion energy that can be achieved is limited to the trapping time interval by escape of accelerated ions as a result of increasing gyroradius. The theory reproduces measurements by the Imp spacecraft of energetic ion spectra reaching well into the several 100 keV energies and being of power law shape and thus tentatively attributed to acceleration during reconnection. This and some other reconnection related mechanisms for ion acceleration has recently been (uncritically) compiled in a long review on acceleration by \citet{birn2012}. Still, the opinions about ion acceleration in reconnection, i.e. in direct relation to reconnection, are controversial. As pointed out above, \citet{oieroset2002} in their observations of reconnection related particle acceleration explicitly decline ion acceleration, which makes sense in view of our recent knowledge about the size of the electron diffusion region. Since ion acceleration requires already fairly high energy ions, it is unlikely that such ions will be trapped around the small electron diffusion region. This makes further ion acceleration unlikely, at least in a small one or few magnetic-islands long reconntion system like the magnetotail. However, in large systems of multi-current layers like the heliosheath or astrophysical systems the Bulanov-Zelenyi mechanism might well work. There, energetic ions may oscillate many times between different stochastically distributed islands and X lines and picking up energy in a stochastic process as has been discussed under extreme conditions \citep[][see below for a critical assessment]{drake2013}.}

{The question of ion acceleration remains to be of vital interest less for the magnetosphere and interplanetary space than in view of observed high energy cosmic ray spectra. These are believed to be generated in relativistic shock acceleration but suffer from the serious problem that shock acceleration requires the presence of a substantial amount of pre-accelerated high energy ion populations \citep[for a review of relativistic shocks and related acceleration the reader may consult][where  it had been suggested that post shock acceleration might include the interaction of many adjacent turbulent current sheets in a large domain possibly undergoing reconnection]{bykov2011}. In fact, ion acceleration near a single current sheet is probably weak for the reason that the plasmoids formed have no strong effect on the non-magnetized ion dynamics in thin collisionless current sheets, a strong argument against the mechanisms \citep{zelenyi2011,birn2012} discussed above. This might be different in fat sheets. In the thin sheet, ion-diffusion-region ions can become accelerated along an externally present sufficiently strong forcing convection field. In this case the ions gain energy over the distance of about twice the ion inertial length.} {This pick-up energy is of the order of $\sim2e|E_y|\lambda_i$.} {In the magnetotail this energy may reach at the very best a few 100 keV only, an energy irrelevant for cosmic rays and marginally relevant only for injection into the radiation belts which adiabatic acceleration during substorm and, for the innermost radiation belt, other nucleonic processes explain sufficiently well.} 

{The arguments that shocks may self-consistently produce their own seed population are widely spread among different models referring to shock surfing, shock ramp instabilities, so-called Bell instability and their variants, upstream plasma instability and turbulence and others more,  but are not completely convincing. Reconnection in the turbulent post-shock plasma has thus been proposed as a possible agent of generating the seed ion population.} 

{\citet{sironi2011} recently performed simulations of pair plasma acceleration \emph{behind ultra-relativistic shocks} including reconnection in a multitude of current layers in the post-shock flow.  Though simulations of this kind do not explicitly refer to ions, the results are nevertheless applicable to infer about ion acceleration for the reason that in ultra-relativistic shocks all particles move at about common velocity $v=c$, and at high Lorentz factors (energies) $\Gamma\sim10^3$ mass differences are less important. \citet{sironi2011} found that reconnection in the turbulent current sheets behind the shock may indeed provide a seed population of electron-positron pairs for further shock-acceleration if injected into the Fermi process. Reconnection in the post shock region seems to generate a \emph{very flat power law spectrum of pairs} close to the extreme marginally allowed slope $s\sim 1.5$ (implying complete inhibition of any heat fluxes), being much flatter than the slope found in the simulations \citep{jaroschek2004a,jaroschek2004b,jaroschek2008} of a single and many current layers. This spectrum does, however, not extend up to infinite energies but obeys a cut-off at $\Gamma_c=\Gamma_0\Sigma^{1/(2-s)}$ where $\Gamma_0$ is the initial Lorentz factor, and $\Sigma=\beta^{-1}_\mathit{tot}$, with $\beta_\mathit{tot}$ the total-plasma-to-magnetic energy ratio (including thermal and kinetic energies). Some of the accelerated energetic pair particles are capable to escape back into the upstream shock region where, after coupling to the flow and returning to the shock, they become additionally accelerated in the ordinary Fermi mechanism when bouncing back and forth across the shock. The final spectrum is found to obey a steeper slope of $s\sim 2.5$ (now naturally allowing for heat fluxes but ruling out higher moments) which extends to high energies of the order of cosmic rays.}

A non-relativistic statistical theory of particle bouncing between merging (interacting) plasmoids (magnetic islands) was recently developed by \citet{drake2013} in view of the role of reconnection in particle acceleration in the heliosheath behind the termination shock in order to explain the anomalous cosmic ray spectrum. \citet{drake2013} give arguments for an omnidirectional ion energy spectral distributions obeying the marginal extremely flat power law $f(\epsilon_i)\sim \epsilon_i^{-1.5}$ corresponding to nonrelativistic velocity distributions $f(v)\sim v^{-5}$ in presence of a self-consistent fire-hose criterion. This theory is based on complete suppression of   heat flux which observations seem to justify being applicable to the heliosphere and have been claimed to be typical for the anomalous cosmic ray component in interplanetary space.

Such extremely flat power law distributions form the high-energy tail of so-called $\kappa$-distributions which at this flatness categorically require absence of any heat flux in the collisionless system. They are the \emph{flattest power laws allowed} under extreme conditions \citep{treumann1999a,treumann1999b,treumann2008,yoon2012} by fundamental physical requirements (particle number and total energy conservation) independent of which mechanism causes the spectrum. They can thus not be taken as typical for acceleration by reconnecting current layers. Any mechanism which suppresses the heat flow will do. Distribution functions in other systems (solar flares, magnetosphere) obey much steeper energy distributions. The extreme power laws of the anomalous cosmic ray component might, however, indeed have been caused in acceleration in the highly extended turbulent heliosheath which is believed to consist of a chain of a multitude of current layers, presumably undergoing multiple reconnection. Here, any distribution of energetic particles interacting with a myriad of plasmoids might have sufficient time to evolve until reaching the final extremely flat power law shape. In any case investigation of particle acceleration is of vital importance not only for space but in particular for astrophysics. }

\section{Conclusions and Outlook}
The picture of collisionless reconnection is on the verge of gradually rounding up. During the last decade computational capacities have grown enormously enabling high resolution and large scale numerical simulations even in three dimensions \citep[for a collection of the  most recent results the reader is referred to the review by][]{karimabadi2013b}. In addition, the availability of ever more sophisticated spacecraft-borne instrumentation as well as multi-spacecraft experiments in near-Earth space, have produced large amounts of high resolution observations and data input allowing for gradually more reliable  tests of existing models and theories of reconnection. By now, it has been unambiguously confirmed that, at least in all space plasmas accessible to spacecraft, reconnection is collisionless. Collisionless reconnection turned out being one of the farthest reaching plasma processes causing magnetic re-configuration and relaxation. It causes conversion of magnetically stored energy into plasma heating and particle acceleration. It enables rapid mixing of plasmas in all collisionless regions where other diffusion models are completely ruled out. Here it provides the basic mechanism of dissipation of excess energy accumulated in complex magnetic configurations respectively electric currents and plasma flows, which are the two equivalent views of electromagnetic energy storage. 

Simulations, observations and measurements combined with numerical particle-in-cell simulations have unambiguously confirmed that any collisionless reconnection takes place in very thin current sheets the active region of which is on the electron inertial scale $|z|\sim\lambda_e\ll\!\!< L$. This scale is microscopic compared to all geometrical scales $L$ in space and presumably also in astrophysical plasmas. Collisionless reconnection is thus driven primarily by electron dynamics. Reconnection is rather a kinetic plasma than a fluid process, even though it can be formally attributed to the additional non-collisional terms occurring in the generalized Ohm's law which cause breaking the frozen-in condition. {\citep[][recently investigated transition from collisionless to collisional reconnection in terms of the Lundqvist number $S$ dependence on the ratio of the Sweet-Parker length $L_\mathit{SP}$ to ion gyroradius $r_i$, finding the Sweet-Parker range below $S\lesssim 10^4$ and, at larger $L_\mathit{SP}$, plasmoid mediated collisional MHD reconnection virtually extending the collisionless into the collision regime.]{daughton2012}} Of these terms the \emph{crucial} ones have been identified to be the \emph{divergence of the electron pressure tensor} term, $\nabla\cdot\mathsf{P}_e$ that, on the electron inertial scale, is taken in the geometrical frame of the current sheet where $\mathsf{P}_e$ possesses non-diagonal elements, as well as the \emph{convective electron-inertial} term, $\mathbf{V}_e\cdot\nabla\mathbf{V}_e,$ which accounts for the nonlinear gradient of the bulk electron speed along the electron flow, all taken in the ion frame. Both terms become large and complement each other around the X line in the electron diffusion region where they generate the reconnection electric field $E_\mathit{rec}\approx E_\|$. In three dimensions an \emph{electron-drag-term}, $-\langle \delta N\delta E\rangle$, in the total electric field including the convection and reconnection electric fields, must be added to these. Its contribution is due to fluctuations $\delta N$ in density-caused three-dimensional structures of the electron current sheets. The required electron pressure anisotropy is generated by driving the electron inflow, either self-consistently or forced from the outside by plasma inflow due to imposed convective electric fields. {This has been shown by simulations \citep{daughton2005,karimabadi2005a,karimabadi2005b} \emph{without artificially distorting} the current sheet letting it evolve self-consistently in the external forcing. Such conditions are naturally realized at the outer boundaries of the magnetosphere and in the geomagnetic tail plasma sheet. Similar processes are expected to work in the solar corona and in discontinuities in the solar wind.} 

Recognition of the micro-physical reasons in the pseudo-non-diagonal electron pressure tensor elements, inertial terms, and electron drag for breaking the frozen-in condition is most important. It resolves the long mystery of the absence of sufficiently large anomalous resistivities and dissipation in collisionless reconnection that was reflected in the astonishingly weak plasma wave activity found near the X line \citep[cf., e.g.,][among others]{bale2002,treumann1990}, a fact which troubled both observers and theoreticians for almost three decades. Very high plasma wave intensities were expected for long near the X point; these expectations had even been summarized in the notion of a reconnection ``fireball'' in the Earth's plasma sheet during substorms. Some residual anomalous dissipation might still be present at the reconnection site; being a byproduct of collisionless reconnection,  it is of little importance in the reconnection process itself, however, while causing several secondary effects. Heating of the plasma in reconnection is entirely due to the dissipation provided by the reconnection electric field. The production of anomalous collisions still remains an interesting problem in order to elucidate its wave and wave-particle-interaction origin, its spatial distribution, contribution to the spatial structure of the reconnection electric field, and its role in heating and particle acceleration. {In spite of this, it seems that the \emph{basic physics of collisionless reconnection} (and possible reconnection in general), i.e. the gross physical mechanism of collisionless reconnection, is by now very close to be understood. As it has turned out, this mechanism is far from the original fluid dynamical and magnetohydrodynamical picture which served as a guide over nearly four decades.} 

Aside from the two kinds of forced and non-forced reconnection, of which the first is probably more frequently realized in nature, an important role must be attributed to the presence of guide fields \citep[as suggested by][]{karimabadi1999}. Guide fields are naturally present in the oblique encounter of two magnetized plasmas of different properties and magnetic fields. Guide fields, being directed along the current in the separating current sheet, act magnetizing on the electrons. Since the original convection electric field is along the guide field, it accelerates the electrons along the guide field. This process is essentially independent of reconnection but amplifies the electron current which, when exceeding the Buneman threshold, excites electron holes in the collisionless current sheet, forming chains of localized, possibly very strong electric fields containing a dilute hot trapped electron component and being accompanied by very fast accelerated electron beams along the guide field. The role of such chains of holes is still widely unexplored. It is, however, suggestive to think of them as a plasma heating and accelerating facility as suggested by particle-in-cell simulations of the generation of electron holes \citep[cf., e.g.][]{newman2001}. In addition, many such holes cause a distinct structuring of the electron density and pressure in the current layer. They contribute to drag, pressure anisotropy and pseudo-viscosity and may, therefore, secondarily affect reconnection. Heating the plasma by holes temporarily switches the hole production rate off until further electron acceleration in the convection electric field re-accelerates them and reproduces a new chain of holes. In the mean time ion-acoustic waves are excited. Hence, the reconnection region in the presence of guide fields is filled with a population of holes propagating on the background of a weak ion-acoustic wave level. In addition, electron holes are a mechanism of cooling the non-trapped passing electron component, transforming it into a hole-cooled but very fast electron beam along the guide field. Such beams readily excite Langmuir turbulence which, in the dense high-$\beta$ current layer, generate second plasma-harmonic electromagnetic radiation. Observation of this radiation maps the reconnection site into remotely detectable radiation as, for instance, emitted from the solar corona during flares, a process of substantial interest in astrophysics but still unexplored in relation to collisionless reconnection. Similarly, electron holes entering the very low density electron exhaust may radiate  also in the electron-cyclotron maser mode if the plasma dilution inside the holes is strong enough for  the electron cyclotron frequency locally to exceed the electron plasma frequency. This radiation might become very intense but remains trapped around the X point, being unable to escape into free space. Such processes have also been identified at the separatrices in reconnection where electron acceleration takes place via Hall current closure and the presence of kinetic Alfv\'en and lower-hybrid-drift wave electric fields. These processes have been exhaustively investigated by simulations \citep[e.g.][and others]{daughton2004,daughton2011,ricci2005,karimabadi2013b}.  

Most of these insights have been generated by two-dimensional particle-in-cell simulations of collisionless reconnection. Sufficiently precise three-dimensional simulations involving large simulation boxes and sufficiently large particle numbers have become available only recently. The first results of such simulations showed that reconnection in three dimensions is very fast though more complicated than in two dimensions. The electron structure of the X line breaks off into several distinct microscopic layers by twisting the current, suggesting that very thin electron current layers are produced and the reconnection is highly structured in the third dimension. Such processes lead to multitudes of localized three-dimensional reconnection sites \citep[similar to those predicted by analytical kinetic theory][]{galeev1975a,galeev1975b} and warped magnetic flux tubes or ropes of their own dynamics. {It has been suggested that reconnection in this way may be the generator of mesoscale plasma turbulence \citep[most recently by huge-scale simulations][]{karimabadi2013b} in the sense that the electron current structure is microscopic on the scale of the X point region, while the flux tubes involved are on scales of the order of the ion inertial scale and larger, becoming macroscopic.} It should be noted here that many such narrow small warped current layers have been identified in Earth's magnetosheath \citep[][]{retino2007} claimed to be caused by the turbulence behind the bow shock and undergoing local reconnection which results in plasma heating. Reconnection in the turbulent magnetosheath has also been reported by \citet{phan2011}. 

{It is believed that reconnection between multiple adjacent antiparallel current layers is responsible for turbulence in the heliosheath and in many extended astrophysical systems obeying the formation of relativistic shocks. Reconnection provides a link to plasma turbulence in this case. Conversely it suggests that any cascading large scale plasma turbulence which produces small-scale magnetic structures and thus small-scale current vortices will necessarily be accompanied by reconnection on the small-scale end of the cascade in twisted and deformed electron-scale current sheets. The main dissipation mechanism on these scales will not have to be searched for in real or anomalous collisions but in turbulent convection-driven electron pseudo-viscosity \citep{hesse1999}, electron inertia and drag-excited collisionless reconnection. This reconnection then provides the natural energy sink for the dissipation of the turbulent energy that is stored in the macroscopic turbulent velocity field. In this respect the recent large-scale brute-force simulations \citep{karimabadi2013a,karimabadi2013b,roytershteyn2013} are of vital not to underestimate importance.}

Pre-existsing meso-scale or MHD turbulence should also affect the evolution of reconnection. The various thoughts on this problem have recently been summarized by \citet{karimabadi2013c}. It might both reduce and increase the global reconnection rate. Several models have been proposed referring to turbulent current broadening and plasmoid interaction, respectively. In the latter case the diffusion layer shrinks under the action of the turbulent plasmoids. Currently no definitive conclusion can be given.Turbulence in three dimensions exhibits chaotic  properties due to field line wandering which results in production of small-scale current layers and flashes of chaoticity-induced reconnection. In this case the collisionless reconnection rate is high in every single reconnection flash while the global volume-averaged reconnection rate will be much less, possibly becoming comparable to the ordinary diffusion rate. Turbulent vortices may, on the other hand, compress a fat current sheet and ignite reconnection \citep{nakamura2013}. Based on a review of available particle simulations, \citet{daughton2012} recently developed a phase diagram for the transition from collisional to collisionless reconnection identifying a large region of plasmoid induced quasi-collisionless reconnection in the collisional regime which arises just due to the compression effect. 

The expected wide distribution of collisionless reconnection in space, and probably allover in plasmas in the universe, poses the question whether the notion  of collisional reconnection makes any sense if collisionless reconnection proceeds on so fast a temporal scale and in so small regions with so high efficiency as suggested by observations and simulations. In fact, if the requirement is held up that, being restricted to current sub-sheets of width of the electron inertial scale, collisional reconnection can compete only if the collisional mean free path $\lambda_\mathit{mfp}\sim\lambda_e$ becomes comparable to the electron inertial length. It is easily shown that this condition imposes a limit on the electron temperature given by $T_e/m_ec^2\lesssim 1/\sqrt{N\lambda_e}$ which for realistic densities $N>10^5$ m$^{-3}$ in space can be satisfied only at almost zero electron temperature.  Relaxing this condition to the ion inertial  length  $\lambda_\mathit{mfp}\sim\lambda_i$ does not help. By this argument, collisional reconnection seems to be irrelevant. However, the argument is based on the assumption that the generalized Ohm's law holds also in highly collisional, i.e. in dense media, which is not given a priori. If all terms in Ohm's law are negligible against the resistive term, reconnection will be dominated completely by collisions and will proceed just on the extra-ordinarily slow classical diffusive time scales under stationary conditions evolving into long Sweet-Parker dissipation regions. It may be speeded up only by turbulent correlations in the conducting fluid. This may be realized under large-scale high density astrophysical conditions.   

Collisionless reconnection has been found an efficient accelerator of particles. In particular, the multiple reconnection in either of the above scenarios is a way of accelerating electrons to high energies. This acceleration is provided by the parallel electric fields which maximize along the separatrices. Another mechanism is provided by the interaction of multiple reconnecting current sheets and in the presence of plasmoids. Trapping between closely spaced large sized plasmoids also accelerates ions. Much insight is expected from future performing of three-dimensional particle-in-cell simulations in newly available large set ups. 

{Summarizing, collisionless reconnection has not only become an indisputable reality, it also has matured to an intermediate state where we begin to understand its mechanism. The following facts are noticeable: In spite of previous reservations based on the apparent weakness of dissipative processes in high temperature dilute collisionless plasmas, collisionless reconnection is a fast process proceeding on the electron-Alfv\'en time-scale. It takes place deep inside thin current sheets of widths well below the ion inertial length $\lambda_i$ inside the narrow electron diffusion layer $|z|\lesssim\lambda_e$ centered at the current sheet. It is unclear whether thicker current sheets $|z|>\lambda_i$ at all undergo reconnection {\citep[see also the discussion by][]{daughton2012,karimabadi2013c}.} The opinions are diverging on this point. Under forcing, thick current sheets experience compression until their thickness locally drops below $\lambda_e$ for becoming thin enough for reconnection to set on. On the other hand belief is going on that large-scale turbulence provides sufficient driving of large (MHD) scale reconnection. This is probably vindicated by recent three-dimensional simulations as we have discussed above. Presumably reconnection sets on once the turbulent cascade  reaches down into the small reconnection scales. Concerning the mechanism of reconnection, current knowledge, strongly supported by two-and three-dimensional particle-in-cell simulations, suggests that the primary reason for onset of collisionless reconnection is the divergence of the electron pseudo-viscosity in the non-diagonal electron pressure tensor elements \citep{hesse1999}, preferentially in anisotropic pressure conditions. This is aided by small-scale convective electron inertia. Other non-linearities also contribute, but anomalous collisions can probably be ruled out because of the lacking wave intensity in all known cases, even in forced and guide-field reconnection.  Lower-hybrid-drift waves, the most viable candidate, contribute as well though not via anomalous dissipation but through their electric fields. The most interesting region is the electron diffusion region. Simulations confirmed that  it is highly asymmetrical consisting of a short inner electron current region where pressure and inertia driven dissipation takes place, and a laterally extended electron jetting (or exhaust) region which is the outflow of electron from the reconnection site caused by the reconnection electric field and magnetic stresses on the weakly re-magnetized electrons. The vertical extension of the exhaust is narrow, but its length reaches several ion inertial lengths. Exhaust electrons flow at super-Alfv\'enic speed until ions become involved and the electron flow slows down to merge into the macroscopic reconnection plasma jets. Reconnection under various different conditions is accompanied by a number of most interesting secondary effects which, however, are not vital for the ignition of reconnection. These are quadrupolar Hall fields generated in the ion diffusion region under symmetric (e.g. in the magnetotail) and dipolar Hall fields generated on the weak field-high density side under asymmetric (e.g. at the magnetopause) external conditions. Closure of Hall fields via field-aligned currents along the separatrices, various kinds of wave instabilities near separatrices and also at the boundaries of the electron diffusion region. The presence of guide fields introduces further obliqueness and twisting of the Hall fields causing asymmetry in reconnection configurations. Guide fields allow for magnetization of electrons. In this way they cause delay of reconnection onset. On the other hand, they stimulate reconnection. Inclusion of normal magnetic field components, which are naturally present in closed magnetic configurations, introduce further variations like the formation of dipolarization fronts, additional particle acceleration, formation of secondary X lines and other side effects. In three dimensions collisionless reconnection exploits the freedom of the new dimension by the tearing mode becoming oblique, depending on the strength of the guide magnetic field. Plasmoids now evolve into twisted flux ropes of complicated three-dimensional structure, while the electron dissipation region becomes multiply structured and shorter than in two dimensions. Moreover, other wave modes can be excited in this case. Finally, large scale three-dimensional simulations demonstrate that there is a close relation between plasma turbulence and collisionless reconnection \citep{karimabadi2013b}. The latter provides the microscopic scales on which dissipation of the turbulent energy takes place after the turbulence has cascaded down to form electron scale-narrow current sheets which reconnect and terminate cascading and further formation of smaller scale structures. This connection has turned out to become one of the most promising fields of contemporary research with probably far reaching consequences.}




\begin{acknowledgements}
This research was part of a Visiting Scientist Program at ISSI, Bern, executed by RT. Hospitality of the ISSI staff is thankfully acknowledged. 
\end{acknowledgements}










\end{document}